\documentclass[a4paper,pre,superscriptaddress,floatfix,nofootinbib,twocolumn]{revtex4-1}

\usepackage{bbold}
\usepackage{bbm}
\usepackage[pdftex]{graphicx}
\usepackage{latexsym,amsmath,verbatim,amssymb,txfonts, mathtools}
\usepackage{color}
\usepackage{rotating}
\usepackage{verbatim}
\usepackage{multirow}
\usepackage[english]{babel}
\usepackage{comment}
\usepackage{hyperref}
\usepackage[normalem]{ulem}
\usepackage{tabularx}

\begin{document}
\title{Modeling resource consumption in the US air transportation system via minimum-cost percolation}

\author{Minsuk Kim}
\affiliation{Center for Complex Networks and Systems Research, Luddy School
  of Informatics, Computing, and Engineering, Indiana University, Bloomington,
  Indiana 47408, USA}
\author{C. Tyler Diggans}
\affiliation{Air Force Research Laboratory's Information Directorate, Rome, NY 13441}
\author{Filippo Radicchi}
\affiliation{Center for Complex Networks and Systems Research, Luddy School
  of Informatics, Computing, and Engineering, Indiana University, Bloomington,
  Indiana 47408, USA}
  \email{f.radicchi@gmail.com}

\begin{abstract}
We introduce a dynamic percolation model aimed at describing the consumption, and eventual exhaustion, of resources in transportation networks. In the model, rational agents progressively consume the edges of a network along demanded minimum-cost paths. As a result, the network undergoes a transition between a percolating phase where it can properly serve demand to a non-percolating phase where demand can no longer be supplied.  We apply the model to a 
weighted, directed, temporal, multi-layer network representation of the air transportation system that can be generated using real schedules of commercial flights operated by US carriers. We study how cooperation among different carriers could improve the ability of the overall air transportation system in serving the demand of passengers, finding that unrestricted cooperation could lead to a $30\%$ efficiency increase compared to the non-cooperative scenario. 
Cooperation would require major airlines to share a significant portion of their market, but it would allow also for
an increased robustness of the system against perturbations causing flight cancellations.
Our findings underscore some key benefits that could emerge by simply promoting code-share arrangements among US airlines without altering their current 
cost of operation.
\end{abstract}


\maketitle

\section*{Introduction}                                                                                                                                                     
The air transportation industry represents one of the major sectors of the US
economy. In $2018$,  civil aviation accounted for $\$1.8$ trillion
in total activity, contributed $5.2\%$ to the gross
domestic product, and supported more than $10.9$ million
jobs~\cite{federal2020economic}.
Still according to $2018$ data, the industry enabled the transport of
nearly $900$ million
passengers on flights covering more than $6$ billion
vehicle miles~\cite{pocket2018}.
The impact of the air transport industry, like so many other areas of the
national economy, was adversely affected by the COVID-19 pandemic, however,         
recent reports indicate that the industry went back to
its pre-pandemic levels~\cite{pocket2023}.


The infrastructure that sustains air transportation is composed of a large number of different
elements, interacting and working together in a
complex manner~\cite{barnhart2003applications}. Individual components
of the system have been studied for a long time, including
flight scheduling~\cite{lohatepanont2004airline}, fleet assignment~\cite{barnhart1998flight, jacobs2008incorporating},
aircraft maintenance routing~\cite{barnhart1998flight}, crew scheduling~\cite{vance1997airline},
management (e.g., revenue~\cite{smith1992yield, belobaba1987air}, irregular operations~\cite{ball2007air}, airside
operations~\cite{hall1999information}, and air traffic flow~\cite{odoni1987flow}).
The most studied of these components seems to be the air transportation network (ATN), sometimes referred
to as the airline network. In the ATN, airports are nodes, and an edge between two
nodes exists whenever there is a direct flight between the two
airports. 
Many of the pioneering theoretical works on the structure of the ATN focused on comparisons between point-to-point and 
hub-and-spoke network configurations~\cite{o1994hub, berry1996airline, cook2008airline}.  More recent works have provided network analysis of real-world data-generated ATNs, e.g., ~\cite{guimera2005worldwide, barrat2004}.
Real ATN structures share common attributes with many other types of real networks: 
they exhibit a broad degree distribution~\cite{barabasi1999emergence},
short average pair-wise distance~\cite{watts1998collective}, large values of the
clustering coefficient~\cite{watts1998collective},
and a strongly modular structure~\cite{fortunato2010community}.

The basic binary representation of edges in ATNs can be enriched in several ways. For example, it is possible to
associate weights to the links depending on the number
of passengers traveling between two airports~\cite{barrat2004}. Also,
one can adopt a multiplex representation of the
network~\cite{boccaletti2014structure, kivela2014multilayer}, with each
layer representing a different air carrier serving different routes
among a common set of airports~\cite{cardillo2013modeling, radicchi2015percolation}.
Finally, the dynamic nature of flight schedules can be used to perform various
types of temporal analyses~\cite{rocha2017dynamics}. For example, several studies
have focused on the evolution of the structure of ATNs over time~\cite{berechman1996analysis, goetz1997geography, button2002debunking,                      
  rey2003structural, burghouwt2001evolution,                                                                                                          
jin2004geographic, burghouwt2005temporal, malighetti2008connectivity,                                                                                  
wang2007china, da2009structural}. 
Dynamical changes of ATNs over short time scales have been considered in the definitions of
time-dependent path lengths, correlations, and centrality
metrics~\cite{zanin2009dynamics, pan2011path}.
Further, time-series analyses of flight data can be used to
study the propagation of flight delays in an ATN~\cite{fleurquin2013systemic}.
Data about the flow of people on ATNs are also useful in many applications, for instance, since the seminal work by Colizza {\it et al.}~\cite{colizza2006role}, passenger data are systematically included in
metapopulation models for epidemic
spreading~\cite{balcan2009multiscale, balcan2011phase,                                                                                                                 
brockmann2013hidden, pastor2015epidemic}.
Further, passenger flow data on ATNs have been studied in papers
focusing on Markovian processes with memory
on networks, leading to suitably adapted centrality
metrics and community detection algorithms~\cite{rosvall2014memory, scholtes2017network,                                                                               
  peixoto2017modelling}.

A large body of literature focuses on the application 
of methods from percolation theory
to ATNs, aiming at characterizing their robustness, where the robustness of a network is defined as its ability  
to remain connected as its nodes or edges
are removed from the system; this being the straight forward adaptation
of percolation to
graphs~\cite{stauffer2014introduction, newman2018networks}. In the context of ATNs, node removal
is used to mimic failure or shutdown (e.g., due to
weather) of airports, and the relative size of the largest
connected component in the graph is used as a proxy
of infrastructure function. These type of percolation models can be studied under either the
isolated or the interdependent network frameworks, and network  robustness is measured by looking at how the
giant component of the network shrinks as a function
of the fraction of nodes removed from the system. Different
strategies for the removal of nodes can be used to model disruptive
situations, but the most studied model is certainly the
one of ordinary percolation where removed nodes are
randomly selected at uniform~\cite{callaway2000network, albert2000error, karrer2014percolation,                                                                                    
  newman2018networks, buldyrev2010catastrophic, radicchi2015percolation}.
In contrast, scenarios of maximal stress are simulated by optimizing the percolation process, thus selecting the nodes that lead to the
quickest dismantling of the network~\cite{shen2012exact,                                                                                                                 
shen2012polynomial, morone2015influence,                                                                                                               
braunstein2016network, clusella2016immunization, osat2017optimal}.

In this paper, we introduce a percolation-based framework to study a rather different aspect
of ATNs: their effectiveness in meeting the demand of
passengers. As a pillar of the framework, we develop the minimum-cost-percolation (MCP) model, 
which is the natural extension of the shortest-path percolation model, recently considered in Ref.~\cite{kim2024shortest}, to weighted, directed, temporal, and multi-layer networks. 
The MCP model mimics how the resources of an infrastructural 
network are consumed, and eventually depleted, by demanding agents~\cite{stauffer2014introduction}. 
We apply the MCP model to data-generated 
representations of the US ATN. The framework enables analysis of how the US air transportation system 
would adapt and perform under various scenarios. For example, two hypothetical scenarios are considered: (i) a scenario of full cooperation where airlines cooperate by 
supplying passengers with multi-carrier itineraries and (ii) a scenario of no cooperation,
where airlines operate independently. These two hypothetical scenarios are contrasted against a realistic, data-inferred scenario of partial cooperation, where some major airlines form 
commercial alliances with minor airlines, and it is found that the scenario of full cooperation could lead to a $31\%$ increased ability of the
ATN to serve demand compared to the non-cooperative setting, and to a $3\%$ increase with respect to the
scenario of partial cooperation. Such improvements appear to be due to the cooperative ATN connecting 
pairs of high-demand airports that may be under-served in the other two scenarios. Of course, cooperation would require
major airlines to share some portion of their market, but with no need of changing their actual schedules.
Also, we consider scenarios of service disruption, for example, by suppressing all flights operated by specific air carriers.
We find that the system would still be able to properly serve demand irrespective of the level of cooperation among air carriers, however,
full cooperation among air carriers would dramatically increase the ability of the network to react to the malfunction of some of its components.
We explicitly consider the hypothetical case where all flights of {\it Delta Air Lines} would be suppressed, finding that full cooperation would 
lead to a $4\%$ and $33\%$ improved ability to supply demand compared to the scenarios of partial or no cooperation, respectively.
Also, we consider the effect that a $50\%$ reduction in the number of aircraft would have on the system, finding that the full-cooperation scenario 
would still correspond to a relative ability of the system to serve demand respectively
$5\%$ and $46\%$ higher than those valid under the two other scenarios of cooperation.




\section*{Results}

\subsection*{Construction of the framework}

\begin{figure*}[!htb]
    \includegraphics[width = 0.7\textwidth]{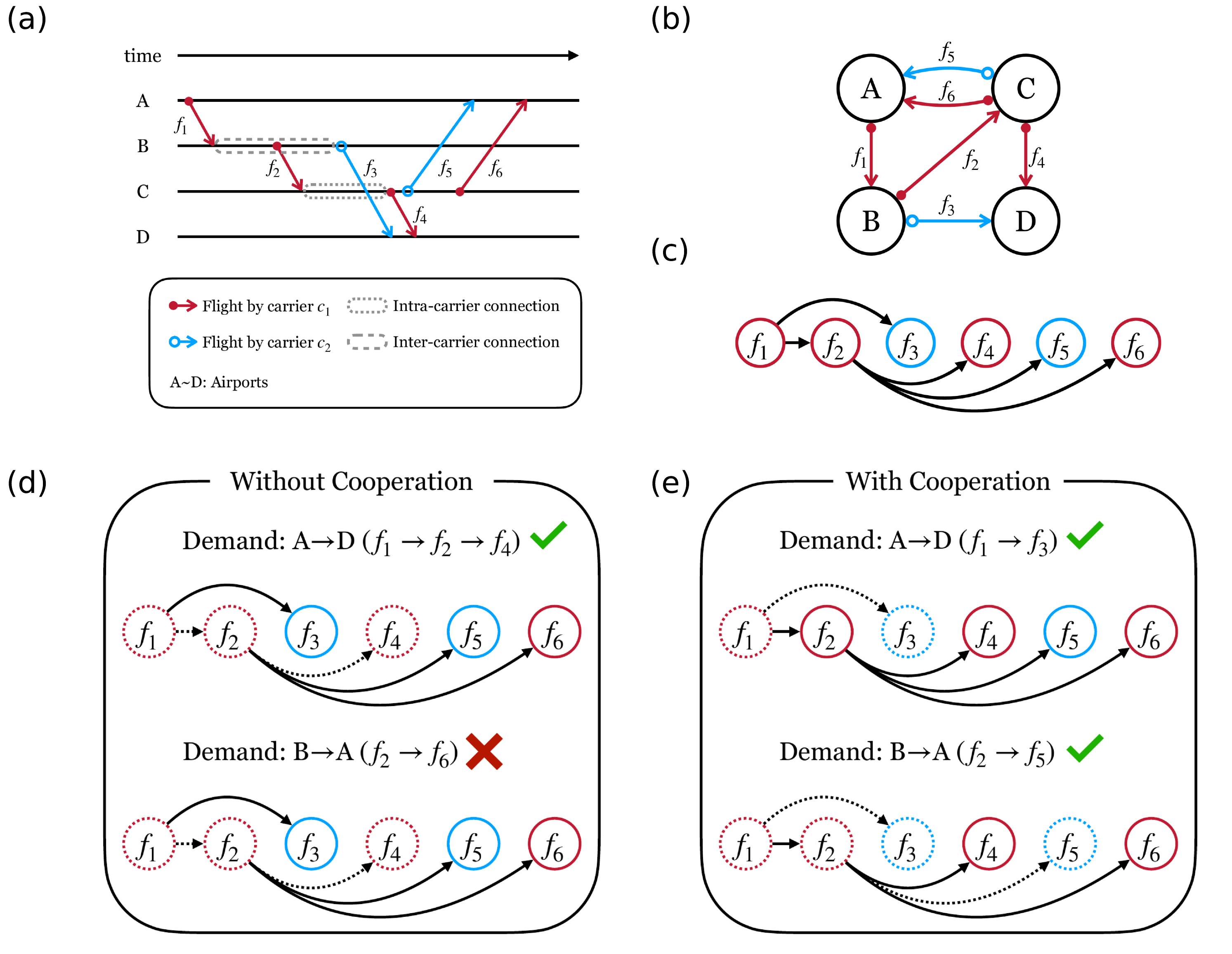}
    \caption{\textbf{Multi-carrier flight-connection network and minimum-cost-percolation model.} {\bf (a)} Schematic diagram of a flight schedule between airports $A, B, C$ and $D$. The schedule is composed of flights $f_1, f_2, f_4$ and $f_6$ operated by carrier $c_1$ and flights $f_3$ and $f_5$ operated by carrier $c_2$. To avoid overcrowding the figure, we highlight only the intra-carrier connection between $f_2$ and $f_4$ and inter-carrier connection between $f_1$ and $f_3$. {\bf (b)} Traditional representation of the airline network derived from (a), where nodes are airports, and directed and weighted edges are flights. {\bf (c)} Multi-carrier flight-connection network derived from (a), where nodes are flights and directed edges denote available connections between flights. {\bf (d)} An illustrative sample application of the minimum-cost-percolation model to the flight schedule of (a) where, for simplicity, each flight has a capacity of only one seat, and the cost function optimized by the agents is taken to be the itinerary duration. Suppose that the demanded itineraries are $A \to D$ and $B \to A$. If cooperation among carriers is not allowed, then the minimum-cost itinerary $A \to D$ is served as $f_1 \to f_2\to f_4$. Having no more available seats, those flights are removed from the flight-connection network and represented as dashed circles; also, all connections of these flights to other flights are removed from the graph. As a result, the second itinerary $B \to A$ can not be supplied. {\bf (e)} Same as in (d), but allowing for cooperation. In this case, both the demanded itineraries can be supplied:  $A \to D$ is served as $f_1 \to f_3$ and $B \to A$ is served as $f_2\to f_5$.
    } 
    \label{fig:1}
\end{figure*}

The minimum-cost-percolation (MCP) model mimics the dynamics of individual agents consuming resources supplied by a transportation network, see Methods for a detailed description and Figure~\ref{fig:1} for an illustration. 

The MCP model requires three main inputs: the topology of the network, the demand of the agents, and the cost function used to evaluate itineraries served by the network. At each stage of the dynamical percolation process, one pair of origin-destination nodes $o \to d$ is selected at random, but in proportion to its weight from the demand set; this represents a desire of the corresponding agent to move in the network along one of the minimum-cost (e.g., fastest, shortest) itineraries that connects $o$ to $d$. If such a path is available to the agent, then the path is supplied, and the capacity of all edges in the supplied path is reduced. If the capacity of an edge is exhausted, then the edge is no longer available, i.e., it is removed from the network, and that resource will no longer be available to the remaining agents. Initially, when no resources are used, the network is in the so-called percolating phase, where the generic agent can find an available path between the nodes $o$ and $d$. However, as agents are supplied with paths, meaning that edges' capacities are progressively reduced and 
edges are ultimately removed from the network, the graph eventually fragments into multiple connected components, displaying a transition to a non-percolating phase, where agents can no longer be supplied with a path that meets their demand. 
In this phase, the resources of the network are considered exhausted. Natural observables to monitor the above-described transition are inspired by percolation theory, for example, the fraction of satisfied agents, which is analogous to the percolation strength~\cite{stauffer2014introduction}.

To explore this framework in the context of ATNs, we feed the MCP model with networks that are generated using freely available data concerning the schedule of commercial flights operated by US air carriers~\cite{bts, faa}, see Methods for details and Figure~\ref{fig:1} for an illustration. For all results reported in the main portion of this paper, we consider flights operated on April 18, 2023, however, in the SM we also report results for daily schedules for April 18, 2019 and November 22, 2023. The two days chosen in April are just regular weekdays (Tuesday and Thursday, respectively); we purposely consider one day before and the other day after the COVID-19 pandemic. The selected day in November is instead the Wednesday immediately before the 2023 Thanksgiving day; this is supposed to be one of the busiest day in the year for the US air transportation system. In all cases, we focus only on flights operated between airports in the contiguous US. Also, for convenience, we represent the infrastructure in terms of flight-connection networks (FCNs). 

In a FCN, flights are nodes, and two flights are connected by a directed edge if they can be used in sequence along an itinerary. This condition is met if the destination airport of the first flight is the same as the departure airport of the second flight, and if the difference between the departure time of the second flight and the arrival time of the first flight is sufficiently large. We opt for a multi-layer representation of the FCN, where each layer contains all flights operated by an air carrier~\cite{cardillo2013modeling}; intra-layer connections are always present if the time gap between two flights exceeds the value of the input parameter $\delta \geq 0$. Additional inter-layer connections are established depending on the scenario of cooperation at hand. In the scenario of no cooperation, inter-layer connections are not allowed at all. If partial cooperation is present, then inter-layer connections among flights operated by commercial partners are allowed. These connections are present whenever their time gap is larger than $\delta$. In the scenario of full cooperation, inter-layer connections are allowed also among air carriers with no partnership; these connections are established if the time gap between two flights is larger than the input parameter $\epsilon \geq \delta$. By construction, the set of edges present in the FCN of full cooperation is a superset of the set of edges present in the FCN of partial cooperation, which is a superset of the set of edges for the FCN with no cooperation. 
In Figure~\ref{fig_sm:fcn-statistics}, we plot the total number of edges that are present in the FCN as a function of the model parameters. As expected, the overall number of edges is minimal when cooperation is not allowed; also, the number of edges decreases as $\delta$ and $\epsilon$ increase. The non-trivial finding is that roughly $70\%$ of the edges are between layers for a rather wide range of values for the parameters $\epsilon$ and $\delta$ when full cooperation is allowed. By contrast, for partial cooperation, only $40\%$ of the connections are among different air carriers.

The scenario of partial cooperation appears to be the one currently adopted in the US market, where air carriers form alliances/partnerships; the scenarios of no cooperation and full cooperation are instead hypothetical scenarios that can be studied using our framework. In all our results reported in the main paper, we set $\delta = 30$ minutes and $\epsilon = 60$ minutes. A minimum gap of $30$ minutes between same-carrier/alliance flights appears to be a realistic setting; for cross-carrier/alliance connections, doubling the minimum time gap seems reasonable to properly account for the additional time required for luggage transfer and terminal changes. Notice that the condition $\epsilon \geq \delta$ {\it de facto} means that changing carrier/alliance during an itinerary incurs a cost, here for simplicity measured in time. Additional results for other choices of the parameters $\delta$ and $\epsilon$ are reported in the SM (See Figures~\ref{fig_sm:fig2-different-epsilon}, \ref{fig_sm:fig3-different-epsilon}, \ref{fig_sm:fig4-different-epsilon}, and \ref{fig_sm:fig5-different-epsilon}).

The FCN can be seen as an adaptation of the so-called vehicle-sharing network, introduced in Ref.~\cite{vazifeh2018addressing} for the study of the minimum-fleet problem in urban mobility, to the MCP framework. A fundamental difference is that the vehicle-sharing network of Ref.~\cite{vazifeh2018addressing} is a representation of the supplied demand (trips represent demanded and supplied itineraries, and a connection between two trips stands for the possibility of the same vehicle to supply both of the itineraries) rather than the infrastructural graph as in our case. Also here, we allow for a multi-layer representation of the system that is absent in  Ref.~\cite{vazifeh2018addressing}.

\subsection*{Validation of the framework}

\begin{figure}[!htb]
    \includegraphics[width = 0.48\textwidth]{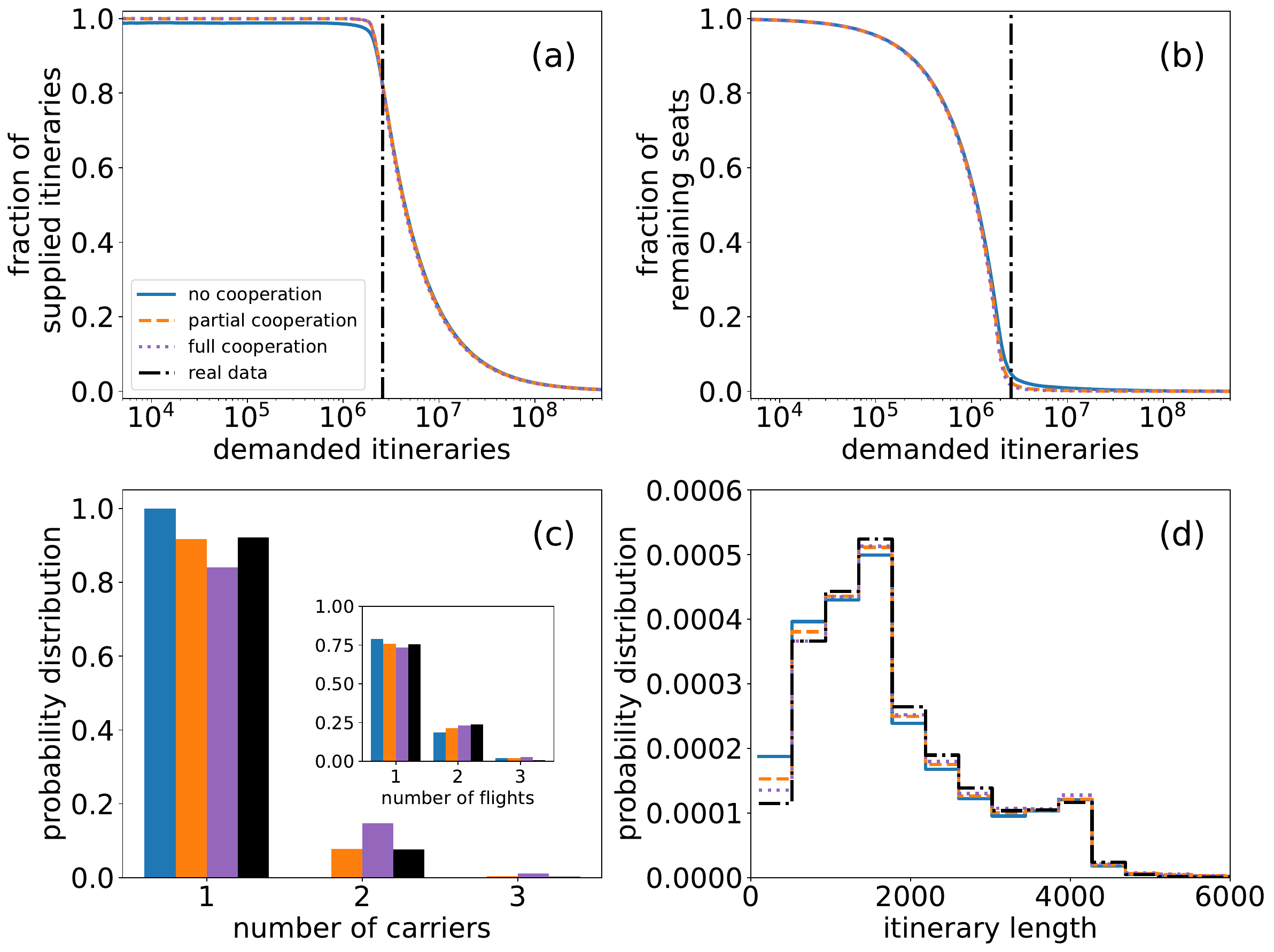}
    \caption{{\bf Validation of the minimum-cost-percolation (MCP) model.}
    We consider all flights operated between pairs of airports in the contiguous United States on April 18, 2023 and estimate demand using data about sold tickets in the second quarter of April 2023.  Results are valid for the MCP model where the cost of an itinerary is given by its length. We consider flight-connection networks under the scenarios of no cooperation (blue solid line), partial cooperation (orange dashed line) and full cooperation (purple dotted line). The minimum gap between connecting flights is $\delta = 30$ minutes if the flights are operated by the same carrier/alliance, and $\epsilon = 60$ minutes if operated by different carriers/alliances. Whenever possible, we compare results of the MCP model with real data (black dashed-dotted line). {\bf (a)} Fraction of supplied itineraries and {\bf(b)} fraction of remaining seats as functions of the raw number of demanded itineraries. The vertical lines denote the estimated number of daily served passengers in April 2023~\cite{april23}. 
    {\bf (c)} Fraction of supplied itineraries composed of flights with given number of operating carriers. The inset displays the distribution of supplied itineraries
    composed by a given number of connecting flights. {\bf (d)} Distribution of the length (measured in km) of the supplied itineraries.
    }
    \label{fig:2}
\end{figure}

We first use available data concerning the actual tickets sold by airline companies in the second quarter of year 2023 as the input demand for the MCP model, see Methods for details.
This type of data represents supply but has been used before as a proxy for demand~\cite{joshi2022equitable}. Figure~\ref{fig:2} summarizes the results for the MCP model given this ``supplied'' demand where the cost function is taken to be the length of the itineraries.
It appears that the infrastructure is able to adequately supply all of the first $2.5 \times 10^6$ demanded itineraries, but after that, a sharp transition occurs, both in the fraction of supplied demand [Figure~\ref{fig:2}(a)] and in the fraction of remaining seats [Figure~\ref{fig:2}(b)]. Further, this result is insensitive to the level of cooperation enabled in the FCN; this holds also for other cost functions (Figures~\ref{fig_sm:fcn-validation-duration-Y2023M4D18} and~\ref{fig_sm:fcn-validation-seats-Y2023M4D18}), 
and for FCNs representing different daily schedules (Figures~\ref{fig_sm:fcn-validation-length-Y2019M4D18} and~\ref{fig_sm:fcn-validation-length-Y2023M11D22}).

These findings are not surprising since the input used for demand is by definition the demand that was effectively served by the infrastructure; the transition occurs when $2.5 \times 10^6$ demanded itineraries are served because this was roughly the number of served passengers on a daily basis in April 2023~\cite{april23}. Similarly, the non-noticeable difference between the various scenarios of cooperation is due to the fact that most of the itineraries are served by individual carriers [Figure~\ref{fig:2}(c)], and so any demand that is not satisfiable by the current infrastructure would not be present in the set of sold tickets. In short, using information of sold tickets to proxy demand is useless for testing the optimality of the infrastructure in supplying the true demand, even more so if the testing includes a comparison between different scenarios of cooperation.

On the positive side, the results reported in Figure~\ref{fig:2} tell us that the MCP model is sufficiently accurate in describing the dynamics of resource consumption in the US air transportation system. In fact, the MCP model is agnostic to anything that regards the real market of airline tickets. Nonetheless, it captures almost perfectly some key statistical properties of the sold itineraries. The distributions of the number of connecting flights [inset of Figure~\ref{fig:2}(c)] and total length [Figure~\ref{fig:2}(d)] of supplied itineraries are almost identical; if only partial cooperation is allowed, then the distribution of the number of operating carriers per itinerary obtained by the MCP model also matches the one computed from real data well [Figure~\ref{fig:2}(c)]. According to the MCP model, supply can be sufficiently well predicted from demand using a Poisson model, see Figure~\ref{fig_sm:supply-demand-DB1B-distance-Y2023M4D18}; also, the demands that are supplied under the various regimes of cooperation are all well correlated one to the other, see Figure~\ref{fig_sm:scatter-plots-DB1B-distance-Y2023M4D18}.
The same qualitative results hold if other cost functions are considered in the MCP model, see  Figures~\ref{fig_sm:supply-demand-DB1B-time-Y2023M4D18}, ~\ref{fig_sm:supply-demand-DB1B-seats-Y2023M4D18}, ~\ref{fig_sm:scatter-plots-DB1B-time-Y2023M4D18}, ~\ref{fig_sm:scatter-plots-DB1B-seats-Y2023M4D18}.

\subsection*{Estimation of the effectiveness of the air transportation system}

We also consider the demand that is generated according to the gravity model of human mobility~\cite{zipf1946p, erlander1990gravity}. This model assumes that the
demand between two locations is proportional to their respective populations, but inversely proportional to their geographical distance. As detailed in the Methods section, to apply this model
in our framework, we leverage additional free data from the NASA's Socioeconomic Data and Applications Center~\cite{GPWv4_2018} and make use of the DBSCAN algorithm~\cite{ester1996density} to
group together airports serving the same metropolitan areas. The other components of the MCP framework (the FCN and cost functions) are identical to those considered previously. 

\begin{figure*}[!htb]
    \includegraphics[width = 0.85\textwidth]{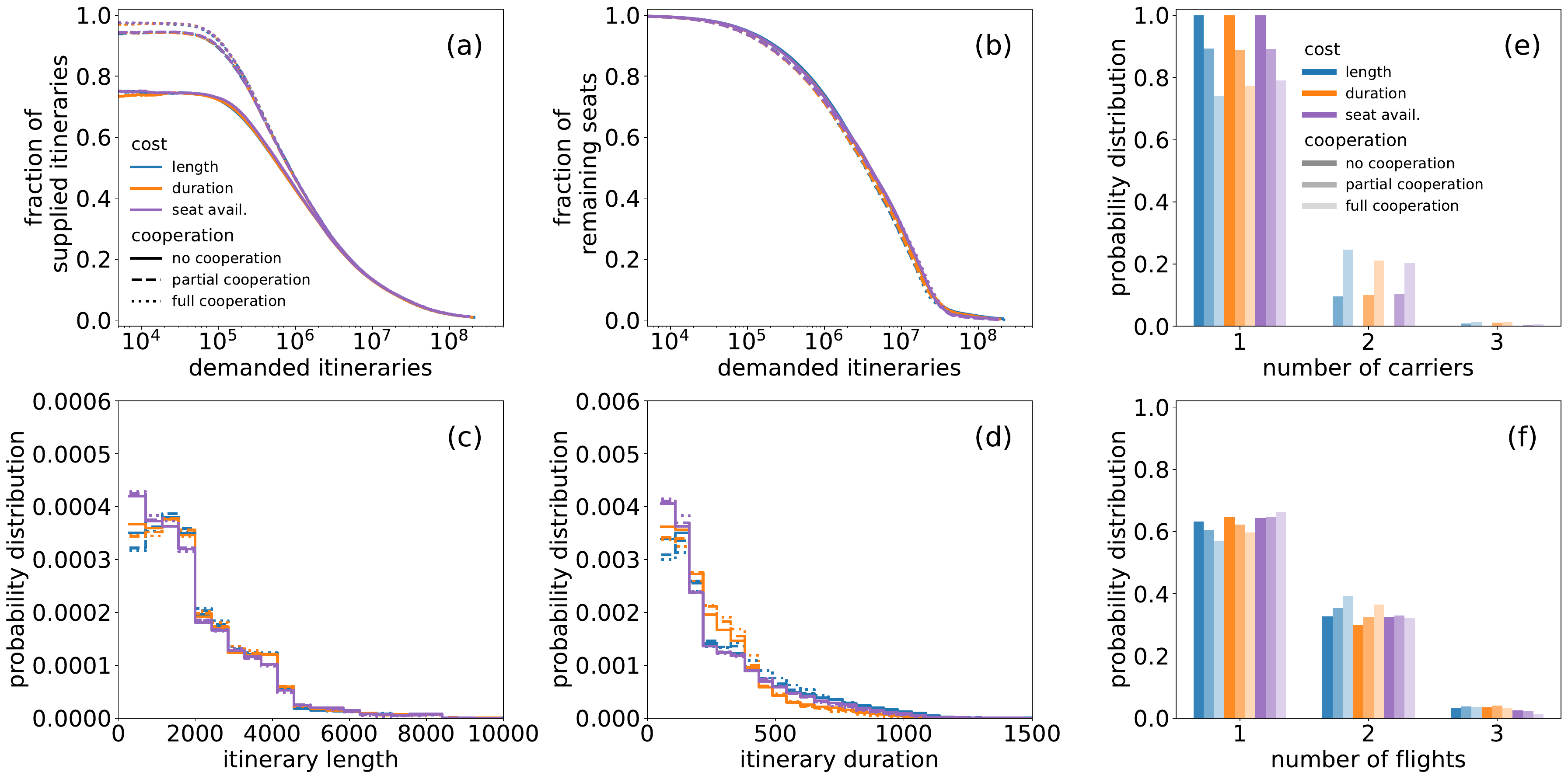}
    \caption{{\bf Cooperation among carriers and performance of the air transportation system.}
    The minimum-cost-percolation model is applied to the flight schedule of April 18, 2023. Here, we proxy demand using the gravity model with parameters $\alpha = \beta = 0.5$ and $\gamma = 1.0$, see Eq.~(\ref{eq:gravity}). The flight-connection network is generated either allowing for no cooperation (solid), partial cooperation (dashed), or full cooperation (dotted). The minimum gap between connecting flights is $\delta = 30$ minutes if the flights are operated by the same carrier/alliance, and $\epsilon = 60$ minutes if operated by different carriers/alliances.  Results for the {\bf (a)} fraction of supplied itineraries and {\bf (b)} fraction of remaining seats as functions of demanded itineraries are included for differing cost functions: length (blue), duration (orange) and seat availability (purple). {\bf (c)} Distributions of the length and {\bf (d)} duration of supplied itineraries. Distributions of {\bf (e)} number of operating carriers and {\bf (f)} number of connecting flights in the supplied itineraries. We use solid colors for no cooperation, partially transparent for partial cooperation, and  light colors for full cooperation.  
    }
    \label{fig:3}
\end{figure*}

Results of the MCP analysis for the gravity-model demand are reported in Figure~\ref{fig:3}, where we note that differences with the results of Figure~\ref{fig:2} are apparent. First of all, there are clear gaps in the ability of the 
network to serve the demand when it operates under the different scenarios of cooperation. 
If no cooperation is allowed, only $74\%$ of the demand can be fulfilled even when all
the resources are available; partial cooperation leads the FCN to serve $94\%$ demand; 
finally, full cooperation allows the FCN to supply $97\%$ of the demand [Figure~\ref{fig:3}(a)]. 
In short, the fully cooperative scenario induces a $[100 \, (97-74)/74] \% = 31\%$ increment in the ability of the FCN to serve demand compared to the non-cooperative scenario; 
if passing from partial to full cooperation, the relative change is instead $3\%$.
At the same time, the transition to the non-percolating regime starts when about $10^5$ agents are supplied with their demand, i.e., much earlier than what was observed in Figure~\ref{fig:2}. The transition is also less pronounced than previously observed; further, when the transition does occur, a great percentage of seats are still available in the network [Figure~\ref{fig:3}(b)]; finally, significant differences are visible between observed and predicted supply (Figure~\ref{fig_sm:supply-demand-gravity_B-distance-Y2023M4D18}). Overall, $10\%$ and $20\%$ of the supplied itineraries are characterized by being served by at least two carriers for the scenarios of partial and full cooperation, respectively [Figure~\ref{fig:3}(e)]; also, about $40\%$ of the supplied itineraries are composed of more than one flight irrespective of the level of allowed cooperation [Figure~\ref{fig:3}(f)]. Once more, these numbers are radically different from those recorded when demand is estimated using actual sold tickets. The distributions of length and duration of the supplied itineraries indicate that many of them include relatively short flights [Figure~\ref{fig:3}(c) and (d)]. 

All the above conclusions are insensitive to the specific cost function used in the MCP model, but vary to a degree in some details. As expected, when agents optimize the length of the itineraries, then the average length of supplied itineraries is smaller than for the other cases [Figures~\ref{fig:3}(c)], however, their typical duration is longer than for the other decision protocols [Figures~\ref{fig:2}(d)]. The opposite conclusions are valid for the MCP model where itineraries are chosen with the goal of optimizing their duration. 
The model where the cost of flights is related to seat availability generates itineraries that are generally long both in time and space, so it appears not as ideal as the other two decision protocols, except perhaps for the airline ticket sales. 

All of the above results are based on a specific choice of parameters for the gravity model. In Figure~\ref{fig_sm:cooperative-fcn-gravity-A-Y2023M4D18}, we consider a different choice for these parameters.  While the qualitative conclusions remain unaffected, several quantitative differences are visible in the outcome of the MCP model including (i) the gap in performance between scenarios with different levels of cooperation, (ii) the location of the transition point where resources of the infrastructure are exhausted, and (iii) the typical length and duration of the supplied itineraries. Finally, all the above also applies in general to other daily schedules, see Figures~\ref{fig_sm:cooperative-fcn-Y2019M4D18}, ~\ref{fig_sm:cooperative-fcn-Y2023M11D22},~\ref{fig_sm:cooperative-fcn-gravity-A-Y2019M4D18} and~\ref{fig_sm:cooperative-fcn-gravity-A-Y2023M11D22}.


\subsection*{Characterization of the cooperative air transportation system}

\begin{figure*}[!htb]
    \centering
    \includegraphics[width=.85\linewidth]{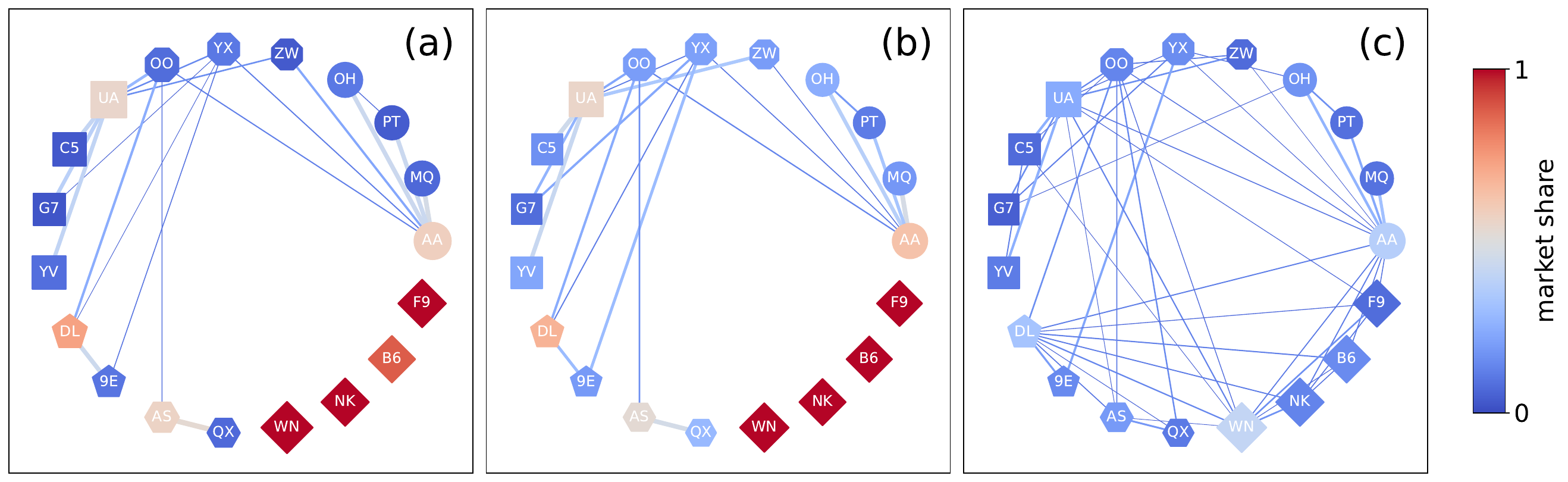}
    \caption{
    {\bf Market-share networks.} {\bf (a)} Each node in the network is an air carrier that operated flights in the contiguous US on April 18, 2023. To build the network we consider data about sold tickets in the second quarter of year 2023, but only those itineraries that are composed of two flights. We establish weighted connections between air carriers depending on the fraction of co-operated itineraries between the two air carriers, see Eq.~(\ref{eq:market-share}).  For visualization purposes we do not display connections whose weight is smaller than $0.1$. 
    The size of the nodes is proportional to the logarithm of the total number of flights operated by the corresponding carrier; their color denotes the fraction of two-flight itineraries that were operated individually by the air carrier. For major airlines and their subsidiaries, we use the same symbols: circles for {\it American Airlines} (AA), squares for {\it United Air Lines} (UA), pentagons for {\it Delta Air Lines} (DL), and hexagons for {\it Alaska Airlines} (AS). Octagons denote minor airlines having non-exclusive agreements with multiple major airlines. All other airlines are represented using diamonds. The list of airlines's codes can be found at \url{https://www.bts.gov/topics/airlines-and-airports/airline-codes}. {\bf (b)} Same visualization as in (a), but results are obtained using the minimum-percolation model that makes use of the gravity model for demand, and itineraries are selected using a minimum-length protocol. The flight-connection network is constructed allowing for partial cooperation between air carriers. {\bf (c)} Same visualization as in (b), but the flight-connection network is constructed allowing for full cooperation between air carriers. 
    }
    \label{fig:4}
\end{figure*}

Interestingly, allowing for full cooperation would require a significant change in the way major airlines share the market of tickets. In Figure~\ref{fig:4}(a), we display the network of cooperation that can be inferred by analyzing real data of sold itineraries composed of exactly two flights [the weighted adjacency matrix of the network is in Figure~\ref{fig_sm:market-share-adjacency-matrix-gravity-B-Y2023M4D18} (a)]. The vast majority of these itineraries are served individually by major airlines. For example, if one flight is operated by {\it Delta Air Lines} (DL), then in $70\%$ of the cases the other flight is operated by DL too. Further, the network structure well reflects established agreements, if any, between airlines. For example, all exclusive subsidiaries of {\it American Airlines} (AA) share their market exclusively with AA. {\it SkyWest Airlines} (OO), {\it Republic Airline} (YX) and {\it Air Wisconsin Airlines} (ZW) uniformly share their market with the three major carriers AA, DL and {\it United Air Lines} (UA). Finally, air carriers such as {\it Spirit Air Lines} (NK) and {\it Southwest Airlines} (WN) appear as isolated nodes since they have their own markets and do not cooperate with other airlines. The analysis of the outcome of the MCP model, where demand is generated according to the gravity model, leads to radically different networks of cooperation between airlines depending of the level of cooperation that is allowed. If only partial cooperation is allowed, then the network of market share is almost identical to the one inferred from the sold tickets, see Figure~\ref{fig:4}(b). However, if full cooperation is enabled, then  the network is characterized by the presence of many more edges than those visible in Figure~\ref{fig:4}(a) and those edges have relatively similar weights; further, no node is disconnected from the rest. Overall, the major airlines that operate most of the flights still dominate, however, those major airlines now share a significant portion of their market with other major airlines. Results of Figure~\ref{fig:4}(c) are valid when the itinerary length is minimized in the MCP model, but similar conclusions can be drawn if other cost functions are considered (Figures~\ref{fig_sm:market-share-network-gravity-B-duration-Y2023M4D18} and ~\ref{fig_sm:market-share-network-gravity-B-seats-Y2023M4D18}) or if different flight schedules are analyzed (Figures~\ref{fig_sm:market-share-network-gravity-B-length-Y2019M4D18}, \ref{fig_sm:market-share-network-gravity-B-length-Y2023M11D22}, \ref{fig_sm:market-share-adjacency-matrix-gravity-B-Y2023M4D18}, \ref{fig_sm:market-share-adjacency-matrix-time-seats-Y2023M4D18}, \ref{fig_sm:market-share-adjacency-matrix-gravity-B-Y2019M4D18}, \ref{fig_sm:market-share-adjacency-matrix-gravity-B-Y2023M11D22}). In particular, these qualitative observations are valid also for the pre-COVID-19 market-share network where the number of operating air carriers was slightly larger than after the pandemic (Figure~\ref{fig_sm:market-share-adjacency-matrix-gravity-B-Y2019M4D18}).

In Figure~\ref{fig_sm:market_share}, we repeat the same analysis as in Figure~\ref{fig:4} feeding the MCP model with the demand estimated from sold tickets, however, the resulting networks appear almost unchanged. The structure of the market-share network emerging from this analysis is due to the radically different way in which the agents exploit the resources of the network depending on the level of cooperation allowed. In the regime of partial cooperation, most of the connections are among same-carrier flights, as the high correlation between the utilization of the FCN in the regimes of no and partial cooperation indicate (Figures~\ref{fig_sm:scatter-plots-DB1B-distance-Y2023M4D18},~\ref{fig_sm:scatter-plots-gravity_B-distance-Y2023M4D18}, ~\ref{fig_sm:scatter-plots-DB1B-time-Y2023M4D18} and~\ref{fig_sm:scatter-plots-DB1B-seats-Y2023M4D18}).
When full cooperation is enabled instead, agents take advantage of very different types of flight connections, many of them being connections available only under the scenario of unrestricted cooperation.

\subsection*{Enhanced robustness induced by unrestricted air-carrier cooperation}



\begin{figure}[!htb]
    \includegraphics[width = 0.48\textwidth]{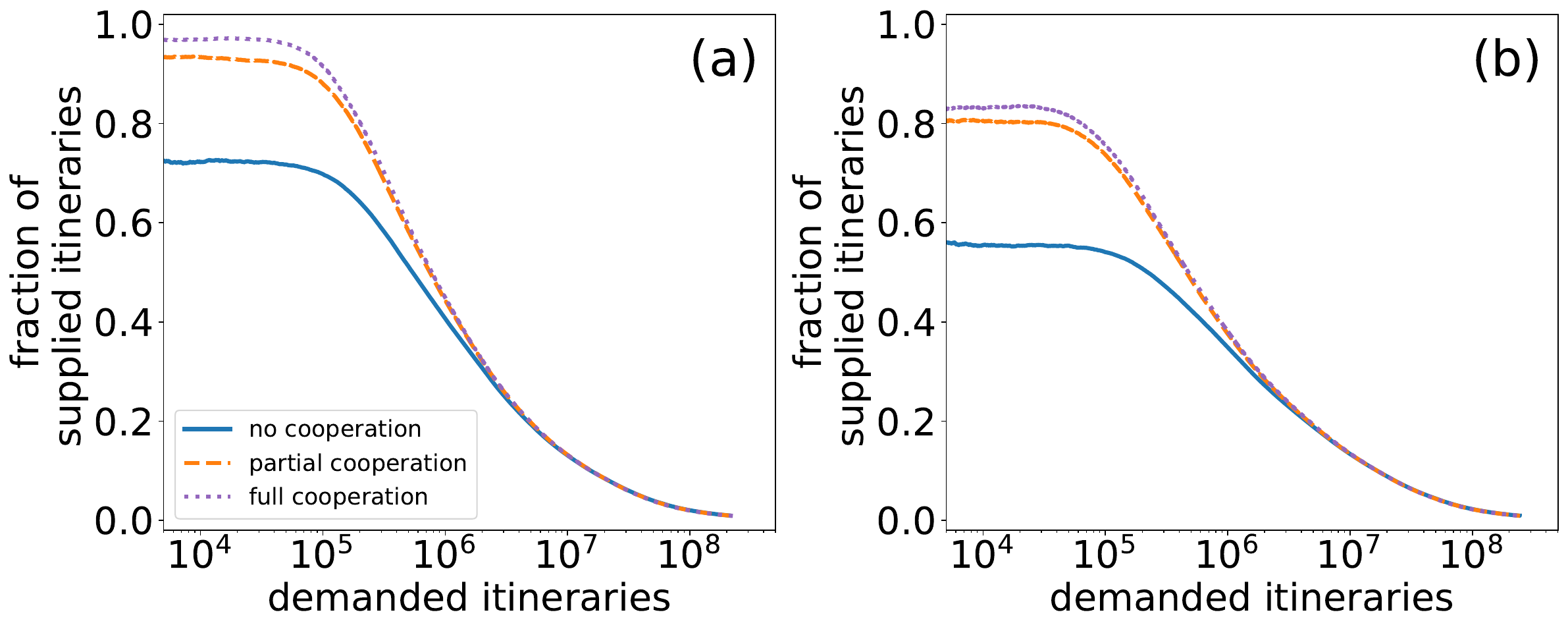}
    \caption{{\bf Cooperation among carriers and robustness of the air transportation system.}
    {\bf (a)} Same as in panel (a) of Figure~\ref{fig:3}, but the flight-connection network is obtained after removing all flights operated by {\it Delta Air Lines} (DL). 
    {\bf (b)} Same as in panel (a), but the flight-connection network is created after removing each aircraft with probability $0.5$.
    }
    \label{fig:5}
\end{figure}

We use the MCP framework to study two different types of perturbations induced in the FCN. In one case, we remove all flights operated by DL.
This serves to emulate the disruption to the service experienced during the so-called  2024 CrowdStrike incident, when many DL flights
were canceled~\cite{deltastrike}. The results displayed in Figure~\ref{fig:5}(a) show how the entire air transportation system would react to such a perturbation depending on the level of
cooperation allowed in the FCN. Canceling all DL flights in the day considered in our analysis would correspond to 
the removal of $14\%$ of $18545$ flights, and a reduction of $19\%$ of $2794066$ total number of available seats.
The effect of such a perturbation is mostly visible in the point at which the transition occurs, whereas plateau values in the operating regime appear almost identical to those observed previously; still, the higher performance of the FCN under full cooperation with respect to other cooperative scenarios is quite apparent. In the second case, we focus on a $50\%$ reduction of the fleet of aircraft operating US domestic flights. 
Even if subject to such a dramatic perturbation, the FCN would still be able to connect a significant portion of the demanded origin-destination pairs irrespective of the regime of cooperation; still a fully cooperative FCN would display a greater performance compared to the FCNs obtained under the other cooperative scenarios. 

\section*{Conclusion}

In this paper, we introduced a computational framework that leverages methods typical of 
percolation theory to simulate the consumption, and eventual depletion, of resources in 
 networks. In the framework, rational agents progressively consume resources following a
minimum-cost decision protocol for the selection of itineraries 
connecting demanded origin-destination pairs, hence the name of minimum-cost-percolation (MCP) framework.
The framework is very general and can be applied to the analysis of various types of data-generated networks, for example, 
transportation and/or delivery networks. As a proof of concept, we applied it to weighted, directed, temporal and multi-layer
network representations of the US air transportation system that can be generated using freely available data. After validating the MCP framework by means of 
recreating statistical patterns typical of real itineraries, we took advantage of its flexibility to study a series of hypothetical scenarios.
We specifically focused our attention on the role of cooperation among air carriers in serving 
demanded itineraries via commercial partnerships. We found that allowing for 
unrestricted cooperation among air carriers could be very beneficial by (i) increasing the percentage 
of reachable origin-destination pairs and (ii) improving the robustness of the system
against disruptive perturbations. Those benefits would emerge by simply encouraging code-sharing agreements among 
airlines without altering their flights schedules, thus
their cost of operation.

Several limitations are present in our study and, here, we discuss only some of the most obvious. 
First, all our results are based on approximations of the ground-truth demand of the US population.
Using supplied demand inferred from data about sold tickets indicates that the current infrastructure is already
operating in a nearly optimal fashion. On the other hand, demand generated according to the 
gravity model indicates that there is still a lot of room for the optimization of the infrastructure.  
The reality should be somewhere in the middle, but identifying exactly where
would require the development of non-trivial models for real demand. Second, our results are obtained under the
assumption that that all agents are identical in the sense that they all optimize the same cost function.
The MCP framework could be easily extended to account for population heterogeneity, however, such an extension would require
access to commercial data that, at the moment, is not freely available.
Finally, our analysis focused solely on the properties of the transportation system, but neglected potential disadvantages in 
airlines' revenues that may emerge from reshaping the network of market share. This is a central aspect that must be accounted for
in any eventual attempt of optimization of the air transportation system that relies on our computational framework.
Properly addressing all the limitations of the MCP framework and then taking advantage of it
in downstream applications are tasks left to future research.

Code to reproduce the results of this paper is available at \url{https://github.com/danielhankim/minimum-cost-percolation-us-airline.git}. 

\begin{acknowledgements}
This research was partially supported by the Air Force Office of Scientific
Research under grant numbers FA9550-24-1-0039 and 23RICOR001. 
The authors also acknowledge support from 
the Air Force Summer Faculty Fellowship Program.
The funders had no role in study design, data collection,
and analysis, the decision to publish, or any opinions, findings, conclusions, or recommendations expressed in the manuscript.
\end{acknowledgements}


\section*{Methods}

\subsection*{Data}

 We fuse data from multiple freely available sources. 
We rely on the Airline On-Time Performance database of the Bureau of Transportation Statistics (BTS)~\cite{bts}, which provides 
the origin $o_f$ and destination $d_f$ airports, 
scheduled departure and arrival times $\tau^{(d)}_f \leq \tau^{(a)}_f$, air carrier $c_f$, and tail number of the aircraft $n_f$
for each commercial flight $f$ operated by a US carrier. To estimate the number of available seats $s_f$ on the flight $f$,
we use the tail number $n_f$ to identify the specifics of the aircraft 
that operated flight $f$ 
in the register of the Federal Aviation Administration (FAA)~\cite{faa}.
We also use the DB1B database from BTS which contains $10\%$ randomly selected itineraries supplied/sold by US air carriers. 
DB1B data are aggregated over quarters of a year, so no detailed temporal information for the itineraries is 
provided. Nonetheless, for each itinerary, one can retrieve some information about the individual flights that compose it, 
including the origin and destination of the flights and operating carriers, as well as the total number of passengers associated with the itinerary.
 We take advantage of data from the OpenFlights database to get the geographical coordinates of the origin and destination airports for each flight to estimate the geographical distance between airports and convert arrival and departure times of planes to Coordinated Universal Time (UTC) standard times~\cite{openflights}. Finally, we use census population data from the NASA's Socioeconomic Data and Applications Center (SEDAC) operated by the Center for International Earth Science Information Network (CIESIN) at Columbia University~\cite{GPWv4_2018}. This data provides estimates of the number of inhabitants in the US for cells of size $30$ km $\times \ 30$ km.

\subsection*{Multi-carrier flight-connection network}


\subsubsection*{Network construction}

 We aggregate information for all individual flights operated in a single day to generate a weighted, directed, time-stamped, and multi-layer graph representing the US air transportation infrastructure for that specific day. Instead of using the typical representation where the infrastructure is seen as a graph where the airports are the nodes and the edges are given by flights connecting pairs of airports~\cite{guimera2005worldwide, colizza2006role}, we map this data into a so-called flight-connection network (FCN) $\mathcal{F}$, wherein the nodes are flights; and further, multiple layers of nodes are present in the FCN, each layer representing a specific air carrier.
 Two flights $f$ and $g$ that are operated by the same air carrier
 are connected by a directed intra-layer edge $f \to g$ if both conditions
\begin{equation}
    d_f = o_g
    \label{eq:cond1}
\end{equation}
and 
\begin{equation}
    \tau^{(d)}_g - \tau^{(a)}_f  > \delta \; ,
    \label{eq:cond2}
\end{equation}
where $\delta \geq 0$ is a tunable parameter, are satisfied.  Essentially, the edge $f \to g$ is present whenever the same passenger can realistically take both flights $f$ and $g$. 

We then distinguish three scenarios: (i) no cooperation, (ii) partial cooperation and (iii) full cooperation.
In the non-cooperative scenario, flights operated by different carriers are not connected, meaning the various layers are independent from eachother. 
In the scenario of partial cooperation, two flights operated by different carriers having commercial partnerships
are connected if both conditions of Eqs.~(\ref{eq:cond1}) and~(\ref{eq:cond2}) are met.
This means that the set of edges of the non-cooperative scenario is a subset of the set of edges of the network
operating with partial cooperation.
Finally, in the scenario of full cooperation, two flights $f$ and $g$ operated by different air carriers
that are not part of the same alliance
are connected by the directed inter-layer edge $f \to g$ if condition~(\ref{eq:cond1}) holds and if condition
\begin{equation}
    \tau^{(d)}_g - \tau^{(a)}_f  > \epsilon 
    \label{eq:cond3}
\end{equation}
is satisfied too, with
$\epsilon  \geq 0$ being a tunable parameter. We always impose $\epsilon \geq \delta$, thus the 
set of edges of the network operating under partial cooperation is a subset of the set of edges
valid for the network of full cooperation.

We build FCNs for a few selected days by focusing only on flights operated between airports in the contiguous US, i.e., 
more than $95\%$ of all domestic flights. While processing data, we exclude flights with no 
tail number reported for the aircraft; these constitute only a negligible fraction of all flights. All results reported in the main paper are based on the schedule of April 18, 2023; after excluding $913$ flights, the network contains $18545$ flights operated between $300$ airports for a total of $\sum_f s_f = 2794066$ seats.
For the choice of the parameters $\delta = 30$ minutes and $\epsilon = 60$ minutes, the FCN contains the following number of edges:
$E^{(nc)} = 813694$ (no cooperation),  $E^{(pc)} = 1464725$ (partial cooperation), and  $E^{(fc)} = 2613625$ (full cooperation).
Results for other days are reported in Table~\ref{tab:fcn-table}. 




\subsubsection*{Commercial partnerships among air carriers}
To infer the existence of commercial partnerships among air carriers, 
we use BTS data about sold tickets.  
For simplicity, we consider sold itineraries composed of only two flights. 
We then count the number $w(c,c')$  of itineraries with one flight operated 
by carrier $c$ and the other by carrier $c'$. We define the proportion of the market shared
by the carrier $c$ with carrier $c'$ as
\begin{equation}
    s(c | c') = \frac{w(c,c')}{\sum_{q} \, w(c,q)} \; .
    \label{eq:market-share}
\end{equation}
To establish the presence of a commercial relationship between carrier $c$ and $c'$, we require that
\begin{equation}
\max \{ s(c|c'), s(c'|c) \} > \theta \; .
 \label{eq:comm-relationship}
\end{equation}
In all our analyses, we set $\theta = 0.1$. This specific choice 
gives rise to the network of shared market appearing in Figure~\ref{fig:4}.
For simplicity of visualization, in that figure, we create symmetric connections between carriers $c$ and $c'$
with weight equal to $\frac{1}{2} \left[ s(c|c') +  s(c'|c) \right]$.
Note that the above quantities are used also to estimate the weight
of the connections in the networks of market share based on the results from the MCP model.

\subsubsection*{Properties of the flight-connection network}

The FCN is a directed acyclic graph, and the acyclic nature of the FCN allows for a more straightforward description of the framework as well as for a simpler computation of the main quantities that are needed in our framework. A directed path $f_1 \to f_2 \to \cdots \to f_\ell$ in the FCN $\mathcal{F}$ identifies a feasible itinerary between the locations $o_{f_1}$ and $d_{f_\ell}$. For compactness of notation, we use $\vec{f} = \left( f_1, f_2, \ldots, f_\ell \right)$ to denote a path, and with $|\vec{f}| = \ell$ the number of its components.
Various metrics can be associated with the path 
$\vec{f}$,
e.g., the number 
$|\vec{f}|$
of connecting flights in the path, the geographical length of the itinerary
\begin{equation}
L \left(\vec{f}\right) = \sum_{i=1}^{|\vec{f}|} \, D (o_{f_i}, d_{f_i}) \; ,
\label{eq:dist}
\end{equation}
where $D (o, d)$ is the geographical distance between airports $o$ and $d$, as well as the total duration of the itinerary
\begin{equation}
T \left(\vec{f}\right) =  \tau^{(a)}_{f_{|\vec{f}|}} - \tau^{(d)}_{f_1} \; ,
\label{eq:time}
\end{equation}

Given a pair of origin and destination airports $o \to d$, one can identify all possible itineraries connecting the two locations, i.e., all paths 
$\vec{f}$
such that $o_{f_1} = o$ and $d_{f_{|\vec{f}|}} = d$. If at least one such path exists, then $d$ is reachable from $o$. And, if $d$ is reachable from $o$, minimum-cost itineraries for the pair $o \to d$ can be defined based on some metric such as their length or duration. We also consider the case where the cost of an itinerary is a function of the available seats on the various legs that compose the itinerary:
\begin{equation}
M \left(\vec{f}\right)
= \sum_{i=1}^{|\vec{f}|} \, \frac{1}{s_{f_i}}  \; .
\label{eq:seats}
\end{equation}
where $s_{f_i}$ is the number of seats available on flight $f_i$. In the above expression, the fact that each flight's contribution is given by the inverse of the number of available seats is a simple, but arbitrary choice.

\subsection*{Demand}

A fundamental ingredient of our framework is modeling how resources are demanded by the population of agents. We represent this ingredient as a set $\mathcal{D}$ of weighted node pairs: the pair $o \to d \in \mathcal{D} $ denotes the labels of the origin and destination nodes demanded by a generic agent; the weight $w_{o \to d}$ reflects the effective demand for the specific pair $o \to d$.

\subsubsection*{Supplied demand}

As mentioned above, the DB1B data contains $10\%$ randomly selected itineraries supplied by US air carriers during the specific quarter at hand, each itinerary weighted by the number of passengers that actually traveled on that itinerary. In our data-driven construction of $\mathcal{D}_{s}$, we simply set $w_{o \to d}$ equal to the total number of passengers who traveled from airport $o$ to airport $d$ as found in the BTS data. We only care about the origin and destination airports of an itinerary, and not about eventual connecting flights present in the itinerary. For example, the itineraries $\tt{IND} \to \tt{ATL} \to \tt{SEA}$ and $\tt{IND} \to \tt{ORD} \to \tt{SEA}$ contribute identically in the construction of  $\mathcal{D}_{s}$ with $o = \tt{IND}$ and $d = \tt{SEA}$.  

We rely on data from the second quarter of 2023 when considering the FCNs representing the schedule of April 18, 2023. For the other days that we consider, we construct demand sets using data from other quarters/years to represent as best as possible the corresponding FCNs. Note that we pre-process data as to consider only pairs of origin-destination airports that are among the airports used in the construction of the FCNs. This corresponds to excluding about $10\%$ of all tickets appearing in the DB1B dataset.

\subsubsection*{Estimated demand}

It is important to note that the definition of $\mathcal{D}_{s}$ tacitly assumes that the effectively supplied demand contained in the DB1B data is representative of the real demand for the US population. This is a weak assumption as it clearly neglects the fact that some demand can not be supplied by the infrastructure. We therefore also consider a different approach where the demand $\mathcal{D}_g$ is estimated by the gravity model for human mobility~\cite{zipf1946p, erlander1990gravity}. 
We use census population data from SEDAC
to obtain estimates of
the number of inhabitants in the US for cells of size $30$ km $\times \ 30$ km. 
Then, we pre-process the set of the US airports using the density-based spatial clustering of applications with noise (DBSCAN) algorithm~\cite{ester1996density}. This algorithm serves to cluster points of data based on their proximity in space, and the outcome of the algorithm depends on a tunable parameter $z_{\text{DBSCAN}}$ that defines the maximum value of the distance for two points to be considered neighbors. In our case, we set $z_{\text{DBSCAN}} = 50$ km to create a partition of all $300$ airports in the contiguous US into a total of $278$ clusters which we call super-airports.  For example, one cluster contains {\tt JFK}, {\tt LGA}, {\tt EWR}, and {\tt HPN}, which all serve the New York City metropolitan area. Similarly, the Los Angeles metropolitan area corresponds to a super-airport with the five airports ({\tt LAX}, {\tt SNA}, {\tt BUR}, {\tt LGB}, and {\tt ONT}); while the Washington DC area is associated with a cluster composed of the three airports ({\tt DCA}, {\tt IAD}, and {\tt BWI}). Generally, most super-airports correspond to one airport only, for example {\tt IND}.  This pre-processing was found to be necessary because if for instance the population is based on simple nearest-airport measures, some very large airports such as {\tt JFK} end up being associated with a very small population.  The center of mass of the airports within a super-airport determines the geographical position of the latter, and we assign the population $m_a$ to each super-airport $a$ by simply associating each census cell to the nearest airport, and then to the corresponding super-airport. 

To generate the demand set $\mathcal{D}_g$, we simply use the gravity model
\begin{equation}
w_{o \to d} \sim 
\left\{
\begin{array}{ll} 
\left( m_o^{\alpha} \, m_d^{\beta} \right) / \left[D(o,d)\right]^\gamma  & \textrm{ if } D(o,d) > z_{\min}
\\
0 & \textrm{ if } D(o,d) \leq  z_{\min}
\end{array}
\right.
\; ,
    \label{eq:gravity}
\end{equation}
where $m_o$ and $m_d$ are the populations associated with $o$ and $d$ respectively, $D(o,d)$ is the geographical distance between $o$ and $d$, and $\alpha$, $\beta$, $\gamma$, and $z_{\min}$ are all tunable parameters of the model. The parameters $\alpha$, $\beta$ and $\gamma$ serve to weigh the importance of the populations and the geographical distance between super-airports, and the parameter $z_{\min}$ serves instead as a hard threshold to define the minimum length of a demanded itinerary. We set $z_{\min} = 300 \,\textrm{km}$ in line with the empirical finding of Ref.~\cite{balcan2009multiscale}. 


\subsection*{Minimum-cost-percolation model}

The minimum-cost-percolation (MCP) model takes three inputs: the initial FCN network $\mathcal{F}^{(1)}$, the demand set $\mathcal{D}$ and the cost function $C(\cdot)$ associated to each itinerary depending on the available resources. Initially, we set the counter for demanded itineraries $t=1$ and we create an empty list $\vec{S}^{(1)}$ to keep track of the served itineraries. We then iterate the following operations:

\begin{enumerate}

    \item We extract at random, but proportionally to its weight $w_{o \to d}$, one element $o \to d$ from the set $\mathcal{D}$. 

    \item We search in $\mathcal{F}^{(t)}$ for all paths whose first flight departs from airport $o$ and the final flight arrives to airport $d$. We associate to each path 
    $\vec{f}$
    a cost that is given by 
    $C\left(\vec{f}\right)$.
    Although we do not explicit such a dependence, we stress that
    the value of the cost function depends not just on the path, but also the  resources that are available
    when itinerary $t$ is demanded.
    
    \item  If at least one path at point 2 is found, we select at random one among the minimum-cost paths, say $\vec{f}^{(t)}$.
    First we add $\vec{f}^{(t)}$ to $\vec{S}^{(t)}$, storing also information about the number of 
    components $|\vec{f}^{(t)}|$, length $L\left(\vec{f}^{(t)} \right)$, duration 
    $D\left(\vec{f}^{(t)}\right)$, and value of the seat-availability-based cost function $M \left( \vec{f}^{(t)} \right)$
    for the selected itinerary $\vec{f}^{(t)}$.
    Then, we reduce the number of available seats on each of the corresponding flights by one, meaning 
    $s_{f_i} \mapsto s_{f_i} - 1$ for all $i =1 , \ldots, |\vec{f}|$.

    \item  If the number of seats for a flight $f$ becomes zero, we delete the flight $f$ and all its connections from $\mathcal{F}^{(t)}$.

    \item  We map $\mathcal{F}^{(t+1)} \mapsto \mathcal{F}^{(t)}$ and $\vec{S}^{(t+1)} \mapsto \vec{S}^{(t)}$, then we increase $t \mapsto t + 1$. 

\end{enumerate}

We end the above procedure after $t_*$ iterations when the graph $\mathcal{F}^{(t_*)}$ does not contain any nodes and edges.

The efficient implementation of the above model is far from being trivial due to the dynamic nature of the costs of paths as the network changes. 
For example, the network exploration at point 2 is performed using a Dijkstra-like algorithm~\cite{dijkstra2022note}. 
This ensures not only that the search is performed efficiently, but also that suitable paths do not contain the same airport more than once. Also, 
if no path is found at point 2, then the corresponding origin-destination is effectively removed from the demand set. This guarantees that non-existing
paths are searched only once, allowing from a great speed-up in the time required to simulate the MCP model. 
We properly keep track of the increments in $t$ by extracting random variables out of a geometric distribution. 

In the application of the MCP model to the US air transportation network, all airports that are in the input FCN are relabeled using the map to their corresponding super-airport; note that the operation does not change the actual schedule of flights, as no new nodes nor edges are added and/or deleted; we still rely on the true length and duration
of the flights. When supplied demand is used as input to the MCP model, airports are mapped to super-airports so that the supplied demand 
is simply given by the sum over all pairs of airports within the corresponding super-airports.

For each value of the control parameter $t$, we generalize some metrics from the theory of network percolation to assess the
performance of the infrastructure in serving the demand. For example, inspired by the 
so-called percolation strength, we measure the fraction of served itineraries
\begin{equation}
    P(t) = \frac{ | \vec{S}^{(t)} | }{t} \; . 
    \label{eq:fraction_served}
\end{equation}
Also, we keep track of the number of remaining seats by 
simply summing the variables $s_f$ over all flights $f$ still present in $\mathcal{F}^{(t)}$.
The fraction of remaining seats is the analogue of the fraction of retained edges in network percolation.
Regarding the statistics of the itineraries served by the infrastructure,
we measure the distributions of the number of legs, number of carriers, duration, and length.
We also estimate the utilization of each flight connection $f \to g$ as
\begin{equation}
    u_{f \to g} = \sum_{\vec{f} \in \vec{S}^{t_*} } \sum_{i=1}^{|\vec{f}|-1}  \delta_{f, f_i} \, \delta_{g, f_{i+1}} \; .
    \label{eq:weight_flight}
\end{equation}
We note that $0 \leq u _{f \to g} \leq \min \{ s_f, s_g \}$.



\bibliography{bibliography}

\begin{thebibliography}{72}%
\makeatletter
\providecommand \@ifxundefined [1]{%
 \@ifx{#1\undefined}
}%
\providecommand \@ifnum [1]{%
 \ifnum #1\expandafter \@firstoftwo
 \else \expandafter \@secondoftwo
 \fi
}%
\providecommand \@ifx [1]{%
 \ifx #1\expandafter \@firstoftwo
 \else \expandafter \@secondoftwo
 \fi
}%
\providecommand \natexlab [1]{#1}%
\providecommand \enquote  [1]{``#1''}%
\providecommand \bibnamefont  [1]{#1}%
\providecommand \bibfnamefont [1]{#1}%
\providecommand \citenamefont [1]{#1}%
\providecommand \href@noop [0]{\@secondoftwo}%
\providecommand \href [0]{\begingroup \@sanitize@url \@href}%
\providecommand \@href[1]{\@@startlink{#1}\@@href}%
\providecommand \@@href[1]{\endgroup#1\@@endlink}%
\providecommand \@sanitize@url [0]{\catcode `\\12\catcode `\$12\catcode
  `\&12\catcode `\#12\catcode `\^12\catcode `\_12\catcode `\%12\relax}%
\providecommand \@@startlink[1]{}%
\providecommand \@@endlink[0]{}%
\providecommand \url  [0]{\begingroup\@sanitize@url \@url }%
\providecommand \@url [1]{\endgroup\@href {#1}{\urlprefix }}%
\providecommand \urlprefix  [0]{URL }%
\providecommand \Eprint [0]{\href }%
\providecommand \doibase [0]{http://dx.doi.org/}%
\providecommand \selectlanguage [0]{\@gobble}%
\providecommand \bibinfo  [0]{\@secondoftwo}%
\providecommand \bibfield  [0]{\@secondoftwo}%
\providecommand \translation [1]{[#1]}%
\providecommand \BibitemOpen [0]{}%
\providecommand \bibitemStop [0]{}%
\providecommand \bibitemNoStop [0]{.\EOS\space}%
\providecommand \EOS [0]{\spacefactor3000\relax}%
\providecommand \BibitemShut  [1]{\csname bibitem#1\endcsname}%
\let\auto@bib@innerbib\@empty
\bibitem [{\citenamefont {Administration}(2020)}]{federal2020economic}%
  \BibitemOpen
  \bibfield  {author} {\bibinfo {author} {\bibfnamefont {F.~A.}\ \bibnamefont
  {Administration}},\ }\href@noop {} {\  (\bibinfo {year} {2020})}\BibitemShut
  {NoStop}%
\bibitem [{\citenamefont {of~Transportation}(2018)}]{pocket2018}%
  \BibitemOpen
  \bibfield  {author} {\bibinfo {author} {\bibfnamefont {U.~D.}\ \bibnamefont
  {of~Transportation}},\ }\href@noop {} {\  (\bibinfo {year}
  {2018})}\BibitemShut {NoStop}%
\bibitem [{\citenamefont {of~Transportation}(2023)}]{pocket2023}%
  \BibitemOpen
  \bibfield  {author} {\bibinfo {author} {\bibfnamefont {U.~D.}\ \bibnamefont
  {of~Transportation}},\ }\href@noop {} {\  (\bibinfo {year}
  {2023})}\BibitemShut {NoStop}%
\bibitem [{\citenamefont {Barnhart}\ \emph {et~al.}(2003)\citenamefont
  {Barnhart}, \citenamefont {Belobaba},\ and\ \citenamefont
  {Odoni}}]{barnhart2003applications}%
  \BibitemOpen
  \bibfield  {author} {\bibinfo {author} {\bibfnamefont {C.}~\bibnamefont
  {Barnhart}}, \bibinfo {author} {\bibfnamefont {P.}~\bibnamefont {Belobaba}},
  \ and\ \bibinfo {author} {\bibfnamefont {A.~R.}\ \bibnamefont {Odoni}},\
  }\href@noop {} {\bibfield  {journal} {\bibinfo  {journal} {Transportation
  science}\ }\textbf {\bibinfo {volume} {37}},\ \bibinfo {pages} {368}
  (\bibinfo {year} {2003})}\BibitemShut {NoStop}%
\bibitem [{\citenamefont {Lohatepanont}\ and\ \citenamefont
  {Barnhart}(2004)}]{lohatepanont2004airline}%
  \BibitemOpen
  \bibfield  {author} {\bibinfo {author} {\bibfnamefont {M.}~\bibnamefont
  {Lohatepanont}}\ and\ \bibinfo {author} {\bibfnamefont {C.}~\bibnamefont
  {Barnhart}},\ }\href@noop {} {\bibfield  {journal} {\bibinfo  {journal}
  {Transportation Science}\ }\textbf {\bibinfo {volume} {38}},\ \bibinfo
  {pages} {19} (\bibinfo {year} {2004})}\BibitemShut {NoStop}%
\bibitem [{\citenamefont {Barnhart}\ \emph {et~al.}(1998)\citenamefont
  {Barnhart}, \citenamefont {Boland}, \citenamefont {Clarke}, \citenamefont
  {Johnson}, \citenamefont {Nemhauser},\ and\ \citenamefont
  {Shenoi}}]{barnhart1998flight}%
  \BibitemOpen
  \bibfield  {author} {\bibinfo {author} {\bibfnamefont {C.}~\bibnamefont
  {Barnhart}}, \bibinfo {author} {\bibfnamefont {N.~L.}\ \bibnamefont
  {Boland}}, \bibinfo {author} {\bibfnamefont {L.~W.}\ \bibnamefont {Clarke}},
  \bibinfo {author} {\bibfnamefont {E.~L.}\ \bibnamefont {Johnson}}, \bibinfo
  {author} {\bibfnamefont {G.~L.}\ \bibnamefont {Nemhauser}}, \ and\ \bibinfo
  {author} {\bibfnamefont {R.~G.}\ \bibnamefont {Shenoi}},\ }\href@noop {}
  {\bibfield  {journal} {\bibinfo  {journal} {Transportation science}\ }\textbf
  {\bibinfo {volume} {32}},\ \bibinfo {pages} {208} (\bibinfo {year}
  {1998})}\BibitemShut {NoStop}%
\bibitem [{\citenamefont {Jacobs}\ \emph {et~al.}(2008)\citenamefont {Jacobs},
  \citenamefont {Smith},\ and\ \citenamefont
  {Johnson}}]{jacobs2008incorporating}%
  \BibitemOpen
  \bibfield  {author} {\bibinfo {author} {\bibfnamefont {T.~L.}\ \bibnamefont
  {Jacobs}}, \bibinfo {author} {\bibfnamefont {B.~C.}\ \bibnamefont {Smith}}, \
  and\ \bibinfo {author} {\bibfnamefont {E.~L.}\ \bibnamefont {Johnson}},\
  }\href@noop {} {\bibfield  {journal} {\bibinfo  {journal} {Transportation
  Science}\ }\textbf {\bibinfo {volume} {42}},\ \bibinfo {pages} {514}
  (\bibinfo {year} {2008})}\BibitemShut {NoStop}%
\bibitem [{\citenamefont {Vance}\ \emph {et~al.}(1997)\citenamefont {Vance},
  \citenamefont {Barnhart}, \citenamefont {Johnson},\ and\ \citenamefont
  {Nemhauser}}]{vance1997airline}%
  \BibitemOpen
  \bibfield  {author} {\bibinfo {author} {\bibfnamefont {P.~H.}\ \bibnamefont
  {Vance}}, \bibinfo {author} {\bibfnamefont {C.}~\bibnamefont {Barnhart}},
  \bibinfo {author} {\bibfnamefont {E.~L.}\ \bibnamefont {Johnson}}, \ and\
  \bibinfo {author} {\bibfnamefont {G.~L.}\ \bibnamefont {Nemhauser}},\
  }\href@noop {} {\bibfield  {journal} {\bibinfo  {journal} {Operations
  Research}\ }\textbf {\bibinfo {volume} {45}},\ \bibinfo {pages} {188}
  (\bibinfo {year} {1997})}\BibitemShut {NoStop}%
\bibitem [{\citenamefont {Smith}\ \emph {et~al.}(1992)\citenamefont {Smith},
  \citenamefont {Leimkuhler},\ and\ \citenamefont {Darrow}}]{smith1992yield}%
  \BibitemOpen
  \bibfield  {author} {\bibinfo {author} {\bibfnamefont {B.~C.}\ \bibnamefont
  {Smith}}, \bibinfo {author} {\bibfnamefont {J.~F.}\ \bibnamefont
  {Leimkuhler}}, \ and\ \bibinfo {author} {\bibfnamefont {R.~M.}\ \bibnamefont
  {Darrow}},\ }\href@noop {} {\bibfield  {journal} {\bibinfo  {journal}
  {interfaces}\ }\textbf {\bibinfo {volume} {22}},\ \bibinfo {pages} {8}
  (\bibinfo {year} {1992})}\BibitemShut {NoStop}%
\bibitem [{\citenamefont {Belobaba}(1987)}]{belobaba1987air}%
  \BibitemOpen
  \bibfield  {author} {\bibinfo {author} {\bibfnamefont {P.}~\bibnamefont
  {Belobaba}},\ }\emph {\bibinfo {title} {Air travel demand and airline seat
  inventory management}},\ \href@noop {} {Ph.D. thesis},\ \bibinfo  {school}
  {Massachusetts Institute of Technology} (\bibinfo {year} {1987})\BibitemShut
  {NoStop}%
\bibitem [{\citenamefont {Ball}\ \emph {et~al.}(2007)\citenamefont {Ball},
  \citenamefont {Barnhart}, \citenamefont {Nemhauser},\ and\ \citenamefont
  {Odoni}}]{ball2007air}%
  \BibitemOpen
  \bibfield  {author} {\bibinfo {author} {\bibfnamefont {M.}~\bibnamefont
  {Ball}}, \bibinfo {author} {\bibfnamefont {C.}~\bibnamefont {Barnhart}},
  \bibinfo {author} {\bibfnamefont {G.}~\bibnamefont {Nemhauser}}, \ and\
  \bibinfo {author} {\bibfnamefont {A.}~\bibnamefont {Odoni}},\ }\href@noop {}
  {\bibfield  {journal} {\bibinfo  {journal} {Handbooks in operations research
  and management science}\ }\textbf {\bibinfo {volume} {14}},\ \bibinfo {pages}
  {1} (\bibinfo {year} {2007})}\BibitemShut {NoStop}%
\bibitem [{\citenamefont {Hall}(1999)}]{hall1999information}%
  \BibitemOpen
  \bibfield  {author} {\bibinfo {author} {\bibfnamefont {W.}~\bibnamefont
  {Hall}},\ }\emph {\bibinfo {title} {Information flows and dynamic
  collaborative decision-making architecture: Increasing the efficiency of
  terminal area operations}},\ \href@noop {} {Ph.D. thesis},\ \bibinfo
  {school} {PhD Dissertation, Operations Research Center, MIT} (\bibinfo {year}
  {1999})\BibitemShut {NoStop}%
\bibitem [{\citenamefont {Odoni}(1987)}]{odoni1987flow}%
  \BibitemOpen
  \bibfield  {author} {\bibinfo {author} {\bibfnamefont {A.~R.}\ \bibnamefont
  {Odoni}},\ }in\ \href@noop {} {\emph {\bibinfo {booktitle} {Flow control of
  congested networks}}}\ (\bibinfo  {publisher} {Springer},\ \bibinfo {year}
  {1987})\ pp.\ \bibinfo {pages} {269--288}\BibitemShut {NoStop}%
\bibitem [{\citenamefont {O'Kelly}\ and\ \citenamefont
  {Miller}(1994)}]{o1994hub}%
  \BibitemOpen
  \bibfield  {author} {\bibinfo {author} {\bibfnamefont {M.~E.}\ \bibnamefont
  {O'Kelly}}\ and\ \bibinfo {author} {\bibfnamefont {H.~J.}\ \bibnamefont
  {Miller}},\ }\href@noop {} {\bibfield  {journal} {\bibinfo  {journal}
  {Journal of Transport Geography}\ }\textbf {\bibinfo {volume} {2}},\ \bibinfo
  {pages} {31} (\bibinfo {year} {1994})}\BibitemShut {NoStop}%
\bibitem [{\citenamefont {Berry}\ \emph {et~al.}(1996)\citenamefont {Berry},
  \citenamefont {Carnall},\ and\ \citenamefont {Spiller}}]{berry1996airline}%
  \BibitemOpen
  \bibfield  {author} {\bibinfo {author} {\bibfnamefont {S.}~\bibnamefont
  {Berry}}, \bibinfo {author} {\bibfnamefont {M.}~\bibnamefont {Carnall}}, \
  and\ \bibinfo {author} {\bibfnamefont {P.~T.}\ \bibnamefont {Spiller}},\
  }\href@noop {} {\emph {\bibinfo {title} {Airline hubs: costs, markups and the
  implications of customer heterogeneity}}},\ \bibinfo {type} {Tech. Rep.}\
  (\bibinfo  {institution} {National Bureau of Economic Research},\ \bibinfo
  {year} {1996})\BibitemShut {NoStop}%
\bibitem [{\citenamefont {Cook}\ and\ \citenamefont
  {Goodwin}(2008)}]{cook2008airline}%
  \BibitemOpen
  \bibfield  {author} {\bibinfo {author} {\bibfnamefont {G.~N.}\ \bibnamefont
  {Cook}}\ and\ \bibinfo {author} {\bibfnamefont {J.}~\bibnamefont {Goodwin}},\
  }\href@noop {} {\bibfield  {journal} {\bibinfo  {journal} {Journal of
  Aviation/Aerospace Education \& Research}\ }\textbf {\bibinfo {volume}
  {17}},\ \bibinfo {pages} {1} (\bibinfo {year} {2008})}\BibitemShut {NoStop}%
\bibitem [{\citenamefont {Guimera}\ \emph {et~al.}(2005)\citenamefont
  {Guimera}, \citenamefont {Mossa}, \citenamefont {Turtschi},\ and\
  \citenamefont {Amaral}}]{guimera2005worldwide}%
  \BibitemOpen
  \bibfield  {author} {\bibinfo {author} {\bibfnamefont {R.}~\bibnamefont
  {Guimera}}, \bibinfo {author} {\bibfnamefont {S.}~\bibnamefont {Mossa}},
  \bibinfo {author} {\bibfnamefont {A.}~\bibnamefont {Turtschi}}, \ and\
  \bibinfo {author} {\bibfnamefont {L.~N.}\ \bibnamefont {Amaral}},\
  }\href@noop {} {\bibfield  {journal} {\bibinfo  {journal} {Proceedings of the
  National Academy of Sciences}\ }\textbf {\bibinfo {volume} {102}},\ \bibinfo
  {pages} {7794} (\bibinfo {year} {2005})}\BibitemShut {NoStop}%
\bibitem [{\citenamefont {Barrat}\ \emph {et~al.}(2004)\citenamefont {Barrat},
  \citenamefont {Barthelemy}, \citenamefont {Pastor-Satorras},\ and\
  \citenamefont {Vespignani}}]{barrat2004}%
  \BibitemOpen
  \bibfield  {author} {\bibinfo {author} {\bibfnamefont {A.}~\bibnamefont
  {Barrat}}, \bibinfo {author} {\bibfnamefont {M.}~\bibnamefont {Barthelemy}},
  \bibinfo {author} {\bibfnamefont {R.}~\bibnamefont {Pastor-Satorras}}, \ and\
  \bibinfo {author} {\bibfnamefont {A.}~\bibnamefont {Vespignani}},\
  }\href@noop {} {\bibfield  {journal} {\bibinfo  {journal} {Proceedings of the
  National Academy of Sciences}\ }\textbf {\bibinfo {volume} {101}},\ \bibinfo
  {pages} {3747} (\bibinfo {year} {2004})}\BibitemShut {NoStop}%
\bibitem [{\citenamefont {Barab{\'a}si}\ and\ \citenamefont
  {Albert}(1999)}]{barabasi1999emergence}%
  \BibitemOpen
  \bibfield  {author} {\bibinfo {author} {\bibfnamefont {A.-L.}\ \bibnamefont
  {Barab{\'a}si}}\ and\ \bibinfo {author} {\bibfnamefont {R.}~\bibnamefont
  {Albert}},\ }\href@noop {} {\bibfield  {journal} {\bibinfo  {journal}
  {science}\ }\textbf {\bibinfo {volume} {286}},\ \bibinfo {pages} {509}
  (\bibinfo {year} {1999})}\BibitemShut {NoStop}%
\bibitem [{\citenamefont {Watts}\ and\ \citenamefont
  {Strogatz}(1998)}]{watts1998collective}%
  \BibitemOpen
  \bibfield  {author} {\bibinfo {author} {\bibfnamefont {D.~J.}\ \bibnamefont
  {Watts}}\ and\ \bibinfo {author} {\bibfnamefont {S.~H.}\ \bibnamefont
  {Strogatz}},\ }\href@noop {} {\bibfield  {journal} {\bibinfo  {journal}
  {nature}\ }\textbf {\bibinfo {volume} {393}},\ \bibinfo {pages} {440}
  (\bibinfo {year} {1998})}\BibitemShut {NoStop}%
\bibitem [{\citenamefont {Fortunato}(2010)}]{fortunato2010community}%
  \BibitemOpen
  \bibfield  {author} {\bibinfo {author} {\bibfnamefont {S.}~\bibnamefont
  {Fortunato}},\ }\href@noop {} {\bibfield  {journal} {\bibinfo  {journal}
  {Physics reports}\ }\textbf {\bibinfo {volume} {486}},\ \bibinfo {pages} {75}
  (\bibinfo {year} {2010})}\BibitemShut {NoStop}%
\bibitem [{\citenamefont {Boccaletti}\ \emph {et~al.}(2014)\citenamefont
  {Boccaletti}, \citenamefont {Bianconi}, \citenamefont {Criado}, \citenamefont
  {Del~Genio}, \citenamefont {G{\'o}mez-Gardenes}, \citenamefont {Romance},
  \citenamefont {Sendina-Nadal}, \citenamefont {Wang},\ and\ \citenamefont
  {Zanin}}]{boccaletti2014structure}%
  \BibitemOpen
  \bibfield  {author} {\bibinfo {author} {\bibfnamefont {S.}~\bibnamefont
  {Boccaletti}}, \bibinfo {author} {\bibfnamefont {G.}~\bibnamefont
  {Bianconi}}, \bibinfo {author} {\bibfnamefont {R.}~\bibnamefont {Criado}},
  \bibinfo {author} {\bibfnamefont {C.~I.}\ \bibnamefont {Del~Genio}}, \bibinfo
  {author} {\bibfnamefont {J.}~\bibnamefont {G{\'o}mez-Gardenes}}, \bibinfo
  {author} {\bibfnamefont {M.}~\bibnamefont {Romance}}, \bibinfo {author}
  {\bibfnamefont {I.}~\bibnamefont {Sendina-Nadal}}, \bibinfo {author}
  {\bibfnamefont {Z.}~\bibnamefont {Wang}}, \ and\ \bibinfo {author}
  {\bibfnamefont {M.}~\bibnamefont {Zanin}},\ }\href@noop {} {\bibfield
  {journal} {\bibinfo  {journal} {Physics Reports}\ }\textbf {\bibinfo {volume}
  {544}},\ \bibinfo {pages} {1} (\bibinfo {year} {2014})}\BibitemShut {NoStop}%
\bibitem [{\citenamefont {Kivel{\"a}}\ \emph {et~al.}(2014)\citenamefont
  {Kivel{\"a}}, \citenamefont {Arenas}, \citenamefont {Barthelemy},
  \citenamefont {Gleeson}, \citenamefont {Moreno},\ and\ \citenamefont
  {Porter}}]{kivela2014multilayer}%
  \BibitemOpen
  \bibfield  {author} {\bibinfo {author} {\bibfnamefont {M.}~\bibnamefont
  {Kivel{\"a}}}, \bibinfo {author} {\bibfnamefont {A.}~\bibnamefont {Arenas}},
  \bibinfo {author} {\bibfnamefont {M.}~\bibnamefont {Barthelemy}}, \bibinfo
  {author} {\bibfnamefont {J.~P.}\ \bibnamefont {Gleeson}}, \bibinfo {author}
  {\bibfnamefont {Y.}~\bibnamefont {Moreno}}, \ and\ \bibinfo {author}
  {\bibfnamefont {M.~A.}\ \bibnamefont {Porter}},\ }\href@noop {} {\bibfield
  {journal} {\bibinfo  {journal} {Journal of complex networks}\ }\textbf
  {\bibinfo {volume} {2}},\ \bibinfo {pages} {203} (\bibinfo {year}
  {2014})}\BibitemShut {NoStop}%
\bibitem [{\citenamefont {Cardillo}\ \emph {et~al.}(2013)\citenamefont
  {Cardillo}, \citenamefont {Zanin}, \citenamefont {G{\'o}mez-Gardenes},
  \citenamefont {Romance}, \citenamefont {del Amo},\ and\ \citenamefont
  {Boccaletti}}]{cardillo2013modeling}%
  \BibitemOpen
  \bibfield  {author} {\bibinfo {author} {\bibfnamefont {A.}~\bibnamefont
  {Cardillo}}, \bibinfo {author} {\bibfnamefont {M.}~\bibnamefont {Zanin}},
  \bibinfo {author} {\bibfnamefont {J.}~\bibnamefont {G{\'o}mez-Gardenes}},
  \bibinfo {author} {\bibfnamefont {M.}~\bibnamefont {Romance}}, \bibinfo
  {author} {\bibfnamefont {A.~J.~G.}\ \bibnamefont {del Amo}}, \ and\ \bibinfo
  {author} {\bibfnamefont {S.}~\bibnamefont {Boccaletti}},\ }\href@noop {}
  {\bibfield  {journal} {\bibinfo  {journal} {The European Physical Journal
  Special Topics}\ }\textbf {\bibinfo {volume} {215}},\ \bibinfo {pages} {23}
  (\bibinfo {year} {2013})}\BibitemShut {NoStop}%
\bibitem [{\citenamefont {Radicchi}(2015)}]{radicchi2015percolation}%
  \BibitemOpen
  \bibfield  {author} {\bibinfo {author} {\bibfnamefont {F.}~\bibnamefont
  {Radicchi}},\ }\href@noop {} {\bibfield  {journal} {\bibinfo  {journal}
  {Nature Physics}\ }\textbf {\bibinfo {volume} {11}},\ \bibinfo {pages} {597}
  (\bibinfo {year} {2015})}\BibitemShut {NoStop}%
\bibitem [{\citenamefont {Rocha}(2017)}]{rocha2017dynamics}%
  \BibitemOpen
  \bibfield  {author} {\bibinfo {author} {\bibfnamefont {L.~E.}\ \bibnamefont
  {Rocha}},\ }\href@noop {} {\bibfield  {journal} {\bibinfo  {journal} {Chinese
  Journal of Aeronautics}\ }\textbf {\bibinfo {volume} {30}},\ \bibinfo {pages}
  {469} (\bibinfo {year} {2017})}\BibitemShut {NoStop}%
\bibitem [{\citenamefont {Berechman}\ and\ \citenamefont
  {de~Wit}(1996)}]{berechman1996analysis}%
  \BibitemOpen
  \bibfield  {author} {\bibinfo {author} {\bibfnamefont {J.}~\bibnamefont
  {Berechman}}\ and\ \bibinfo {author} {\bibfnamefont {J.}~\bibnamefont
  {de~Wit}},\ }\href@noop {} {\bibfield  {journal} {\bibinfo  {journal}
  {Journal of Transport Economics and Policy}\ ,\ \bibinfo {pages} {251}}
  (\bibinfo {year} {1996})}\BibitemShut {NoStop}%
\bibitem [{\citenamefont {Goetz}\ and\ \citenamefont
  {Sutton}(1997)}]{goetz1997geography}%
  \BibitemOpen
  \bibfield  {author} {\bibinfo {author} {\bibfnamefont {A.~R.}\ \bibnamefont
  {Goetz}}\ and\ \bibinfo {author} {\bibfnamefont {C.~J.}\ \bibnamefont
  {Sutton}},\ }\href@noop {} {\bibfield  {journal} {\bibinfo  {journal} {Annals
  of the Association of American Geographers}\ }\textbf {\bibinfo {volume}
  {87}},\ \bibinfo {pages} {238} (\bibinfo {year} {1997})}\BibitemShut
  {NoStop}%
\bibitem [{\citenamefont {Button}(2002)}]{button2002debunking}%
  \BibitemOpen
  \bibfield  {author} {\bibinfo {author} {\bibfnamefont {K.}~\bibnamefont
  {Button}},\ }\href@noop {} {\bibfield  {journal} {\bibinfo  {journal}
  {Journal of air transport management}\ }\textbf {\bibinfo {volume} {8}},\
  \bibinfo {pages} {177} (\bibinfo {year} {2002})}\BibitemShut {NoStop}%
\bibitem [{\citenamefont {Rey}(2003)}]{rey2003structural}%
  \BibitemOpen
  \bibfield  {author} {\bibinfo {author} {\bibfnamefont {M.~B.}\ \bibnamefont
  {Rey}},\ }\href@noop {} {\bibfield  {journal} {\bibinfo  {journal} {Journal
  of Air Transport Management}\ }\textbf {\bibinfo {volume} {9}},\ \bibinfo
  {pages} {195} (\bibinfo {year} {2003})}\BibitemShut {NoStop}%
\bibitem [{\citenamefont {Burghouwt}\ and\ \citenamefont
  {Hakfoort}(2001)}]{burghouwt2001evolution}%
  \BibitemOpen
  \bibfield  {author} {\bibinfo {author} {\bibfnamefont {G.}~\bibnamefont
  {Burghouwt}}\ and\ \bibinfo {author} {\bibfnamefont {J.}~\bibnamefont
  {Hakfoort}},\ }\href@noop {} {\bibfield  {journal} {\bibinfo  {journal}
  {Journal of Air Transport Management}\ }\textbf {\bibinfo {volume} {7}},\
  \bibinfo {pages} {311} (\bibinfo {year} {2001})}\BibitemShut {NoStop}%
\bibitem [{\citenamefont {Jin}\ \emph {et~al.}(2004)\citenamefont {Jin},
  \citenamefont {Wang},\ and\ \citenamefont {Liu}}]{jin2004geographic}%
  \BibitemOpen
  \bibfield  {author} {\bibinfo {author} {\bibfnamefont {F.}~\bibnamefont
  {Jin}}, \bibinfo {author} {\bibfnamefont {F.}~\bibnamefont {Wang}}, \ and\
  \bibinfo {author} {\bibfnamefont {Y.}~\bibnamefont {Liu}},\ }\href@noop {}
  {\bibfield  {journal} {\bibinfo  {journal} {The Professional Geographer}\
  }\textbf {\bibinfo {volume} {56}},\ \bibinfo {pages} {471} (\bibinfo {year}
  {2004})}\BibitemShut {NoStop}%
\bibitem [{\citenamefont {Burghouwt}\ and\ \citenamefont
  {De~Wit}(2005)}]{burghouwt2005temporal}%
  \BibitemOpen
  \bibfield  {author} {\bibinfo {author} {\bibfnamefont {G.}~\bibnamefont
  {Burghouwt}}\ and\ \bibinfo {author} {\bibfnamefont {J.}~\bibnamefont
  {De~Wit}},\ }\href@noop {} {\bibfield  {journal} {\bibinfo  {journal}
  {Journal of Air Transport Management}\ }\textbf {\bibinfo {volume} {11}},\
  \bibinfo {pages} {185} (\bibinfo {year} {2005})}\BibitemShut {NoStop}%
\bibitem [{\citenamefont {Malighetti}\ \emph {et~al.}(2008)\citenamefont
  {Malighetti}, \citenamefont {Paleari},\ and\ \citenamefont
  {Redondi}}]{malighetti2008connectivity}%
  \BibitemOpen
  \bibfield  {author} {\bibinfo {author} {\bibfnamefont {P.}~\bibnamefont
  {Malighetti}}, \bibinfo {author} {\bibfnamefont {S.}~\bibnamefont {Paleari}},
  \ and\ \bibinfo {author} {\bibfnamefont {R.}~\bibnamefont {Redondi}},\
  }\href@noop {} {\bibfield  {journal} {\bibinfo  {journal} {Journal of Air
  Transport Management}\ }\textbf {\bibinfo {volume} {14}},\ \bibinfo {pages}
  {53} (\bibinfo {year} {2008})}\BibitemShut {NoStop}%
\bibitem [{\citenamefont {Wang}\ and\ \citenamefont
  {Jin}(2007)}]{wang2007china}%
  \BibitemOpen
  \bibfield  {author} {\bibinfo {author} {\bibfnamefont {J.}~\bibnamefont
  {Wang}}\ and\ \bibinfo {author} {\bibfnamefont {F.}~\bibnamefont {Jin}},\
  }\href@noop {} {\bibfield  {journal} {\bibinfo  {journal} {Eurasian Geography
  and Economics}\ }\textbf {\bibinfo {volume} {48}},\ \bibinfo {pages} {469}
  (\bibinfo {year} {2007})}\BibitemShut {NoStop}%
\bibitem [{\citenamefont {da~Rocha}(2009)}]{da2009structural}%
  \BibitemOpen
  \bibfield  {author} {\bibinfo {author} {\bibfnamefont {L.~E.}\ \bibnamefont
  {da~Rocha}},\ }\href@noop {} {\bibfield  {journal} {\bibinfo  {journal}
  {Journal of Statistical Mechanics: Theory and Experiment}\ }\textbf {\bibinfo
  {volume} {2009}},\ \bibinfo {pages} {P04020} (\bibinfo {year}
  {2009})}\BibitemShut {NoStop}%
\bibitem [{\citenamefont {Zanin}\ \emph {et~al.}(2009)\citenamefont {Zanin},
  \citenamefont {Lacasa},\ and\ \citenamefont {Cea}}]{zanin2009dynamics}%
  \BibitemOpen
  \bibfield  {author} {\bibinfo {author} {\bibfnamefont {M.}~\bibnamefont
  {Zanin}}, \bibinfo {author} {\bibfnamefont {L.}~\bibnamefont {Lacasa}}, \
  and\ \bibinfo {author} {\bibfnamefont {M.}~\bibnamefont {Cea}},\ }\href@noop
  {} {\bibfield  {journal} {\bibinfo  {journal} {Chaos: An Interdisciplinary
  Journal of Nonlinear Science}\ }\textbf {\bibinfo {volume} {19}},\ \bibinfo
  {pages} {023111} (\bibinfo {year} {2009})}\BibitemShut {NoStop}%
\bibitem [{\citenamefont {Pan}\ and\ \citenamefont
  {Saram{\"a}ki}(2011)}]{pan2011path}%
  \BibitemOpen
  \bibfield  {author} {\bibinfo {author} {\bibfnamefont {R.~K.}\ \bibnamefont
  {Pan}}\ and\ \bibinfo {author} {\bibfnamefont {J.}~\bibnamefont
  {Saram{\"a}ki}},\ }\href@noop {} {\bibfield  {journal} {\bibinfo  {journal}
  {Physical Review E}\ }\textbf {\bibinfo {volume} {84}},\ \bibinfo {pages}
  {016105} (\bibinfo {year} {2011})}\BibitemShut {NoStop}%
\bibitem [{\citenamefont {Fleurquin}\ \emph {et~al.}(2013)\citenamefont
  {Fleurquin}, \citenamefont {Ramasco},\ and\ \citenamefont
  {Eguiluz}}]{fleurquin2013systemic}%
  \BibitemOpen
  \bibfield  {author} {\bibinfo {author} {\bibfnamefont {P.}~\bibnamefont
  {Fleurquin}}, \bibinfo {author} {\bibfnamefont {J.~J.}\ \bibnamefont
  {Ramasco}}, \ and\ \bibinfo {author} {\bibfnamefont {V.~M.}\ \bibnamefont
  {Eguiluz}},\ }\href@noop {} {\bibfield  {journal} {\bibinfo  {journal}
  {Scientific Reports}\ }\textbf {\bibinfo {volume} {3}},\ \bibinfo {pages}
  {1159} (\bibinfo {year} {2013})}\BibitemShut {NoStop}%
\bibitem [{\citenamefont {Colizza}\ \emph {et~al.}(2006)\citenamefont
  {Colizza}, \citenamefont {Barrat}, \citenamefont {Barth{\'e}lemy},\ and\
  \citenamefont {Vespignani}}]{colizza2006role}%
  \BibitemOpen
  \bibfield  {author} {\bibinfo {author} {\bibfnamefont {V.}~\bibnamefont
  {Colizza}}, \bibinfo {author} {\bibfnamefont {A.}~\bibnamefont {Barrat}},
  \bibinfo {author} {\bibfnamefont {M.}~\bibnamefont {Barth{\'e}lemy}}, \ and\
  \bibinfo {author} {\bibfnamefont {A.}~\bibnamefont {Vespignani}},\
  }\href@noop {} {\bibfield  {journal} {\bibinfo  {journal} {Proceedings of the
  National Academy of Sciences}\ }\textbf {\bibinfo {volume} {103}},\ \bibinfo
  {pages} {2015} (\bibinfo {year} {2006})}\BibitemShut {NoStop}%
\bibitem [{\citenamefont {Balcan}\ \emph {et~al.}(2009)\citenamefont {Balcan},
  \citenamefont {Colizza}, \citenamefont {Gon{\c{c}}alves}, \citenamefont {Hu},
  \citenamefont {Ramasco},\ and\ \citenamefont
  {Vespignani}}]{balcan2009multiscale}%
  \BibitemOpen
  \bibfield  {author} {\bibinfo {author} {\bibfnamefont {D.}~\bibnamefont
  {Balcan}}, \bibinfo {author} {\bibfnamefont {V.}~\bibnamefont {Colizza}},
  \bibinfo {author} {\bibfnamefont {B.}~\bibnamefont {Gon{\c{c}}alves}},
  \bibinfo {author} {\bibfnamefont {H.}~\bibnamefont {Hu}}, \bibinfo {author}
  {\bibfnamefont {J.~J.}\ \bibnamefont {Ramasco}}, \ and\ \bibinfo {author}
  {\bibfnamefont {A.}~\bibnamefont {Vespignani}},\ }\href@noop {} {\bibfield
  {journal} {\bibinfo  {journal} {Proceedings of the national academy of
  sciences}\ }\textbf {\bibinfo {volume} {106}},\ \bibinfo {pages} {21484}
  (\bibinfo {year} {2009})}\BibitemShut {NoStop}%
\bibitem [{\citenamefont {Balcan}\ and\ \citenamefont
  {Vespignani}(2011)}]{balcan2011phase}%
  \BibitemOpen
  \bibfield  {author} {\bibinfo {author} {\bibfnamefont {D.}~\bibnamefont
  {Balcan}}\ and\ \bibinfo {author} {\bibfnamefont {A.}~\bibnamefont
  {Vespignani}},\ }\href@noop {} {\bibfield  {journal} {\bibinfo  {journal}
  {Nature physics}\ }\textbf {\bibinfo {volume} {7}},\ \bibinfo {pages} {581}
  (\bibinfo {year} {2011})}\BibitemShut {NoStop}%
\bibitem [{\citenamefont {Brockmann}\ and\ \citenamefont
  {Helbing}(2013)}]{brockmann2013hidden}%
  \BibitemOpen
  \bibfield  {author} {\bibinfo {author} {\bibfnamefont {D.}~\bibnamefont
  {Brockmann}}\ and\ \bibinfo {author} {\bibfnamefont {D.}~\bibnamefont
  {Helbing}},\ }\href@noop {} {\bibfield  {journal} {\bibinfo  {journal}
  {science}\ }\textbf {\bibinfo {volume} {342}},\ \bibinfo {pages} {1337}
  (\bibinfo {year} {2013})}\BibitemShut {NoStop}%
\bibitem [{\citenamefont {Pastor-Satorras}\ \emph {et~al.}(2015)\citenamefont
  {Pastor-Satorras}, \citenamefont {Castellano}, \citenamefont {Van~Mieghem},\
  and\ \citenamefont {Vespignani}}]{pastor2015epidemic}%
  \BibitemOpen
  \bibfield  {author} {\bibinfo {author} {\bibfnamefont {R.}~\bibnamefont
  {Pastor-Satorras}}, \bibinfo {author} {\bibfnamefont {C.}~\bibnamefont
  {Castellano}}, \bibinfo {author} {\bibfnamefont {P.}~\bibnamefont
  {Van~Mieghem}}, \ and\ \bibinfo {author} {\bibfnamefont {A.}~\bibnamefont
  {Vespignani}},\ }\href@noop {} {\bibfield  {journal} {\bibinfo  {journal}
  {Reviews of modern physics}\ }\textbf {\bibinfo {volume} {87}},\ \bibinfo
  {pages} {925} (\bibinfo {year} {2015})}\BibitemShut {NoStop}%
\bibitem [{\citenamefont {Rosvall}\ \emph {et~al.}(2014)\citenamefont
  {Rosvall}, \citenamefont {Esquivel}, \citenamefont {Lancichinetti},
  \citenamefont {West},\ and\ \citenamefont {Lambiotte}}]{rosvall2014memory}%
  \BibitemOpen
  \bibfield  {author} {\bibinfo {author} {\bibfnamefont {M.}~\bibnamefont
  {Rosvall}}, \bibinfo {author} {\bibfnamefont {A.~V.}\ \bibnamefont
  {Esquivel}}, \bibinfo {author} {\bibfnamefont {A.}~\bibnamefont
  {Lancichinetti}}, \bibinfo {author} {\bibfnamefont {J.~D.}\ \bibnamefont
  {West}}, \ and\ \bibinfo {author} {\bibfnamefont {R.}~\bibnamefont
  {Lambiotte}},\ }\href@noop {} {\bibfield  {journal} {\bibinfo  {journal}
  {Nature communications}\ }\textbf {\bibinfo {volume} {5}},\ \bibinfo {pages}
  {4630} (\bibinfo {year} {2014})}\BibitemShut {NoStop}%
\bibitem [{\citenamefont {Scholtes}(2017)}]{scholtes2017network}%
  \BibitemOpen
  \bibfield  {author} {\bibinfo {author} {\bibfnamefont {I.}~\bibnamefont
  {Scholtes}},\ }in\ \href@noop {} {\emph {\bibinfo {booktitle} {Proceedings of
  the 23rd ACM SIGKDD International Conference on Knowledge Discovery and Data
  Mining}}}\ (\bibinfo {organization} {ACM},\ \bibinfo {year} {2017})\ pp.\
  \bibinfo {pages} {1037--1046}\BibitemShut {NoStop}%
\bibitem [{\citenamefont {Peixoto}\ and\ \citenamefont
  {Rosvall}(2017)}]{peixoto2017modelling}%
  \BibitemOpen
  \bibfield  {author} {\bibinfo {author} {\bibfnamefont {T.~P.}\ \bibnamefont
  {Peixoto}}\ and\ \bibinfo {author} {\bibfnamefont {M.}~\bibnamefont
  {Rosvall}},\ }\href@noop {} {\bibfield  {journal} {\bibinfo  {journal}
  {Nature communications}\ }\textbf {\bibinfo {volume} {8}},\ \bibinfo {pages}
  {582} (\bibinfo {year} {2017})}\BibitemShut {NoStop}%
\bibitem [{\citenamefont {Stauffer}\ and\ \citenamefont
  {Aharony}(1985)}]{stauffer2014introduction}%
  \BibitemOpen
  \bibfield  {author} {\bibinfo {author} {\bibfnamefont {D.}~\bibnamefont
  {Stauffer}}\ and\ \bibinfo {author} {\bibfnamefont {A.}~\bibnamefont
  {Aharony}},\ }\href@noop {} {\emph {\bibinfo {title} {Introduction to
  percolation theory}}}\ (\bibinfo  {publisher} {CRC press},\ \bibinfo {year}
  {1985})\BibitemShut {NoStop}%
\bibitem [{\citenamefont {Newman}(2018)}]{newman2018networks}%
  \BibitemOpen
  \bibfield  {author} {\bibinfo {author} {\bibfnamefont {M.}~\bibnamefont
  {Newman}},\ }\href@noop {} {\emph {\bibinfo {title} {Networks}}}\ (\bibinfo
  {publisher} {Oxford university press},\ \bibinfo {year} {2018})\BibitemShut
  {NoStop}%
\bibitem [{\citenamefont {Callaway}\ \emph {et~al.}(2000)\citenamefont
  {Callaway}, \citenamefont {Newman}, \citenamefont {Strogatz},\ and\
  \citenamefont {Watts}}]{callaway2000network}%
  \BibitemOpen
  \bibfield  {author} {\bibinfo {author} {\bibfnamefont {D.~S.}\ \bibnamefont
  {Callaway}}, \bibinfo {author} {\bibfnamefont {M.~E.}\ \bibnamefont
  {Newman}}, \bibinfo {author} {\bibfnamefont {S.~H.}\ \bibnamefont
  {Strogatz}}, \ and\ \bibinfo {author} {\bibfnamefont {D.~J.}\ \bibnamefont
  {Watts}},\ }\href@noop {} {\bibfield  {journal} {\bibinfo  {journal}
  {Physical review letters}\ }\textbf {\bibinfo {volume} {85}},\ \bibinfo
  {pages} {5468} (\bibinfo {year} {2000})}\BibitemShut {NoStop}%
\bibitem [{\citenamefont {Albert}\ \emph {et~al.}(2000)\citenamefont {Albert},
  \citenamefont {Jeong},\ and\ \citenamefont {Barab{\'a}si}}]{albert2000error}%
  \BibitemOpen
  \bibfield  {author} {\bibinfo {author} {\bibfnamefont {R.}~\bibnamefont
  {Albert}}, \bibinfo {author} {\bibfnamefont {H.}~\bibnamefont {Jeong}}, \
  and\ \bibinfo {author} {\bibfnamefont {A.-L.}\ \bibnamefont {Barab{\'a}si}},\
  }\href@noop {} {\bibfield  {journal} {\bibinfo  {journal} {nature}\ }\textbf
  {\bibinfo {volume} {406}},\ \bibinfo {pages} {378} (\bibinfo {year}
  {2000})}\BibitemShut {NoStop}%
\bibitem [{\citenamefont {Karrer}\ \emph {et~al.}(2014)\citenamefont {Karrer},
  \citenamefont {Newman},\ and\ \citenamefont
  {Zdeborov{\'a}}}]{karrer2014percolation}%
  \BibitemOpen
  \bibfield  {author} {\bibinfo {author} {\bibfnamefont {B.}~\bibnamefont
  {Karrer}}, \bibinfo {author} {\bibfnamefont {M.~E.}\ \bibnamefont {Newman}},
  \ and\ \bibinfo {author} {\bibfnamefont {L.}~\bibnamefont {Zdeborov{\'a}}},\
  }\href@noop {} {\bibfield  {journal} {\bibinfo  {journal} {Physical review
  letters}\ }\textbf {\bibinfo {volume} {113}},\ \bibinfo {pages} {208702}
  (\bibinfo {year} {2014})}\BibitemShut {NoStop}%
\bibitem [{\citenamefont {Buldyrev}\ \emph {et~al.}(2010)\citenamefont
  {Buldyrev}, \citenamefont {Parshani}, \citenamefont {Paul}, \citenamefont
  {Stanley},\ and\ \citenamefont {Havlin}}]{buldyrev2010catastrophic}%
  \BibitemOpen
  \bibfield  {author} {\bibinfo {author} {\bibfnamefont {S.~V.}\ \bibnamefont
  {Buldyrev}}, \bibinfo {author} {\bibfnamefont {R.}~\bibnamefont {Parshani}},
  \bibinfo {author} {\bibfnamefont {G.}~\bibnamefont {Paul}}, \bibinfo {author}
  {\bibfnamefont {H.~E.}\ \bibnamefont {Stanley}}, \ and\ \bibinfo {author}
  {\bibfnamefont {S.}~\bibnamefont {Havlin}},\ }\href@noop {} {\bibfield
  {journal} {\bibinfo  {journal} {Nature}\ }\textbf {\bibinfo {volume} {464}},\
  \bibinfo {pages} {1025} (\bibinfo {year} {2010})}\BibitemShut {NoStop}%
\bibitem [{\citenamefont {Shen}\ \emph {et~al.}(2012)\citenamefont {Shen},
  \citenamefont {Smith},\ and\ \citenamefont {Goli}}]{shen2012exact}%
  \BibitemOpen
  \bibfield  {author} {\bibinfo {author} {\bibfnamefont {S.}~\bibnamefont
  {Shen}}, \bibinfo {author} {\bibfnamefont {J.~C.}\ \bibnamefont {Smith}}, \
  and\ \bibinfo {author} {\bibfnamefont {R.}~\bibnamefont {Goli}},\ }\href@noop
  {} {\bibfield  {journal} {\bibinfo  {journal} {Discrete Optimization}\
  }\textbf {\bibinfo {volume} {9}},\ \bibinfo {pages} {172} (\bibinfo {year}
  {2012})}\BibitemShut {NoStop}%
\bibitem [{\citenamefont {Shen}\ and\ \citenamefont
  {Smith}(2012)}]{shen2012polynomial}%
  \BibitemOpen
  \bibfield  {author} {\bibinfo {author} {\bibfnamefont {S.}~\bibnamefont
  {Shen}}\ and\ \bibinfo {author} {\bibfnamefont {J.~C.}\ \bibnamefont
  {Smith}},\ }\href@noop {} {\bibfield  {journal} {\bibinfo  {journal}
  {Networks}\ }\textbf {\bibinfo {volume} {60}},\ \bibinfo {pages} {103}
  (\bibinfo {year} {2012})}\BibitemShut {NoStop}%
\bibitem [{\citenamefont {Morone}\ and\ \citenamefont
  {Makse}(2015)}]{morone2015influence}%
  \BibitemOpen
  \bibfield  {author} {\bibinfo {author} {\bibfnamefont {F.}~\bibnamefont
  {Morone}}\ and\ \bibinfo {author} {\bibfnamefont {H.~A.}\ \bibnamefont
  {Makse}},\ }\href@noop {} {\bibfield  {journal} {\bibinfo  {journal}
  {Nature}\ }\textbf {\bibinfo {volume} {524}},\ \bibinfo {pages} {65}
  (\bibinfo {year} {2015})}\BibitemShut {NoStop}%
\bibitem [{\citenamefont {Braunstein}\ \emph {et~al.}(2016)\citenamefont
  {Braunstein}, \citenamefont {Dall’Asta}, \citenamefont {Semerjian},\ and\
  \citenamefont {Zdeborov{\'a}}}]{braunstein2016network}%
  \BibitemOpen
  \bibfield  {author} {\bibinfo {author} {\bibfnamefont {A.}~\bibnamefont
  {Braunstein}}, \bibinfo {author} {\bibfnamefont {L.}~\bibnamefont
  {Dall’Asta}}, \bibinfo {author} {\bibfnamefont {G.}~\bibnamefont
  {Semerjian}}, \ and\ \bibinfo {author} {\bibfnamefont {L.}~\bibnamefont
  {Zdeborov{\'a}}},\ }\href@noop {} {\bibfield  {journal} {\bibinfo  {journal}
  {Proceedings of the National Academy of Sciences}\ }\textbf {\bibinfo
  {volume} {113}},\ \bibinfo {pages} {12368} (\bibinfo {year}
  {2016})}\BibitemShut {NoStop}%
\bibitem [{\citenamefont {Clusella}\ \emph {et~al.}(2016)\citenamefont
  {Clusella}, \citenamefont {Grassberger}, \citenamefont {P{\'e}rez-Reche},\
  and\ \citenamefont {Politi}}]{clusella2016immunization}%
  \BibitemOpen
  \bibfield  {author} {\bibinfo {author} {\bibfnamefont {P.}~\bibnamefont
  {Clusella}}, \bibinfo {author} {\bibfnamefont {P.}~\bibnamefont
  {Grassberger}}, \bibinfo {author} {\bibfnamefont {F.~J.}\ \bibnamefont
  {P{\'e}rez-Reche}}, \ and\ \bibinfo {author} {\bibfnamefont {A.}~\bibnamefont
  {Politi}},\ }\href@noop {} {\bibfield  {journal} {\bibinfo  {journal}
  {Physical review letters}\ }\textbf {\bibinfo {volume} {117}},\ \bibinfo
  {pages} {208301} (\bibinfo {year} {2016})}\BibitemShut {NoStop}%
\bibitem [{\citenamefont {Osat}\ \emph {et~al.}(2017)\citenamefont {Osat},
  \citenamefont {Faqeeh},\ and\ \citenamefont {Radicchi}}]{osat2017optimal}%
  \BibitemOpen
  \bibfield  {author} {\bibinfo {author} {\bibfnamefont {S.}~\bibnamefont
  {Osat}}, \bibinfo {author} {\bibfnamefont {A.}~\bibnamefont {Faqeeh}}, \ and\
  \bibinfo {author} {\bibfnamefont {F.}~\bibnamefont {Radicchi}},\ }\href@noop
  {} {\bibfield  {journal} {\bibinfo  {journal} {Nature Communications}\
  }\textbf {\bibinfo {volume} {8}},\ \bibinfo {pages} {1540} (\bibinfo {year}
  {2017})}\BibitemShut {NoStop}%
\bibitem [{\citenamefont {Kim}\ and\ \citenamefont
  {Radicchi}(2024)}]{kim2024shortest}%
  \BibitemOpen
  \bibfield  {author} {\bibinfo {author} {\bibfnamefont {M.}~\bibnamefont
  {Kim}}\ and\ \bibinfo {author} {\bibfnamefont {F.}~\bibnamefont {Radicchi}},\
  }\href@noop {} {\bibfield  {journal} {\bibinfo  {journal} {Physical Review
  Letters}\ }\textbf {\bibinfo {volume} {133}},\ \bibinfo {pages} {047402}
  (\bibinfo {year} {2024})}\BibitemShut {NoStop}%
\bibitem [{bts()}]{bts}%
  \BibitemOpen
  \href@noop {} {\enquote {\bibinfo {title} {{Bureau of Transportation, Airline
  Origin and Destination Survey}},}\ }\bibinfo {howpublished}
  {\url{https://www.transtats.bts.gov}}\BibitemShut {NoStop}%
\bibitem [{faa()}]{faa}%
  \BibitemOpen
  \href@noop {} {\enquote {\bibinfo {title} {Federal aviation administration
  registry},}\ }\bibinfo {howpublished}
  {\url{https://registry.faa.gov/aircraftinquiry/search/nnumberinquiry}}\BibitemShut
  {NoStop}%
\bibitem [{\citenamefont {Vazifeh}\ \emph {et~al.}(2018)\citenamefont
  {Vazifeh}, \citenamefont {Santi}, \citenamefont {Resta}, \citenamefont
  {Strogatz},\ and\ \citenamefont {Ratti}}]{vazifeh2018addressing}%
  \BibitemOpen
  \bibfield  {author} {\bibinfo {author} {\bibfnamefont {M.~M.}\ \bibnamefont
  {Vazifeh}}, \bibinfo {author} {\bibfnamefont {P.}~\bibnamefont {Santi}},
  \bibinfo {author} {\bibfnamefont {G.}~\bibnamefont {Resta}}, \bibinfo
  {author} {\bibfnamefont {S.~H.}\ \bibnamefont {Strogatz}}, \ and\ \bibinfo
  {author} {\bibfnamefont {C.}~\bibnamefont {Ratti}},\ }\href@noop {}
  {\bibfield  {journal} {\bibinfo  {journal} {Nature}\ }\textbf {\bibinfo
  {volume} {557}},\ \bibinfo {pages} {534} (\bibinfo {year}
  {2018})}\BibitemShut {NoStop}%
\bibitem [{apr()}]{april23}%
  \BibitemOpen
  \href@noop {} {\enquote {\bibinfo {title} {{April 2023 U.S. Airline Traffic
  Data Up 7.8\% from the Same Month Last Year }},}\ }\bibinfo {howpublished}
  {\url{https://www.bts.gov/newsroom/april-2023-us-airline-traffic-data-78-same-month-last-year}}\BibitemShut
  {NoStop}%
\bibitem [{\citenamefont {Eskenazi}\ \emph {et~al.}(2023)\citenamefont
  {Eskenazi}, \citenamefont {Joshi}, \citenamefont {Butler},\ and\
  \citenamefont {Ryerson}}]{joshi2022equitable}%
  \BibitemOpen
  \bibfield  {author} {\bibinfo {author} {\bibfnamefont {A.~G.}\ \bibnamefont
  {Eskenazi}}, \bibinfo {author} {\bibfnamefont {A.~P.}\ \bibnamefont {Joshi}},
  \bibinfo {author} {\bibfnamefont {L.~G.}\ \bibnamefont {Butler}}, \ and\
  \bibinfo {author} {\bibfnamefont {M.~S.}\ \bibnamefont {Ryerson}},\ }\href
  {\doibase https://doi.org/10.1016/j.compenvurbsys.2023.101973} {\bibfield
  {journal} {\bibinfo  {journal} {Computers, Environment and Urban Systems}\
  }\textbf {\bibinfo {volume} {102}},\ \bibinfo {pages} {101973} (\bibinfo
  {year} {2023})}\BibitemShut {NoStop}%
\bibitem [{\citenamefont {Zipf}(1946)}]{zipf1946p}%
  \BibitemOpen
  \bibfield  {author} {\bibinfo {author} {\bibfnamefont {G.~K.}\ \bibnamefont
  {Zipf}},\ }\href@noop {} {\bibfield  {journal} {\bibinfo  {journal} {American
  sociological review}\ }\textbf {\bibinfo {volume} {11}},\ \bibinfo {pages}
  {677} (\bibinfo {year} {1946})}\BibitemShut {NoStop}%
\bibitem [{\citenamefont {Erlander}\ and\ \citenamefont
  {Stewart}(1990)}]{erlander1990gravity}%
  \BibitemOpen
  \bibfield  {author} {\bibinfo {author} {\bibfnamefont {S.}~\bibnamefont
  {Erlander}}\ and\ \bibinfo {author} {\bibfnamefont {N.~F.}\ \bibnamefont
  {Stewart}},\ }\href@noop {} {\emph {\bibinfo {title} {The gravity model in
  transportation analysis: theory and extensions}}},\ Vol.~\bibinfo {volume}
  {3}\ (\bibinfo  {publisher} {Vsp},\ \bibinfo {year} {1990})\BibitemShut
  {NoStop}%
\bibitem [{\citenamefont {for International Earth Science Information Network
  CIESIN Columbia~University}(2018)}]{GPWv4_2018}%
  \BibitemOpen
  \bibfield  {author} {\bibinfo {author} {\bibfnamefont {C.}~\bibnamefont {for
  International Earth Science Information Network CIESIN
  Columbia~University}},\ }\href {https://doi.org/10.7927/H4F47M65} {\enquote
  {\bibinfo {title} {Gridded population of the world, version 4 (gpwv4):
  Population density adjusted to match 2015 revision un wpp country totals,
  revision 11},}\ } (\bibinfo {year} {2018}),\ \bibinfo {note} {accessed:
  2024-06-07}\BibitemShut {NoStop}%
\bibitem [{\citenamefont {Ester}\ \emph {et~al.}(1996)\citenamefont {Ester},
  \citenamefont {Kriegel}, \citenamefont {Sander}, \citenamefont {Xu} \emph
  {et~al.}}]{ester1996density}%
  \BibitemOpen
  \bibfield  {author} {\bibinfo {author} {\bibfnamefont {M.}~\bibnamefont
  {Ester}}, \bibinfo {author} {\bibfnamefont {H.-P.}\ \bibnamefont {Kriegel}},
  \bibinfo {author} {\bibfnamefont {J.}~\bibnamefont {Sander}}, \bibinfo
  {author} {\bibfnamefont {X.}~\bibnamefont {Xu}},  \emph {et~al.},\ }in\
  \href@noop {} {\emph {\bibinfo {booktitle} {kdd}}},\ Vol.~\bibinfo {volume}
  {96}\ (\bibinfo {year} {1996})\ pp.\ \bibinfo {pages} {226--231}\BibitemShut
  {NoStop}%
\bibitem [{del()}]{deltastrike}%
  \BibitemOpen
  \href@noop {} {\enquote {\bibinfo {title} {{2024 Delta Air Lines
  disruption}},}\ }\bibinfo {howpublished}
  {\url{https://en.wikipedia.org/wiki/2024_Delta_Air_Lines_disruption}}\BibitemShut
  {NoStop}%
\bibitem [{ope()}]{openflights}%
  \BibitemOpen
  \href@noop {} {\enquote {\bibinfo {title} {{Open Flights}},}\ }\bibinfo
  {howpublished} {\url{https://openflights.org}}\BibitemShut {NoStop}%
\bibitem [{\citenamefont {Dijkstra}(2022)}]{dijkstra2022note}%
  \BibitemOpen
  \bibfield  {author} {\bibinfo {author} {\bibfnamefont {E.~W.}\ \bibnamefont
  {Dijkstra}},\ }in\ \href@noop {} {\emph {\bibinfo {booktitle} {Edsger Wybe
  Dijkstra: his life, work, and legacy}}}\ (\bibinfo {year} {2022})\ pp.\
  \bibinfo {pages} {287--290}\BibitemShut {NoStop}%
\end{thebibliography}%

\newpage
\clearpage
\onecolumngrid

\renewcommand{\theequation}{S\arabic{equation}}
\setcounter{equation}{0}
\renewcommand{\thefigure}{S\arabic{figure}}
\setcounter{figure}{0}
\renewcommand{\thetable}{S\arabic{table}}
\setcounter{table}{0}
\setcounter{page}{1}

\section*{Supplementary Material}


\begin{figure*}[!htb]
    \includegraphics[width=0.75\linewidth]{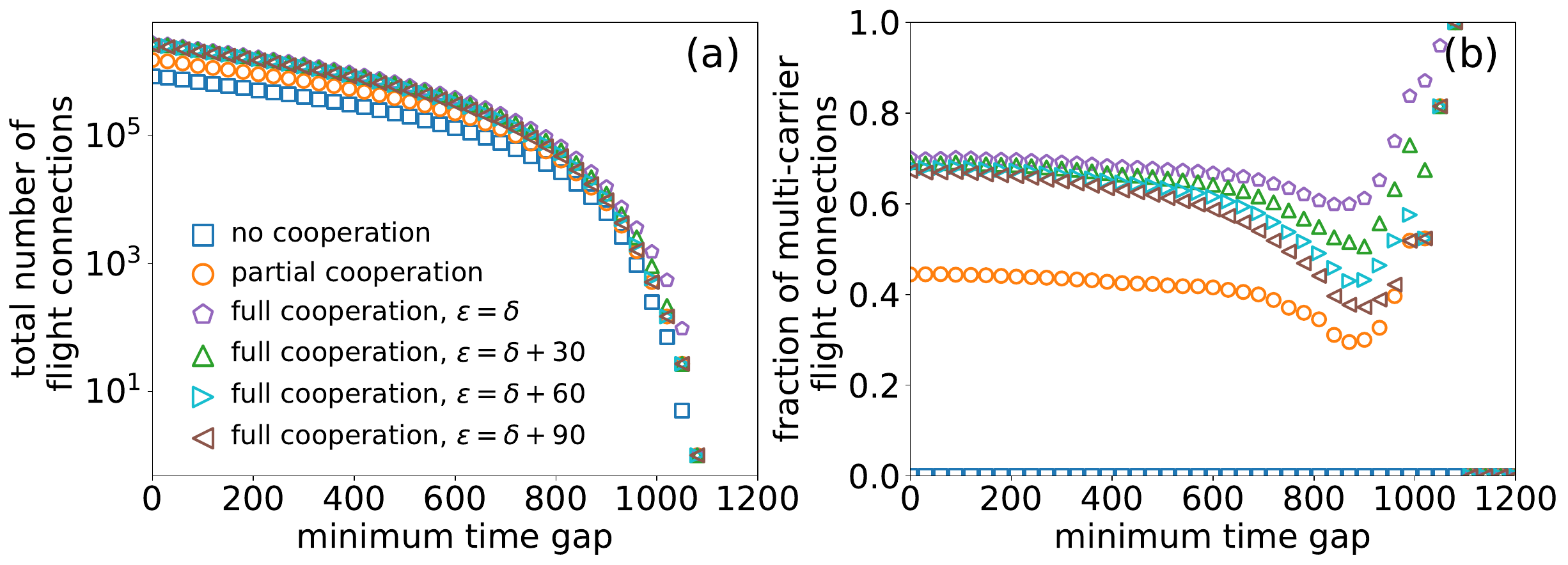}
    \caption{
    {\bf Properties of the flight-connection networks (FCNs).}
    {\bf (a)} Total number of connections between flights as a function the minimum time gap allowed for connections between same-carrier flights, i.e., $\delta$. Time is measured in minutes.
    Different colors/shapes of the symbols represent different settings used to establish connections between multi-carrier/alliance flights. If no cooperation is allowed, then no inter-carrier connections are permitted. For partial cooperation, commercial partners have connected flights as long as their time gap is greater than $\delta$.
    For full cooperation, inter-carrier/alliance connections are allowed only if the minimum time gap between flights exceeds the value of the parameter $\epsilon$. The various FCNs are constructed using data about flights operated between pairs of airports in the contiguous United States on April 18, 2023.
    {\bf (b)} Same data as in (a), but we display the proportion of multi-carrier connections in the FCN as a function the minimum time gap allowed for connections between same-carrier flights.
    }
    \label{fig_sm:fcn-statistics}
\end{figure*}


\begin{table*}[!htb]
\renewcommand{\arraystretch}{1.2}
\centering
\begin{tabularx}{\textwidth}{XXXXXXXXXXXX}
 \hline
 \hline
 Date  & $\tilde{N}$ & $\bar{N}$ & $\hat{N}$ & $N$ & $S$  & $C$ & $\tilde{V}$ & $V$ & $E^{(nc)}$ & $E^{(pc)}$ & $E^{(fc)}$\\ 
 \hline
 2023-04-18 & {$19458$} & {$19360$} & {$18640$} & {$18545$} & {$2794066$} & {$19$} & {$300$} & {$278$} & ${81369}$ & $1464725$ & $2613625$\\ 
 
 \hline
 2019-04-18 & {$23283$} & {$22740$} & {$22512$} & {$21971$} & {$2981889$} & {$22$} & {$314$} & {$290$} & ${1015579}$ & $2081466$ & $3703999$\\ 
 
 \hline
 2023-11-22 & {$21316$} & {$20824$} & {$20439$} & {$19947$} & {$3033540$} & {$19$} & {$300$} & {$278$} & ${939715}$ & $1625289$ & $2981685$\\ 
 \hline 
 \hline 
\end{tabularx}
\caption{
\textbf{Flight-connection networks for different daily schedules.}
From left to right, we report the day of the schedule considered, the original number of flights $\tilde{N}$, the number of flights after discarding flights with missing aircraft information ($\bar{N}$), the number of flights after discarding flights involving at least one airport outside of the contiguous US ($\hat{N}$), the number of flights after discarding flights either with missing aircraft information or flights involving at least one airport outside outside of the contiguous US ($N$), the total number of seats available on these latter flights ($S$), the number of operating carriers ($C$), the number of airports ($\tilde{V}$), the number of super-airports after the application of DBSCAN ($V$), the numbers of connections between flights without cooperation ($E^{(nc)}$), with partial cooperation ($E^{(pc)}$), and with full cooperation ($E^{(fc)}$). The latter are obtained by setting the parameters $\delta = 30$ minutes and $\epsilon = 60$ minutes while constructing the flight-connection networks.}
\label{tab:fcn-table}
\end{table*}


\begin{figure*}[!htb]
    \includegraphics[width = 0.65\textwidth]{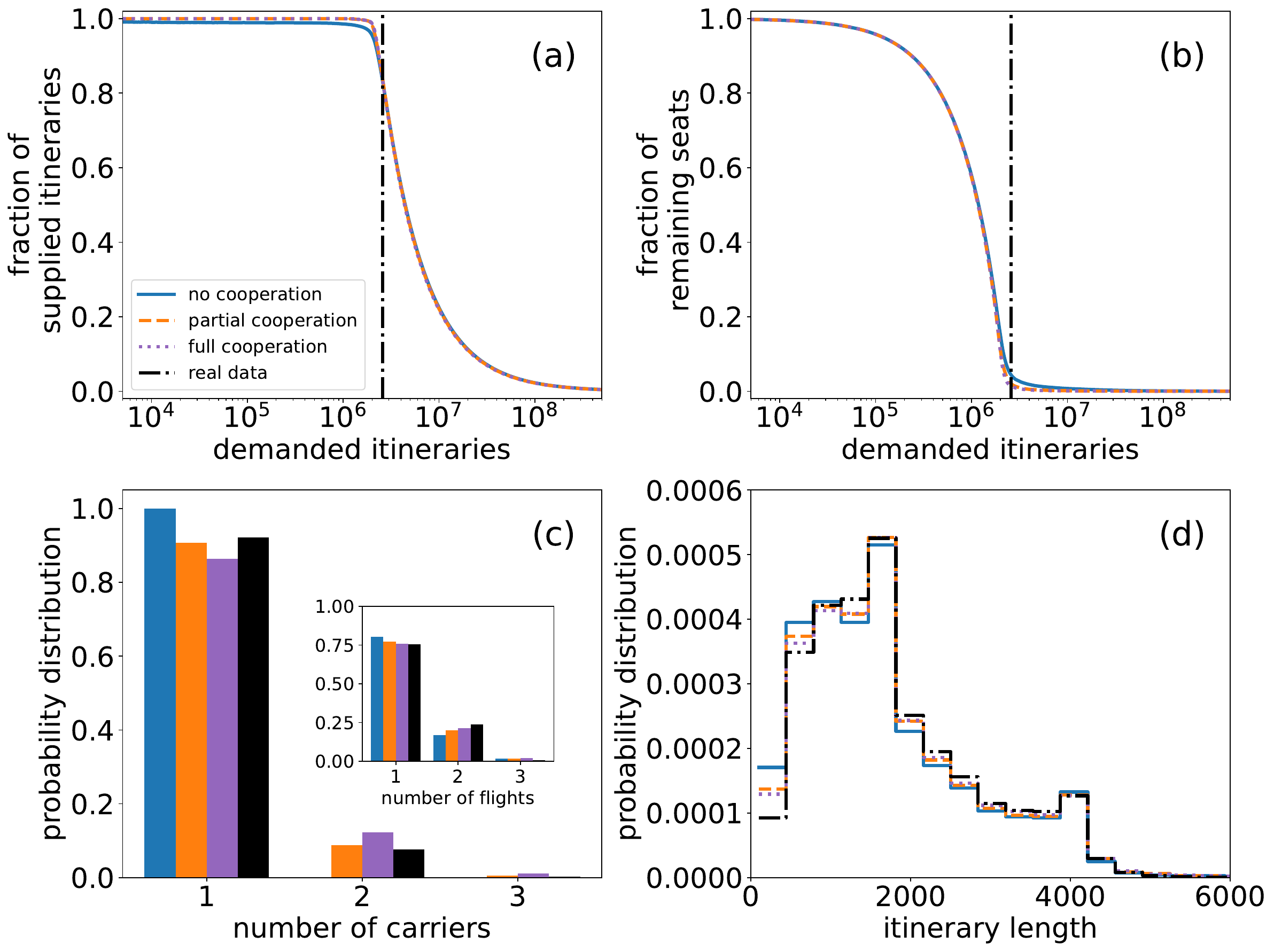}
    \caption{{\bf Validation of the minimum-cost-percolation (MCP) model.}
    Same as in Figure~\ref{fig:2} of the main paper, but results for the MCP model are obtained using the duration of the itineraries as the cost function to be optimized by the agents.
    }
    \label{fig_sm:fcn-validation-duration-Y2023M4D18}
\end{figure*}

\begin{figure*}[!htb]
    \includegraphics[width = 0.65\textwidth]{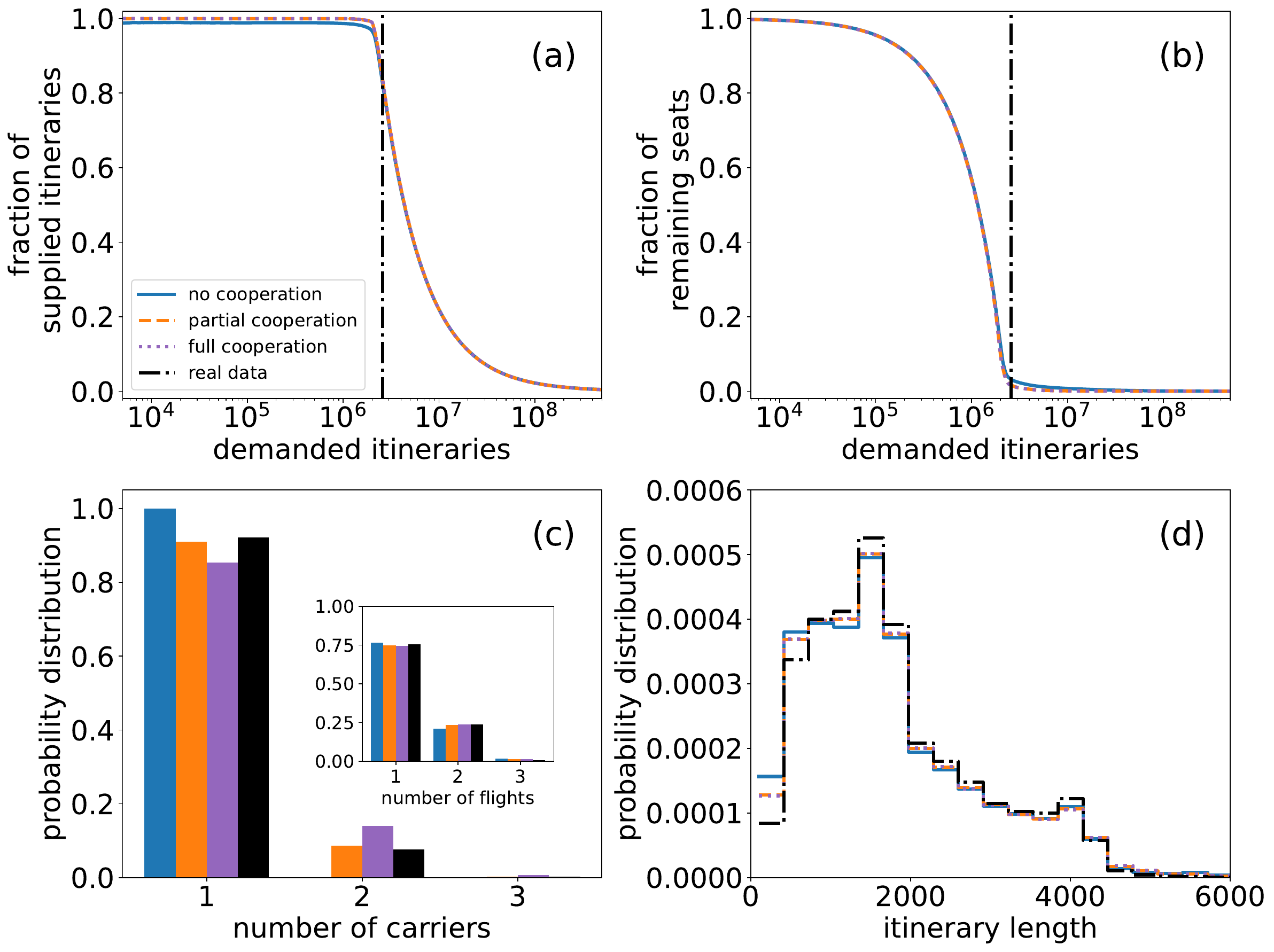}
    \caption{{\bf Validation of the minimum-cost-percolation (MCP) model.}
    Same as in Figure~\ref{fig:2} of the main paper, but results for the MCP model are obtained using the seat availability of the itineraries as the cost function to be optimized by the agents.
    }
    \label{fig_sm:fcn-validation-seats-Y2023M4D18}
\end{figure*}

\clearpage

\begin{figure*}[!htb]
    \includegraphics[width = 0.65\textwidth]{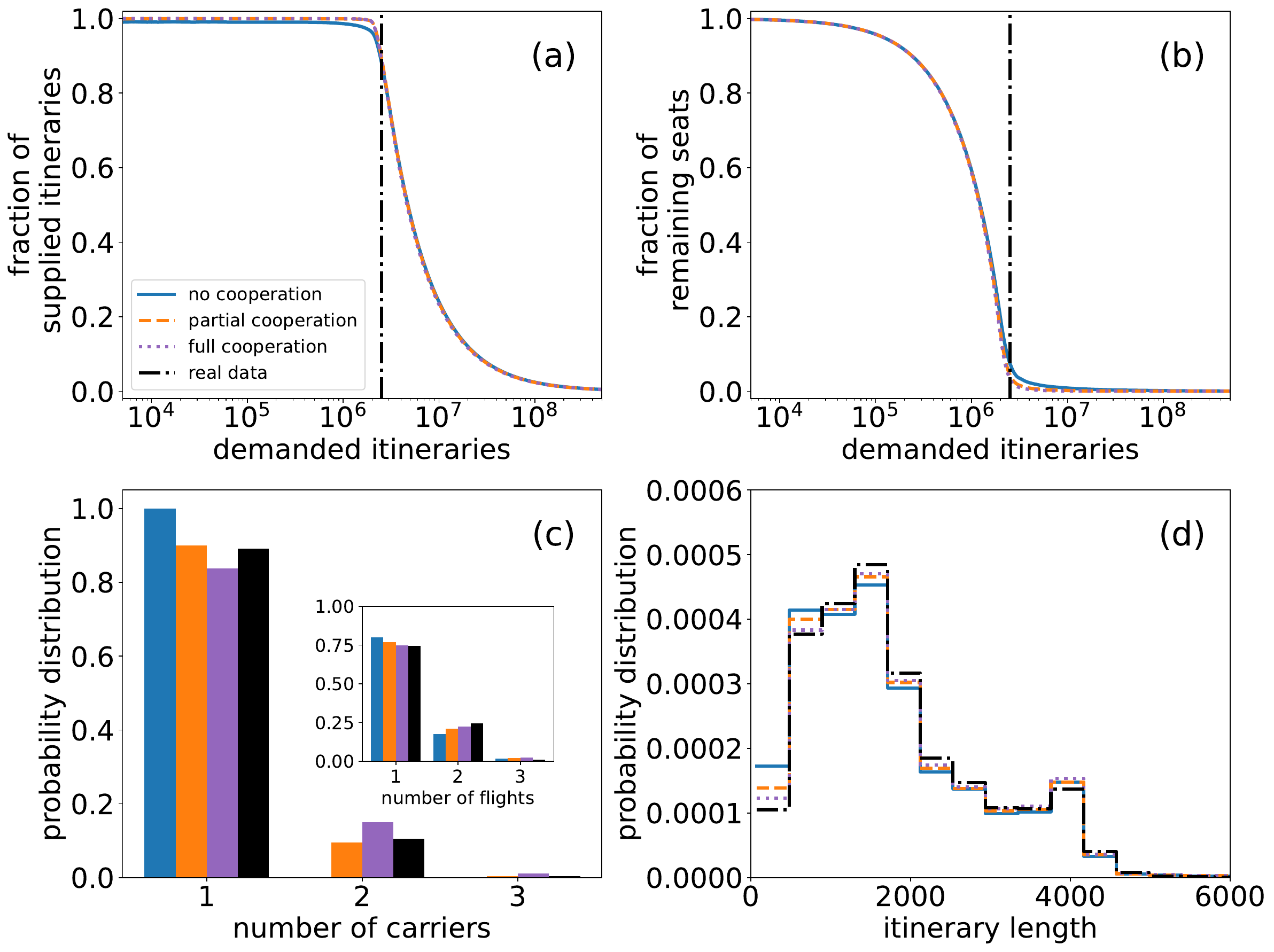}
    \caption{{\bf Validation of the minimum-cost-percolation (MCP) model.}
    Same as in Figure~\ref{fig:2}, but obtained from the flight schedule of April 18, 2019, using the length of itineraries as the cost function and data about sold tickets of the second quarter of 2019.
    The vertical line in panels (a) and (b) corresponds to the average number of daily served passengers for April 2019, see \url{https://www.bts.gov/newsroom/estimated-april-2019-us-airline-traffic-data}.
        }
    \label{fig_sm:fcn-validation-length-Y2019M4D18}
\end{figure*}


\begin{figure*}[!htb]
    \includegraphics[width = 0.65\textwidth]{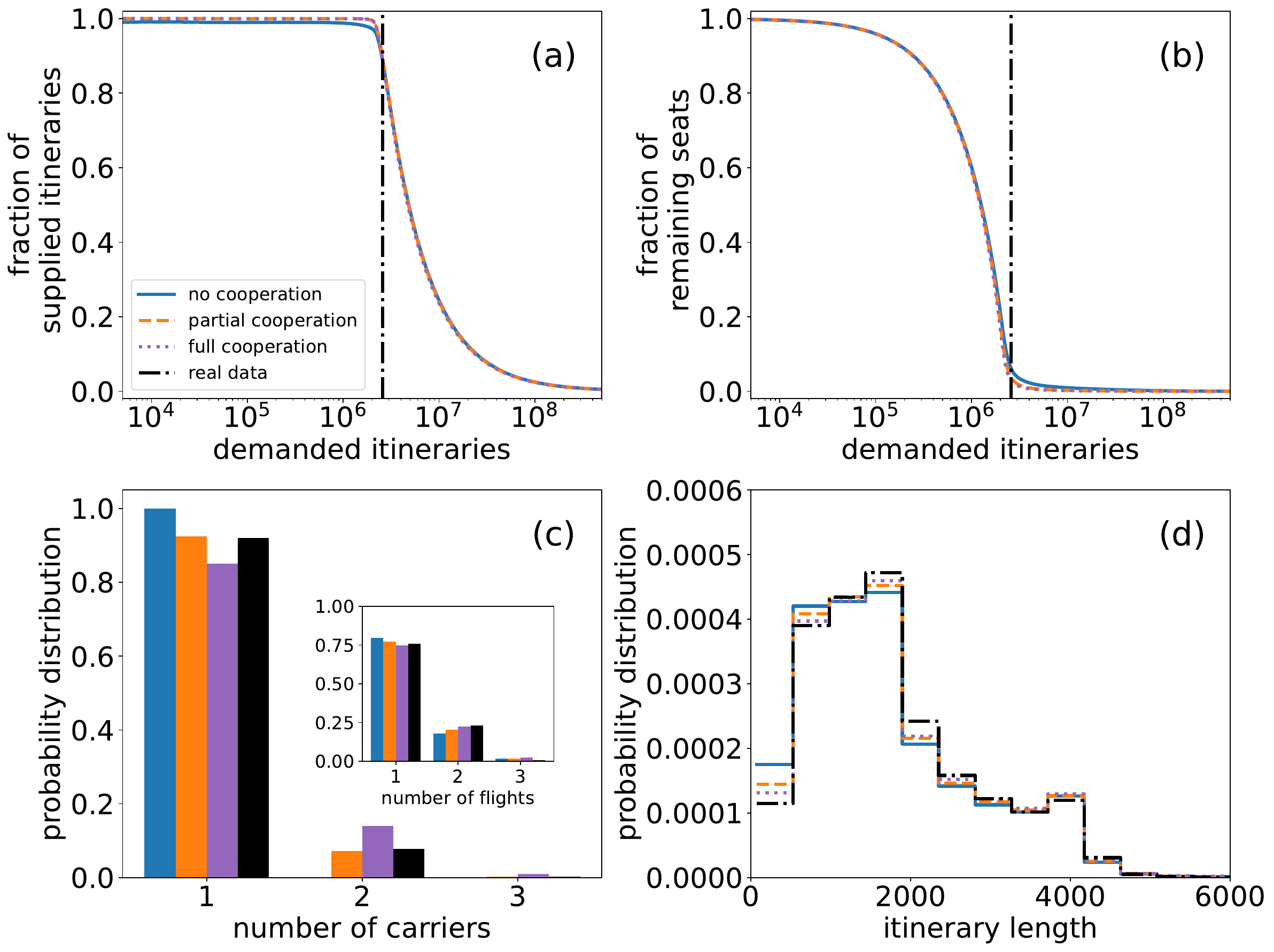}
    \caption{{\bf Validation of the minimum-cost-percolation (MCP) model.}
    Same as in Figure~\ref{fig:2}, but obtained from the flight schedule of November 22, 2023, using the distance of itineraries as the cost function and data about sold tickets of the fourth quarter of 2023.
    The vertical line in panels (a) and (b) corresponds to the average number of daily served passengers for November 2023, see \url{https://www.bts.gov/newsroom/november-2023-us-airline-traffic-data-81-same-month-2022}.
    }
    \label{fig_sm:fcn-validation-length-Y2023M11D22}
\end{figure*}

\clearpage

\begin{figure*}[!htb]
    \includegraphics[width=0.65\textwidth]{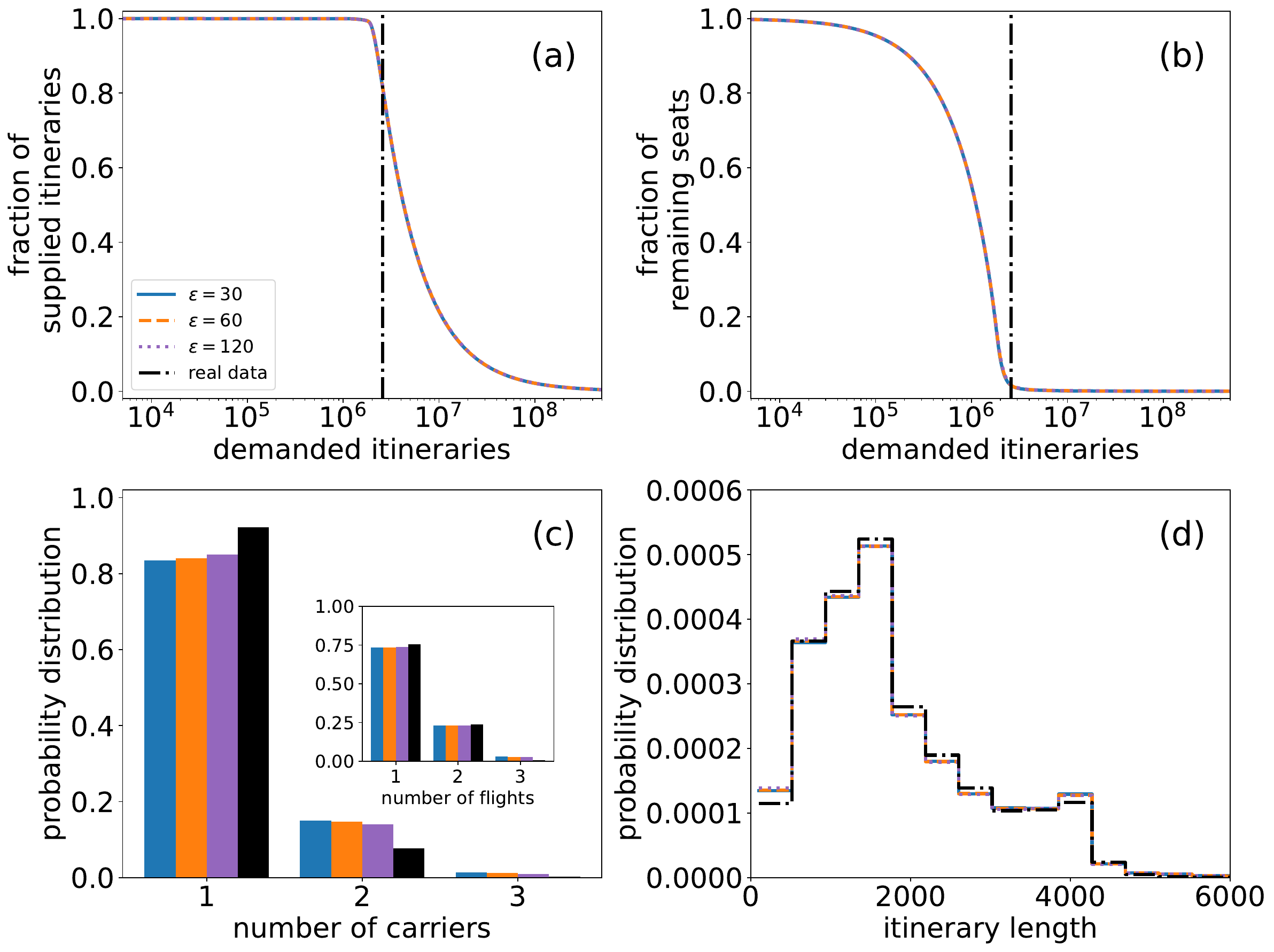}
    \caption{
    {\bf Cooperation among carriers and performance of the air transportation system.}
    Similar to Figure~\ref{fig:2}, but only considering FCNs from the flight schedule of April 18, 2023, considering full cooperation, using $\delta=30$ and different values of $\epsilon$.
    }
    \label{fig_sm:fig2-different-epsilon}
\end{figure*}


\begin{figure*}[!htb]
    \includegraphics[width=0.75\textwidth]{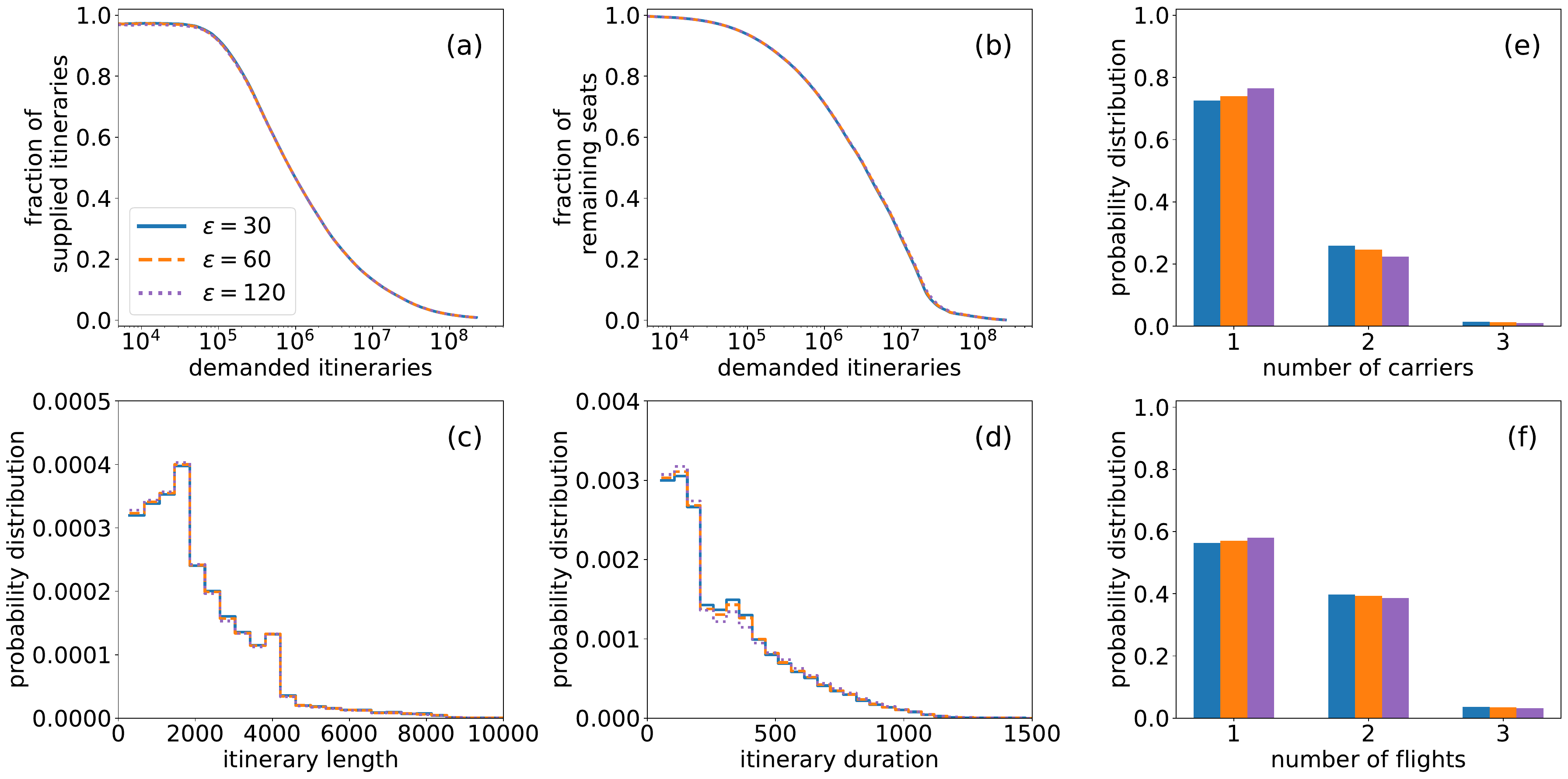}
    \caption{
    {\bf Cooperation among carriers and performance of the air transportation system.}
    Similar to Figure~\ref{fig:3}, but only considering FCNs from the flight schedule of April 18, 2023, considering full cooperation, using $\delta=30$ and different values of $\epsilon$.
    }
    \label{fig_sm:fig3-different-epsilon}
\end{figure*}

\clearpage

\begin{figure*}[!htb]
    \includegraphics[width=0.75\textwidth]{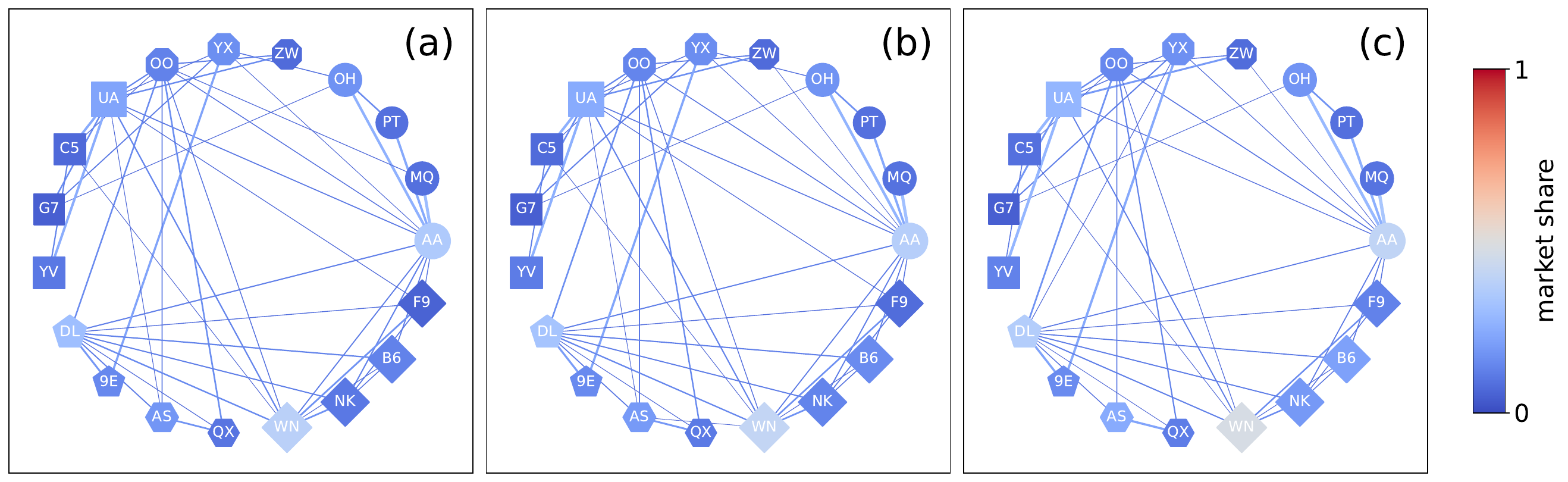}
    \caption{
    {\bf Market-share networks.}
    Similar to Figure~\ref{fig:4}, but only considering FCNs from the flight schedule of April 18, 2023, considering full cooperation, using $\delta=30$ and \textbf{(a)} $\epsilon=30$, \textbf{(b)} $\epsilon=60$, and \textbf{(c)} $\epsilon=120$.
    }
    \label{fig_sm:fig4-different-epsilon}
\end{figure*}


\begin{figure*}[!htb]
    \includegraphics[width=0.65\textwidth]{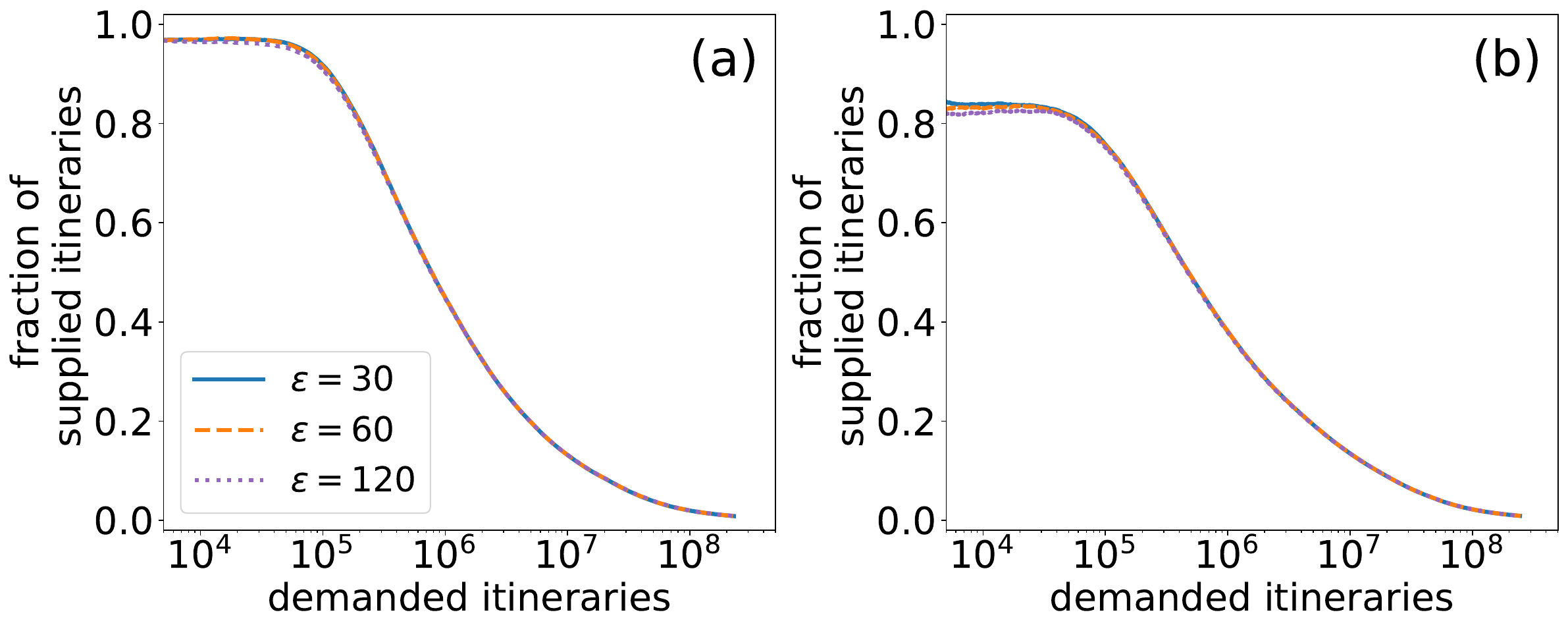}
    \caption{
    {\bf Cooperation among carriers and robustness of the air transportation system.}
    Similar to Figure~\ref{fig:5}, but only considering FCNs from the flight schedule of April 18, 2023, considering full cooperation, using $\delta=30$ and different values of $\epsilon$.
    }
    \label{fig_sm:fig5-different-epsilon}
\end{figure*}

\clearpage

\begin{figure*}[!htb]
    \includegraphics[width = 0.75\textwidth]{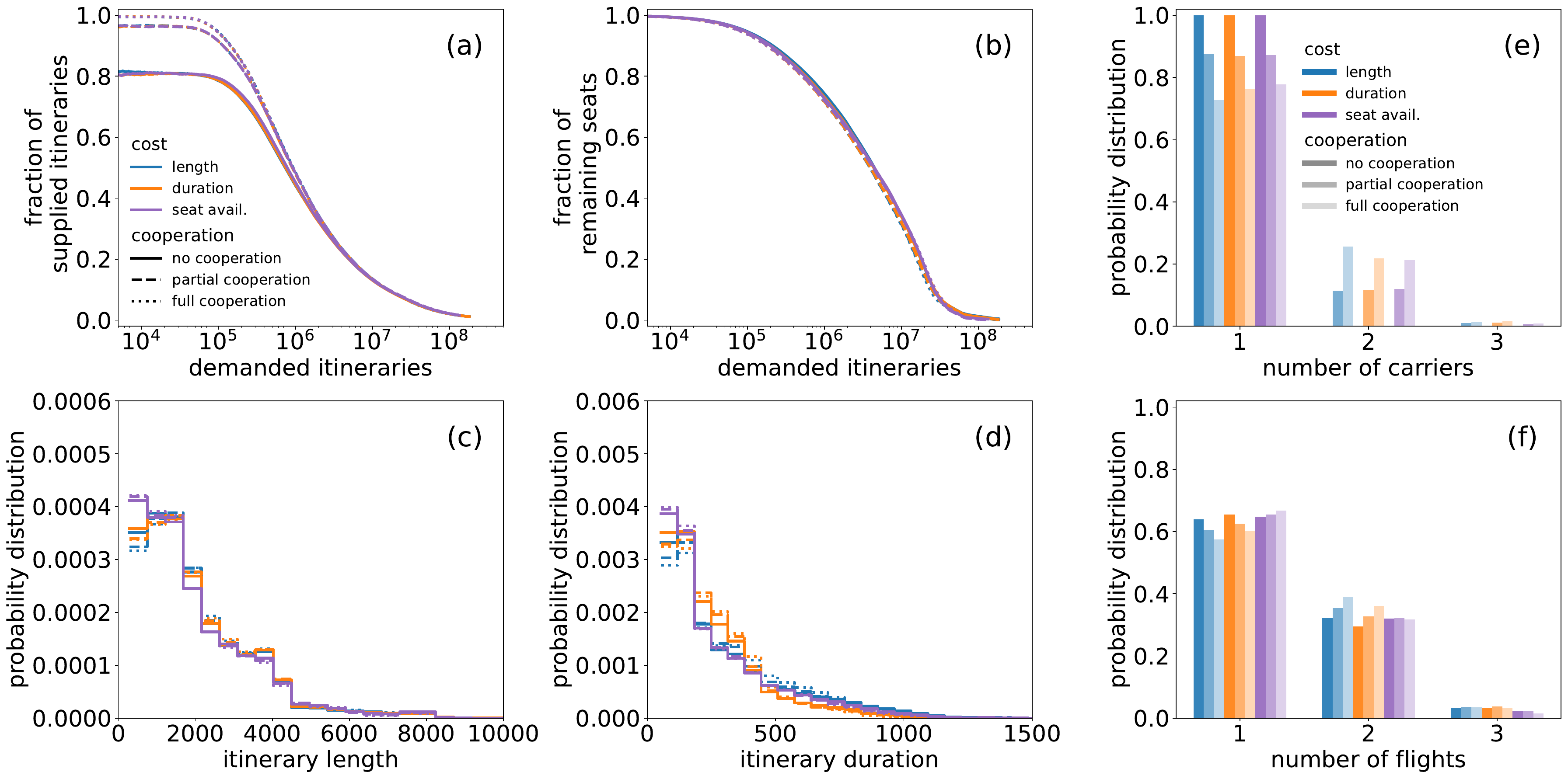}
    \caption{{\bf Cooperation among carriers and performance of the air transportation system.}
    Same as in Figure~\ref{fig:3}, but obtained from the flight schedule of April 18, 2019.
    }
    \label{fig_sm:cooperative-fcn-Y2019M4D18}
\end{figure*}


\begin{figure*}[!htb]
    \includegraphics[width = 0.75\textwidth]{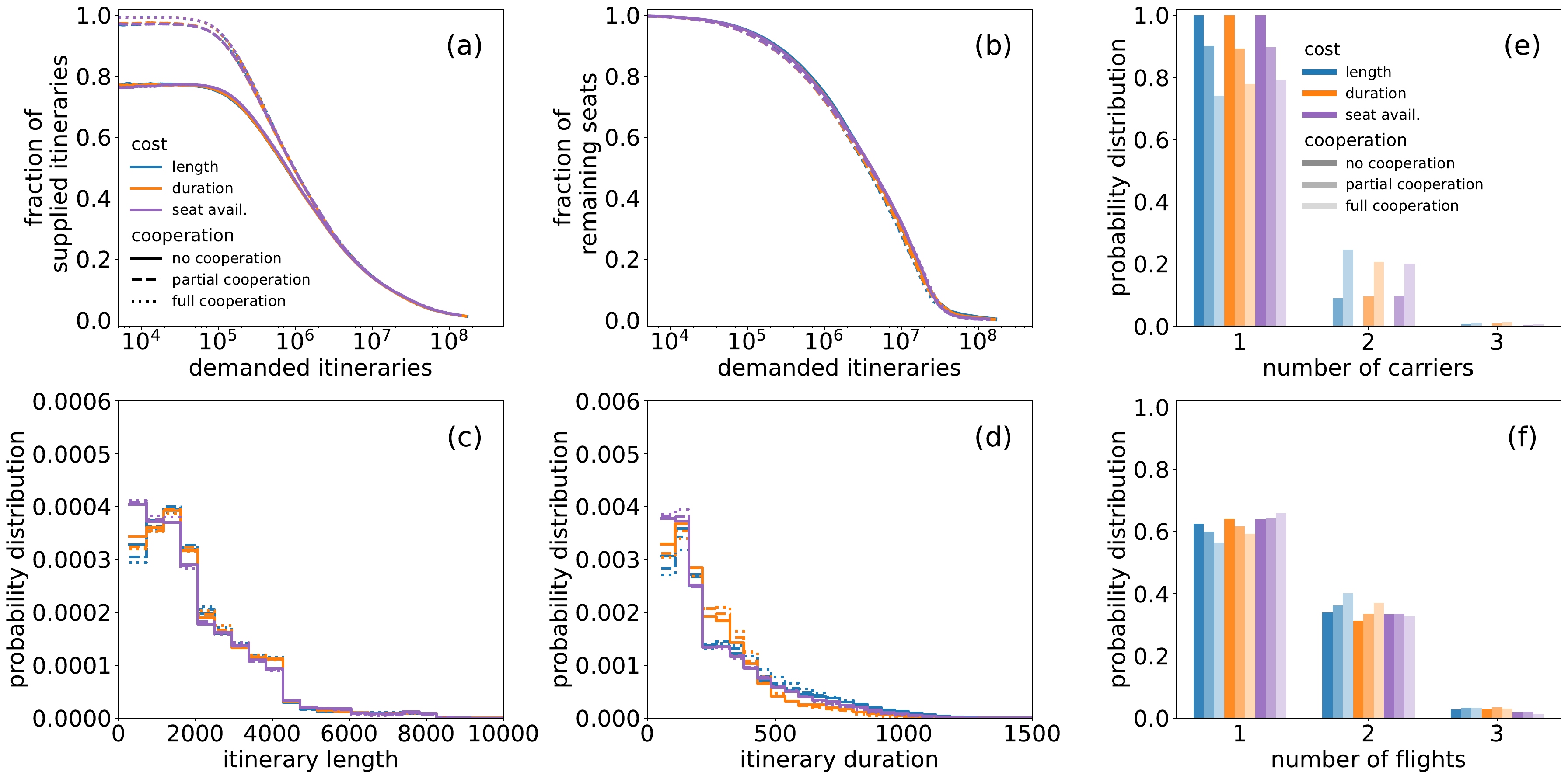}
    \caption{{\bf Cooperation among carriers and performance of the air transportation system.}
    Same as in Figure~\ref{fig:3}, but obtained from the flight schedule of November 22, 2023.
    }
    \label{fig_sm:cooperative-fcn-Y2023M11D22}
\end{figure*}

\clearpage

\begin{figure*}[!htb]
    \includegraphics[width = 0.75\textwidth]{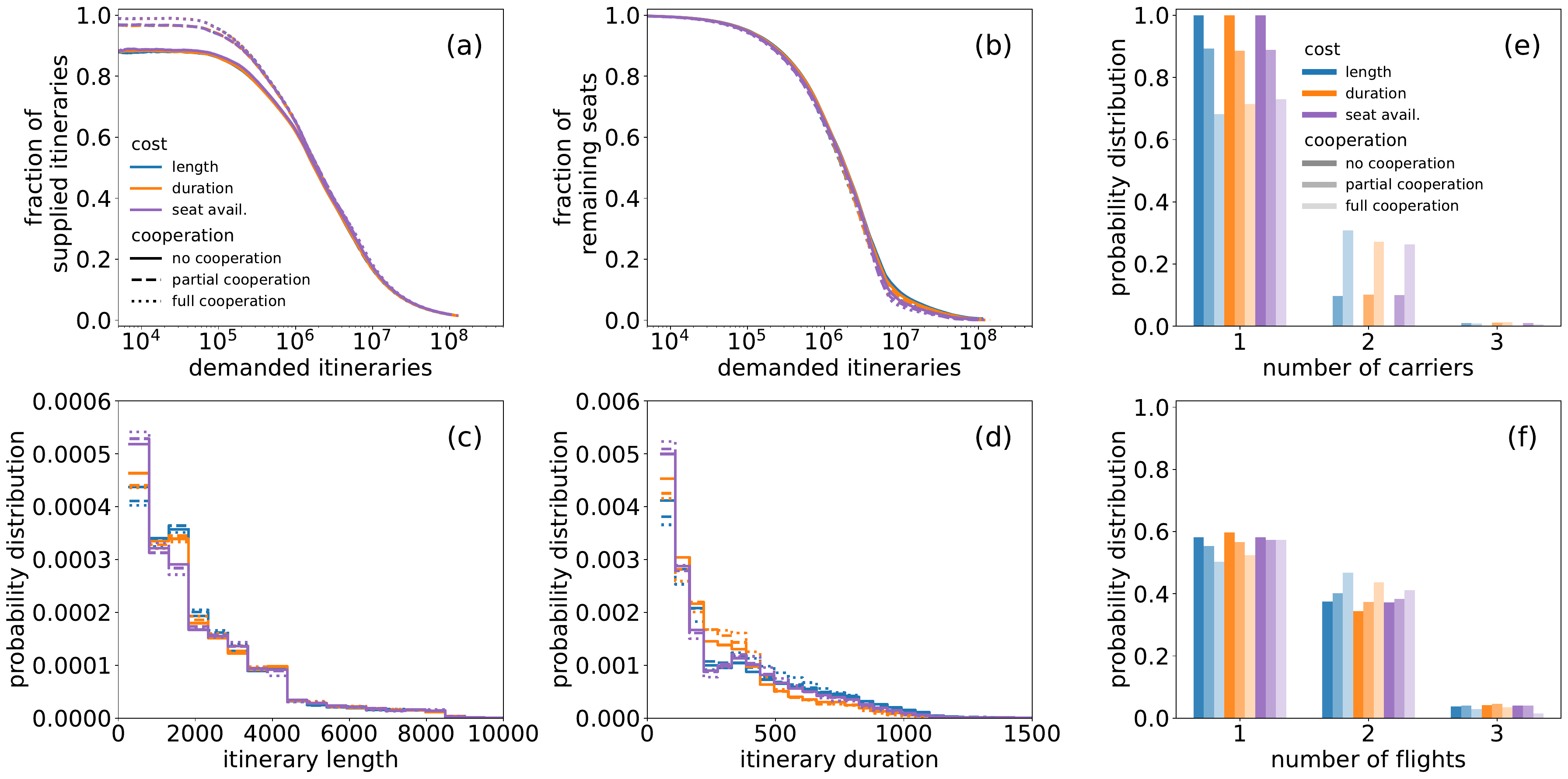}
    \caption{{\bf Cooperation among carriers and performance of the air transportation system.}
    Same as in Figure~\ref{fig:3}, but using demand based on the gravity model with parameters $\alpha=\beta=1.0$ and $\gamma=2.0$, see Eq.(~\ref{eq:gravity}).
    }
    \label{fig_sm:cooperative-fcn-gravity-A-Y2023M4D18}
\end{figure*}


\begin{figure*}[!htb]
    \includegraphics[width = 0.75\textwidth]{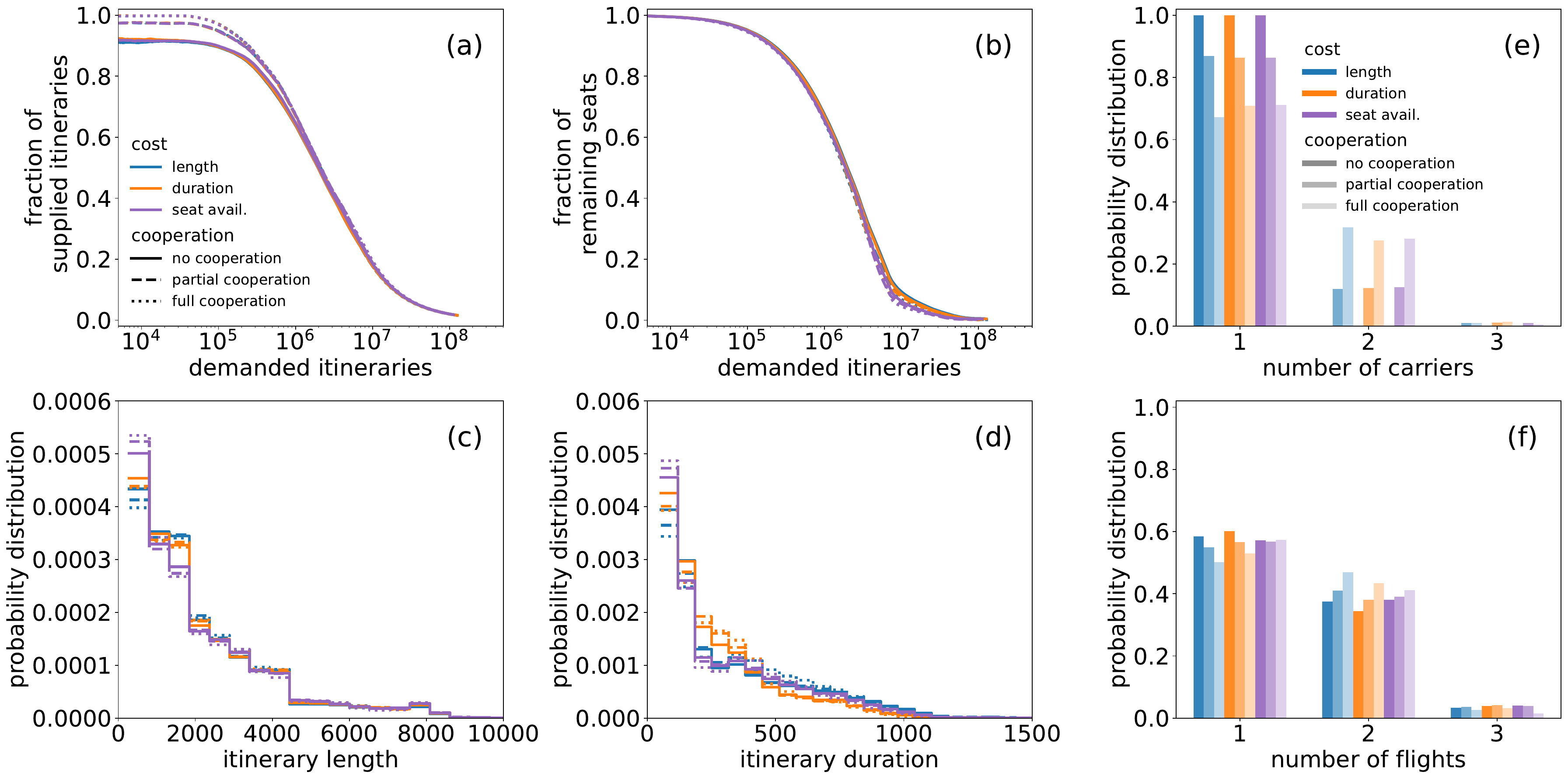}
    \caption{{\bf Cooperation among carriers and performance of the air transportation system.}
    Same as in Figure~\ref{fig:3}, but using demand based on the gravity model with parameters $\alpha=\beta=1.0$ and $\gamma=2.0$, see Eq.(~\ref{eq:gravity}). FCN is generated from the flight schedule of April 18, 2019.
    }
    \label{fig_sm:cooperative-fcn-gravity-A-Y2019M4D18}
\end{figure*}

\clearpage

\begin{figure*}[!htb]
    \includegraphics[width = 0.75\textwidth]{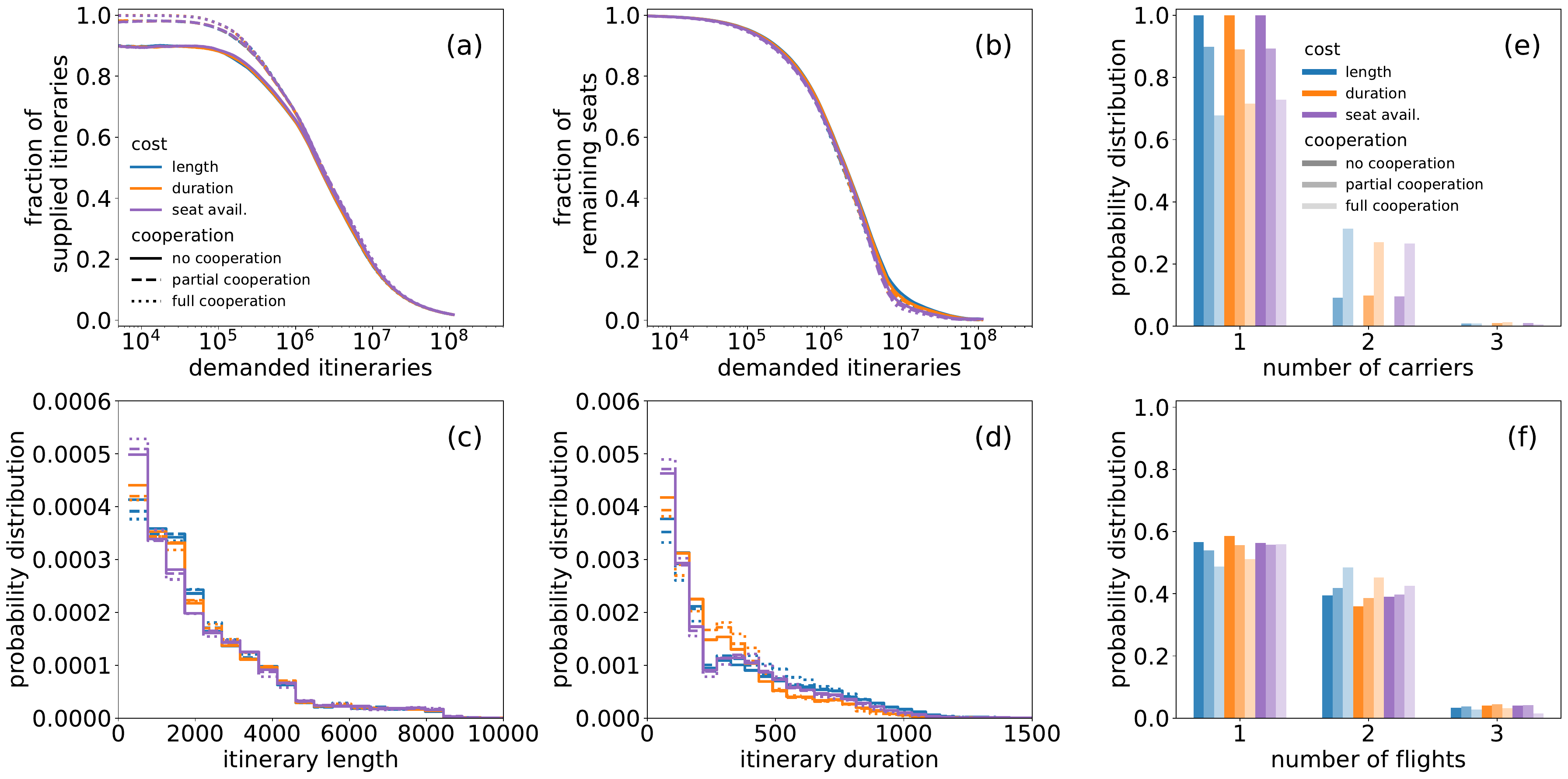}
    \caption{{\bf Cooperation among carriers and performance of the air transportation system.}
    Same as in Figure~\ref{fig:3}, but using demand based on the gravity model with parameters $\alpha=\beta=1.0$ and $\gamma=2.0$, see Eq.(~\ref{eq:gravity}). FCN is generated from the flight schedule of November 22, 2023.
    }
    \label{fig_sm:cooperative-fcn-gravity-A-Y2023M11D22}
\end{figure*}


\begin{figure*}[!htb]
    \includegraphics[width=0.75\textwidth]{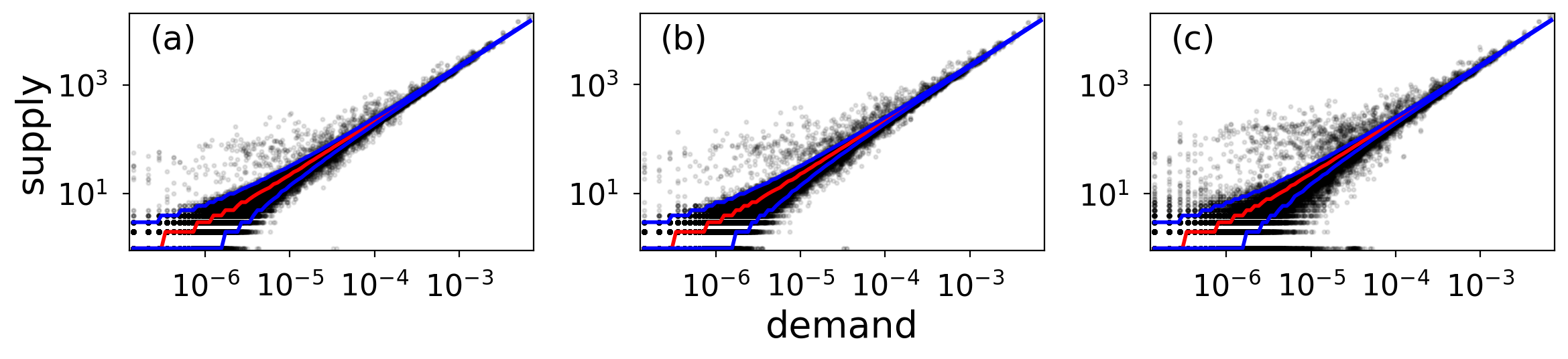}
    \caption{
    {\bf Supply {\it vs.} demand.}
    {\bf (a)} For each origin-destination pair $o \to d$ with non-null demand $w_{o \to d} > 0$, we count the number of agents $r_{o \to d}$ that were supplied with an itinerary
    connecting $o \to d$. We use here the same set of data as in Figure~\ref{fig:2} of the main paper, where demand is proxied by sold tickets and
    the minimum-cost-percolation model utilizes the length of the itinerary as the cost function to be minimized. The flight-connection network is
    obtained with no cooperation allowed among airline carriers. In the scatter plot, each point is one of a pair $o \to d$; its abscissa value is
    given by $w_{o \to d} / \sum_{r \to s} w_{r \to s}$; the ordinate value is instead $r_{o \to d} + 1$. As reference curves, we also display the median (red) and 95\% confidence intervals (blue) of the Poisson distribution obtained when success probability is given by the abscissa values and the total number of events is given by the total number of agents $R$ supplied by an itinerary, here $R =2147880$ in the specific case of these simulations. Clearly, we add one to the median and confidence values
    to make them compatible with visualized supply.
    $v = 8\%$ of the data points fall outside the $95\%$ confidence intervals.
    {\bf (b)} Same as in (a), but for the flight-connection network operating under the scenario of partial cooperation. Here, $R = 2195248$ and $v = 9\%$. {\bf (c)} Same as in (b), but for the flight-connection network operating under the scenario of full cooperation. Here, $R = 2246649$ and $v = 16\%$.
    }
    \label{fig_sm:supply-demand-DB1B-distance-Y2023M4D18}
\end{figure*}


\begin{figure*}
    \includegraphics[width=0.75\textwidth]{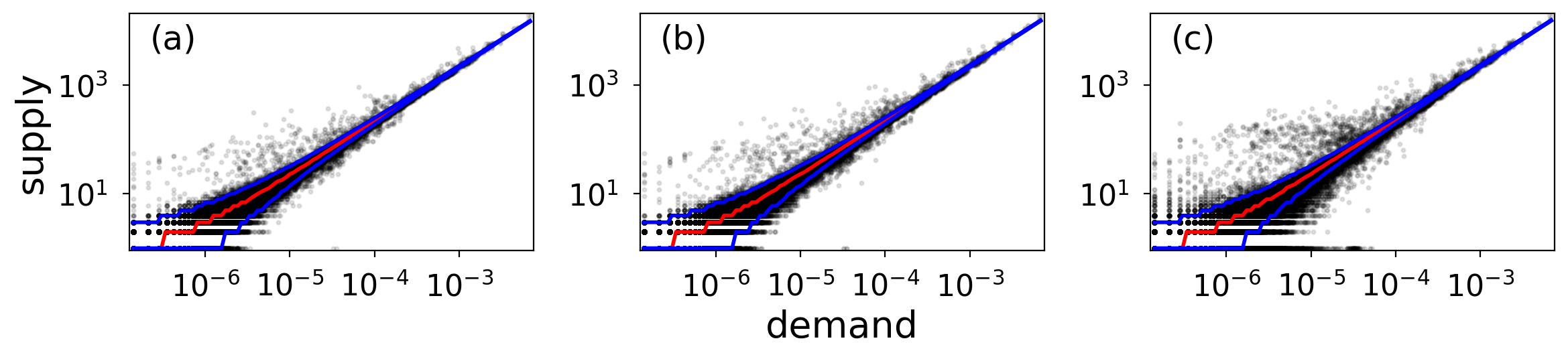}
    \caption{
    {\bf Supply {\it vs.} demand.}
    Same as Figure~\ref{fig_sm:supply-demand-DB1B-distance-Y2023M4D18}, but the minimum-cost-percolation model utilizes the duration of the itinerary as the cost function to be minimized. \textbf{(a)} $R=2189357$ and $v=7\%$, \textbf{(b)} $R=2215595$ and $v=9\%$, and \textbf{(c)} $R=2261038$ and $v=15\%$.
    }
    \label{fig_sm:supply-demand-DB1B-time-Y2023M4D18}
\end{figure*}


\begin{figure*}
    \includegraphics[width=0.75\textwidth]{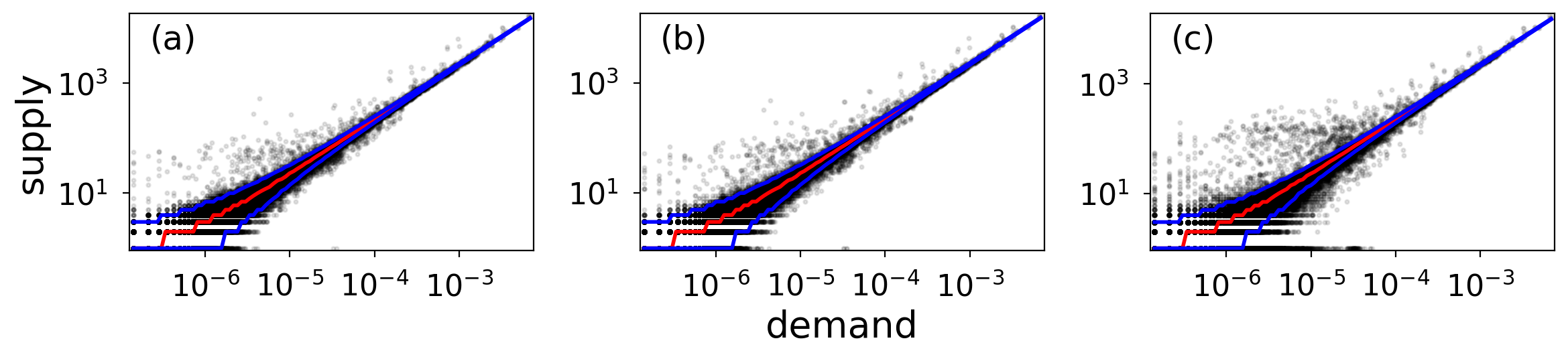}
    \caption{
    {\bf Supply {\it vs.} demand.}
    Same as Figure~\ref{fig_sm:supply-demand-DB1B-distance-Y2023M4D18}, but the minimum-cost-percolation model utilizes the seat-availability-cost of the itinerary as the cost function to be minimized. \textbf{(a)} $R=2193718$ and $v=8\%$, \textbf{(b)} $R=2194463$ and $v=8\%$, and \textbf{(c)} $R=2210792$ and $v=12\%$.
    }
    \label{fig_sm:supply-demand-DB1B-seats-Y2023M4D18}
\end{figure*}

\begin{figure*}[!htb]
    \centering
    \includegraphics[width=0.75\textwidth]{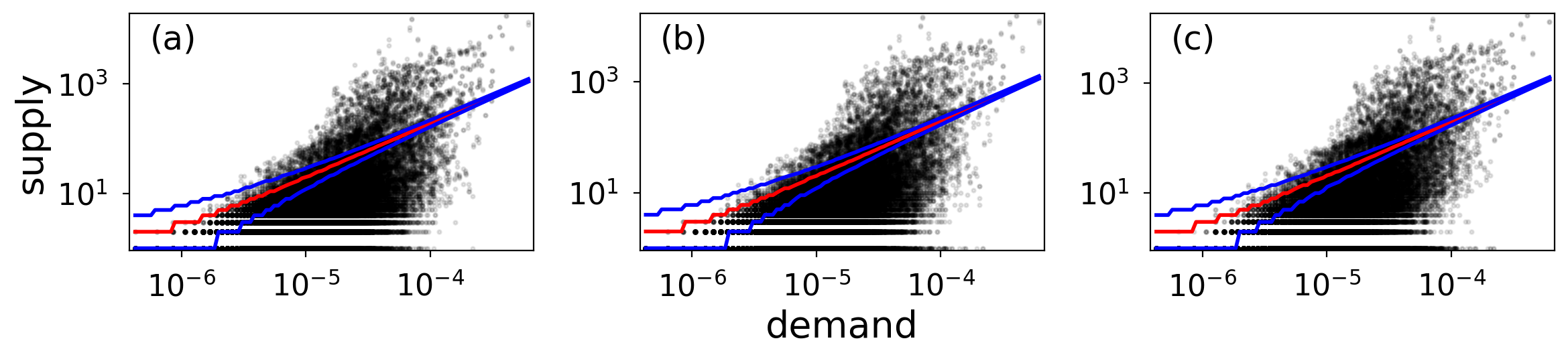}
    \caption{
    \textbf{Supply \textit{vs.} demand.}
    Same analysis as in Figure~\ref{fig_sm:supply-demand-DB1B-distance-Y2023M4D18}, but based on results from Figure~\ref{fig:3} of the main paper, where demand is estimated using the gravity model. Results remain valid for the cost function defined by itinerary length. We find that
    \textbf{(a)} for no cooperation, $R = 1904518$ and $v = 73\%$, 
    \textbf{(b)} for partial cooperation $R = 1932995$ and $v = 73\%$,
    and \textbf{(c)} for full cooperation $R = 1963913$ and $v = 77\%$.
    }
    \label{fig_sm:supply-demand-gravity_B-distance-Y2023M4D18}
\end{figure*}

\clearpage


\begin{figure*}[!htb]
    \centering
    \includegraphics[width=.75\linewidth]{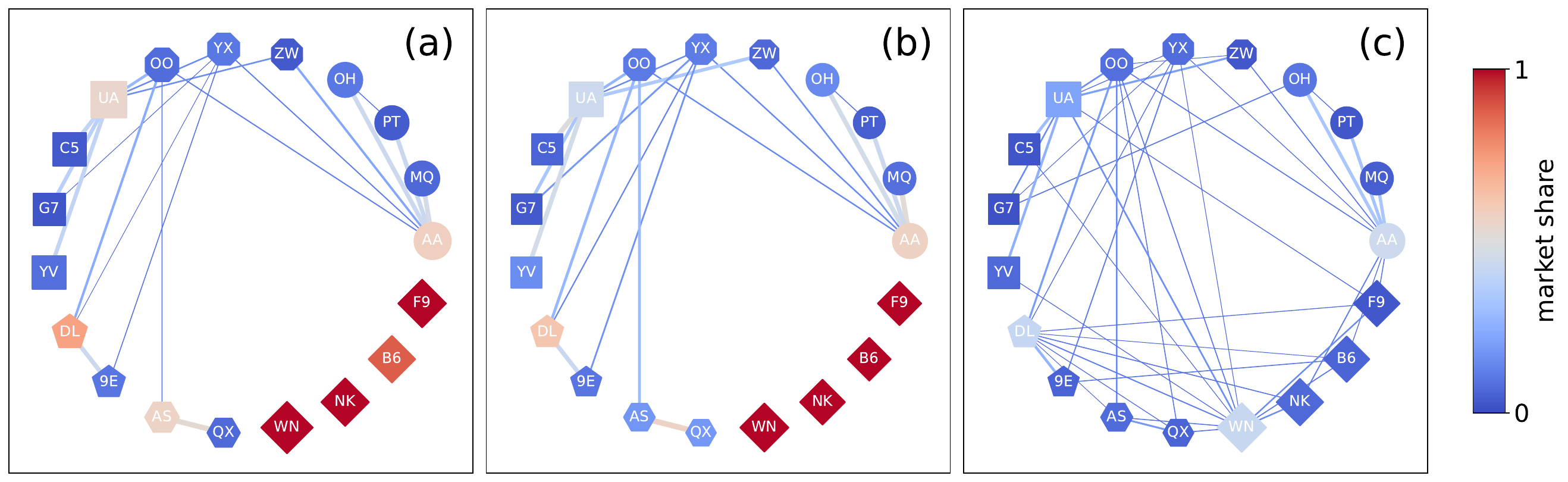}
    \caption{
    {\bf Market-share networks.} {\bf (a)} Same as in Figure~\ref{fig:4}(a). {\bf (b)} Same as in Figure~\ref{fig:4}(b), but obtained using sold tickets as a proxy for demand.
    {\bf (c)} Same as in Figure~\ref{fig:4}(c), but obtained using sold tickets as a proxy for demand.
    }
    \label{fig_sm:market_share}
\end{figure*}


\begin{figure*}[!htb]
    \centering
    \includegraphics[width=.75\linewidth]{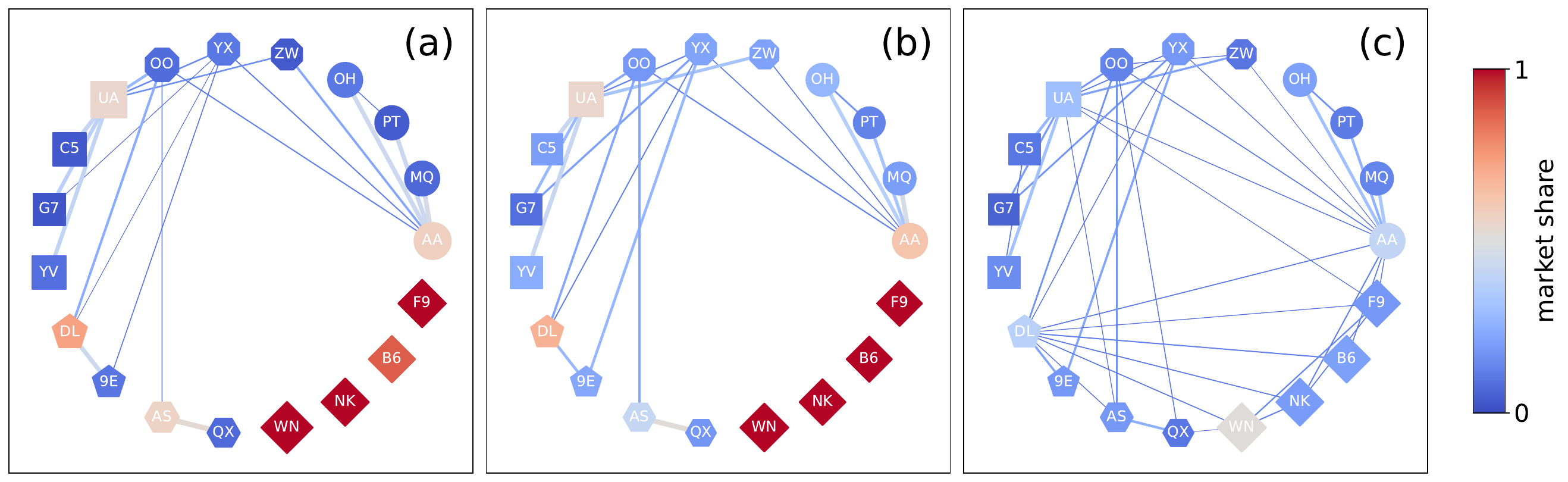}
    \caption{
    {\bf Market-share networks.} {\bf (a)} Same as in Figure~\ref{fig:4}(a). {\bf (b)} Same as in Figure~\ref{fig:4}(b), but considering itineraries selected using a minimum-duration protocol.
    {\bf (c)} Same as in Figure~\ref{fig:4}(c), but considering itineraries selected using a minimum-duration protocol.
    }
    \label{fig_sm:market-share-network-gravity-B-duration-Y2023M4D18}
\end{figure*}


\begin{figure*}[!htb]
    \centering
    \includegraphics[width=.75\linewidth]{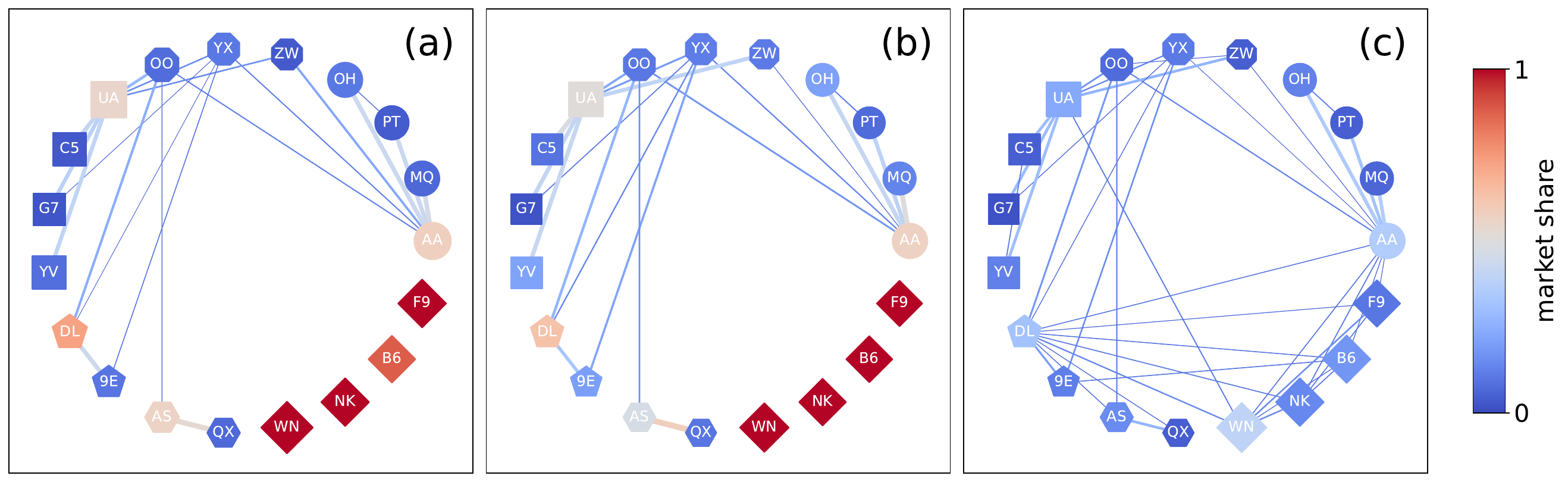}
    \caption{
    {\bf Market-share networks.} {\bf (a)} Same as in Figure~\ref{fig:4}(a). {\bf (b)} Same as in Figure~\ref{fig:4}(b), but considering itineraries selected using a minimum-seat-availability-cost protocol.
    {\bf (c)} Same as in Figure~\ref{fig:4}(c), but considering itineraries selected using a minimum-seat-availability-cost protocol.
    }
    \label{fig_sm:market-share-network-gravity-B-seats-Y2023M4D18}
\end{figure*}


\begin{figure*}[!htb]
    \centering
    \includegraphics[width=.75\linewidth]{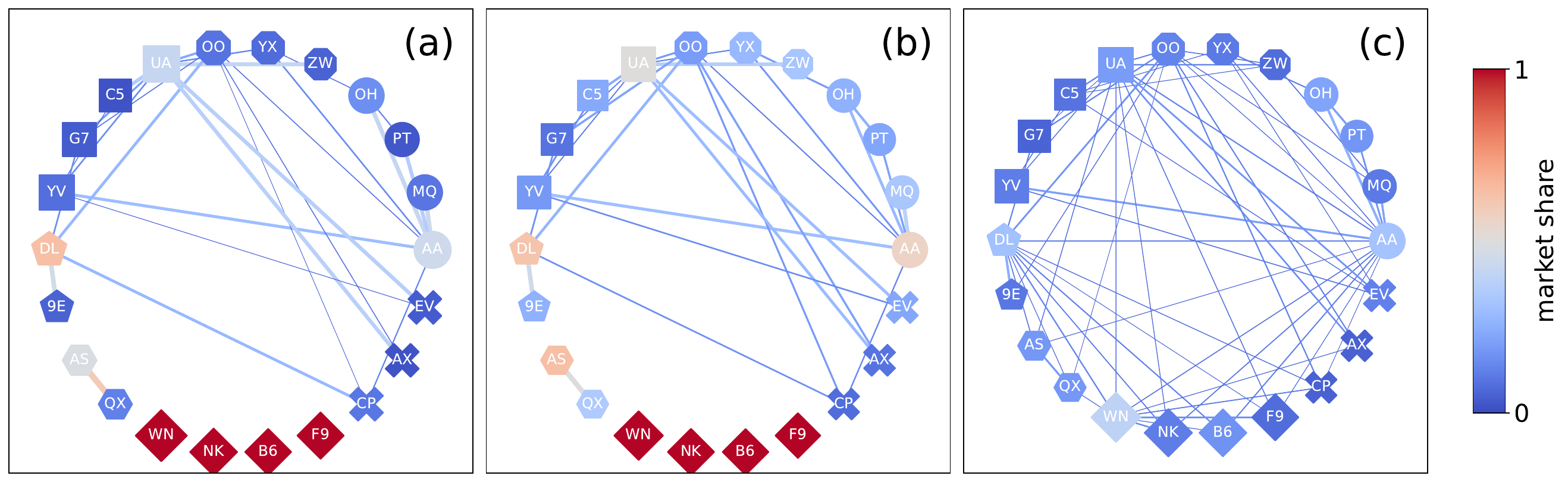}
    \caption{
    {\bf Market-share networks.}
    {\bf (a)} Same as in Figure~\ref{fig:4}(a), but using sold tickets of the second quarter in 2019. 
    {\bf (b)} Same as in Figure~\ref{fig:4}(b), but considering itineraries selected using a minimum-length protocol using FCN generated from flight schedules on April 18, 2019.
    {\bf (c)} Same as in Figure~\ref{fig:4}(c), but considering itineraries selected using a minimum-length protocol using FCN generated from flight schedules on April 18, 2019.
    Note that additional carriers were operating flights in 2019, e.g., \textit{Compass Airlines} (CP), \textit{Trans State Airlines} (AX), and \textit{ExpressJet Airlines} (EV). However, these carriers ceased their operations and are not operating any flights in 2023.}
    \label{fig_sm:market-share-network-gravity-B-length-Y2019M4D18}
\end{figure*}

\begin{figure*}[!htb]
    \centering
    \includegraphics[width=.75\linewidth]{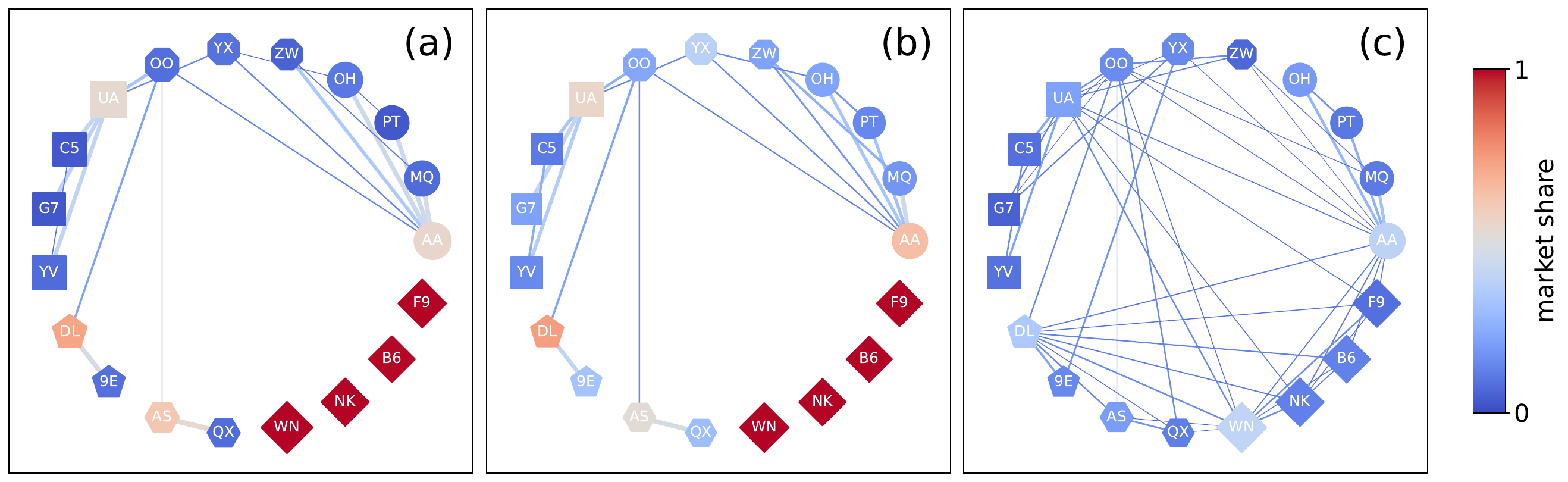}
    \caption{
    {\bf Market-share networks.} 
    {\bf (a)} Same as in Figure~\ref{fig:4}(a), but using sold tickets of the fourth quarter in 2023. 
    {\bf (b)} Same as in Figure~\ref{fig:4}(b), but considering itineraries selected using a minimum-length protocol using FCN generated from flight schedules on November 22, 2023.
    {\bf (c)} Same as in Figure~\ref{fig:4}(c), but considering itineraries selected using a minimum-length protocol using FCN generated from flight schedules on November 22, 2023.
    }
    \label{fig_sm:market-share-network-gravity-B-length-Y2023M11D22}
\end{figure*}

\begin{figure*}[!htb]
    \includegraphics[width=0.75\textwidth]{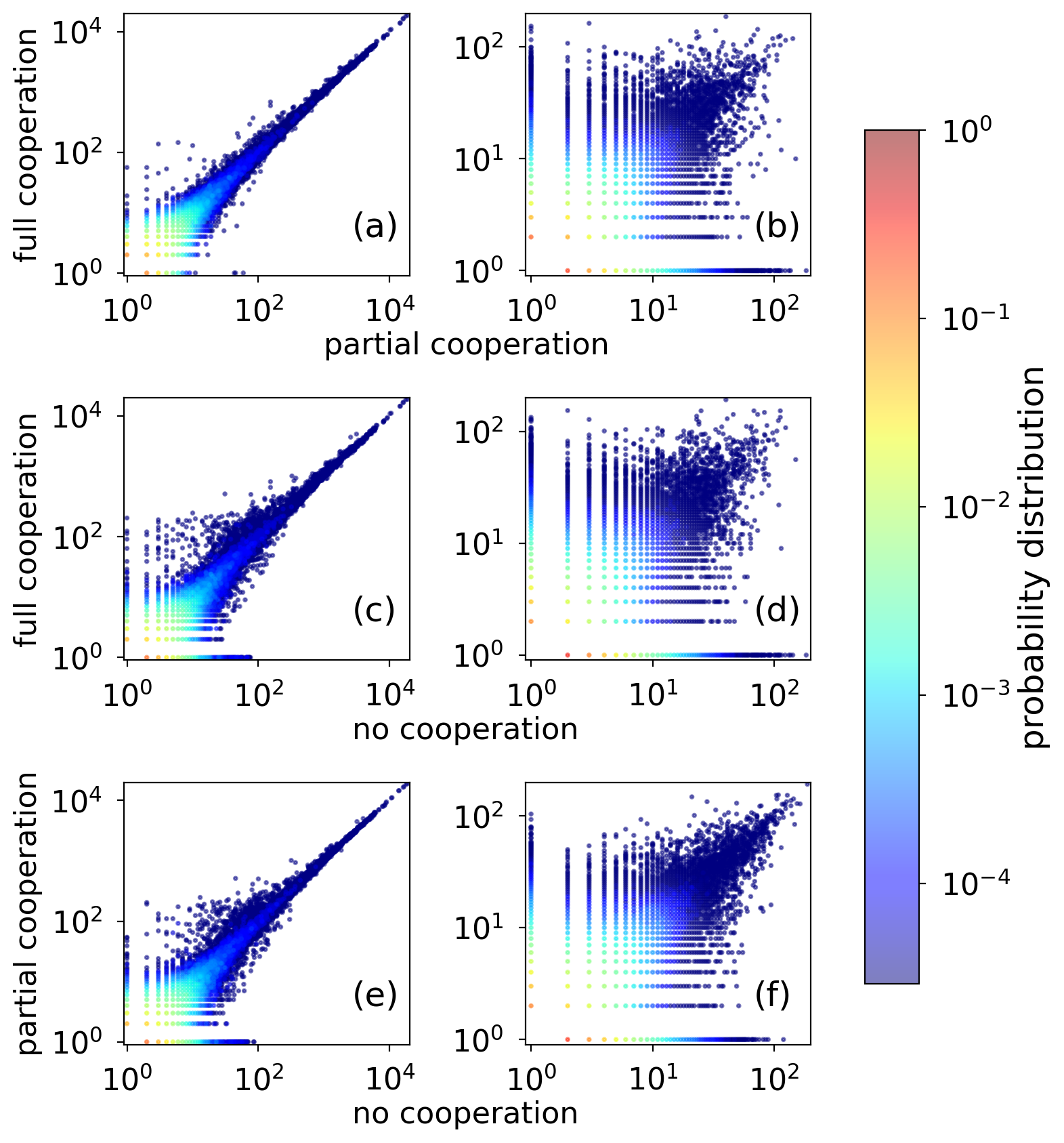}
    \caption{
    {\bf Utilization of the flight-connection network.}
    {\bf (a)} For each origin-destination pair $o \to d$, we count the number of agents $r^{(pc)}_{o \to d}$ and $r^{(fc)}_{o \to d}$ that were supplied with an itinerary
    connecting $o \to d$ under the scenarios of partial cooperation and full cooperation, respectively. 
    We use here the same set data as in Figure~\ref{fig:2} of the main paper, where demand is proxied by sold tickets and
    the minimum-cost-percolation model utilizes the length of the itinerary as the cost function to be minimized. 
    We then plot $r^{(fc)}_{o \to d}+1$ {\it vs.} $r^{(pc)}_{o \to d}+1$. We  color each point in the graph depending on the actual fraction of points with such specific abscissa and ordinate values that are in the sample and the total number of points in the sample. Note that only pairs with for which $\min \{ r^{(fc)}_{o \to d}, r^{(pc)}_{o \to d} \} > 0$ are considered in this analysis. {\bf (b)} We use the same data as in (a), but we estimate the utilization of the connection between flights $f$ and $g$, i.e., $u_{f \to g}$ as defined in Eq.~(\ref{eq:weight_flight}). Also here, we compare for each specific pair $f \to g$, the metric in the scenarios of partial and full cooperation, respectively $u^{(pc)}_{f \to g}$
    $u^{(fc)}_{f \to g}$, and the plot $u^{(fc)}_{f \to g} + 1$ {\it vs.} $u^{(pc)}_{f \to g} + 1$.  We color each point in the graph depending on the actual fraction of points with such specific abscissa and ordinate values that are in the sample and the total number of points in the sample. Note that only pairs with for which $\min \{ u^{(fc)}_{f \to g}, u^{(pc)}_{f \to g} \} > 0$ are considered in this analysis. {\bf (c)} and {\bf (d)} Same as in (a) and (b), respectively, but here the comparison is between the scenarios of full cooperation and no cooperation. {\bf (e)} and {\bf (f)} Same as in (a) and (b), respectively, but here the comparison is between the scenarios of partial cooperation and no cooperation.
    }
    \label{fig_sm:scatter-plots-DB1B-distance-Y2023M4D18}
\end{figure*}

\clearpage

\begin{figure*}[!htb]
    \includegraphics[width=0.75\textwidth]{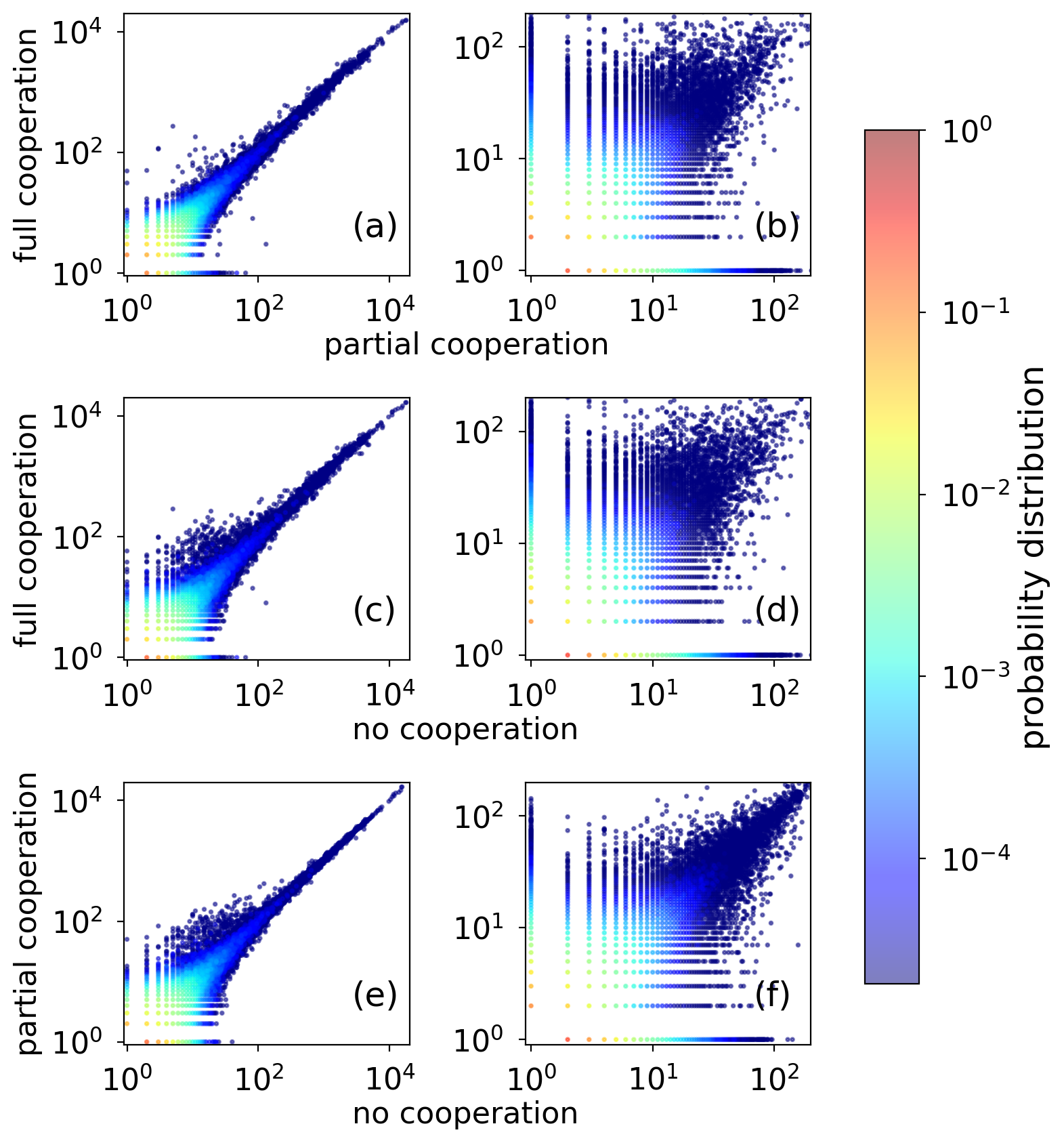}
    \caption{
    {\bf Utilization of the flight-connection network.}
     Same analysis as in Figure~\ref{fig_sm:scatter-plots-DB1B-distance-Y2023M4D18}, but for the results of Figure~\ref{fig:3} of the main paper, where demand is estimated according to the gravity model with parameters $\alpha=\beta=0.5$ and $\gamma=1.0$ and the minimum-cost-percolation model utilizes the length of the itinerary as the cost function to be minimized. 
    }
    \label{fig_sm:scatter-plots-gravity_B-distance-Y2023M4D18}
\end{figure*}

\begin{figure*}[!htb]
    \includegraphics[width=0.75\textwidth]{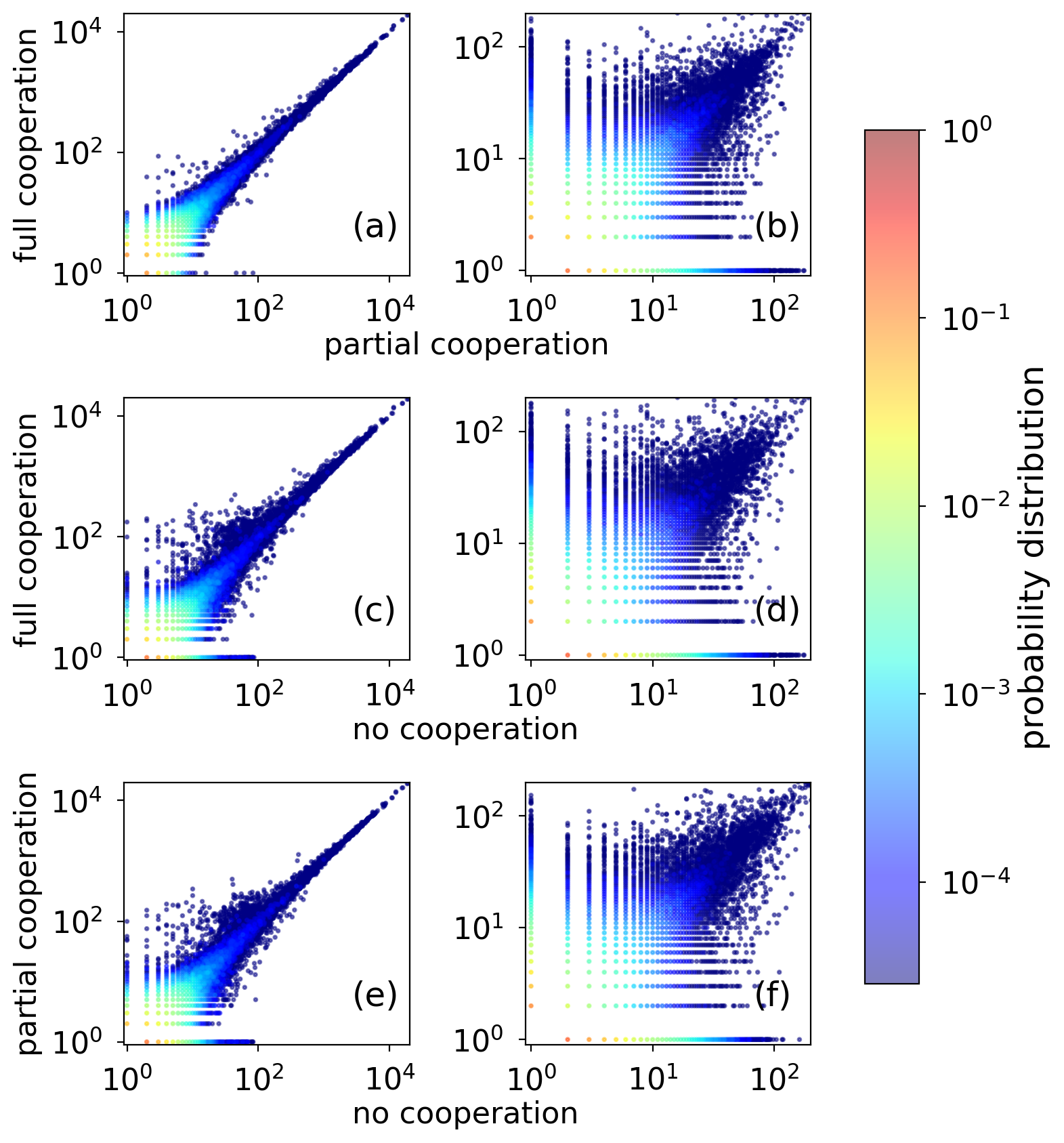}
    \caption{
    {\bf Utilization of the flight-connection network.}
    Same as Figure~\ref{fig_sm:scatter-plots-DB1B-distance-Y2023M4D18}, but the minimum-cost-percolation model utilizes the duration of the itinerary as the cost function to be minimized. 
    }
    \label{fig_sm:scatter-plots-DB1B-time-Y2023M4D18}
\end{figure*}

\clearpage

\begin{figure*}[!htb]
    \includegraphics[width=0.75\textwidth]{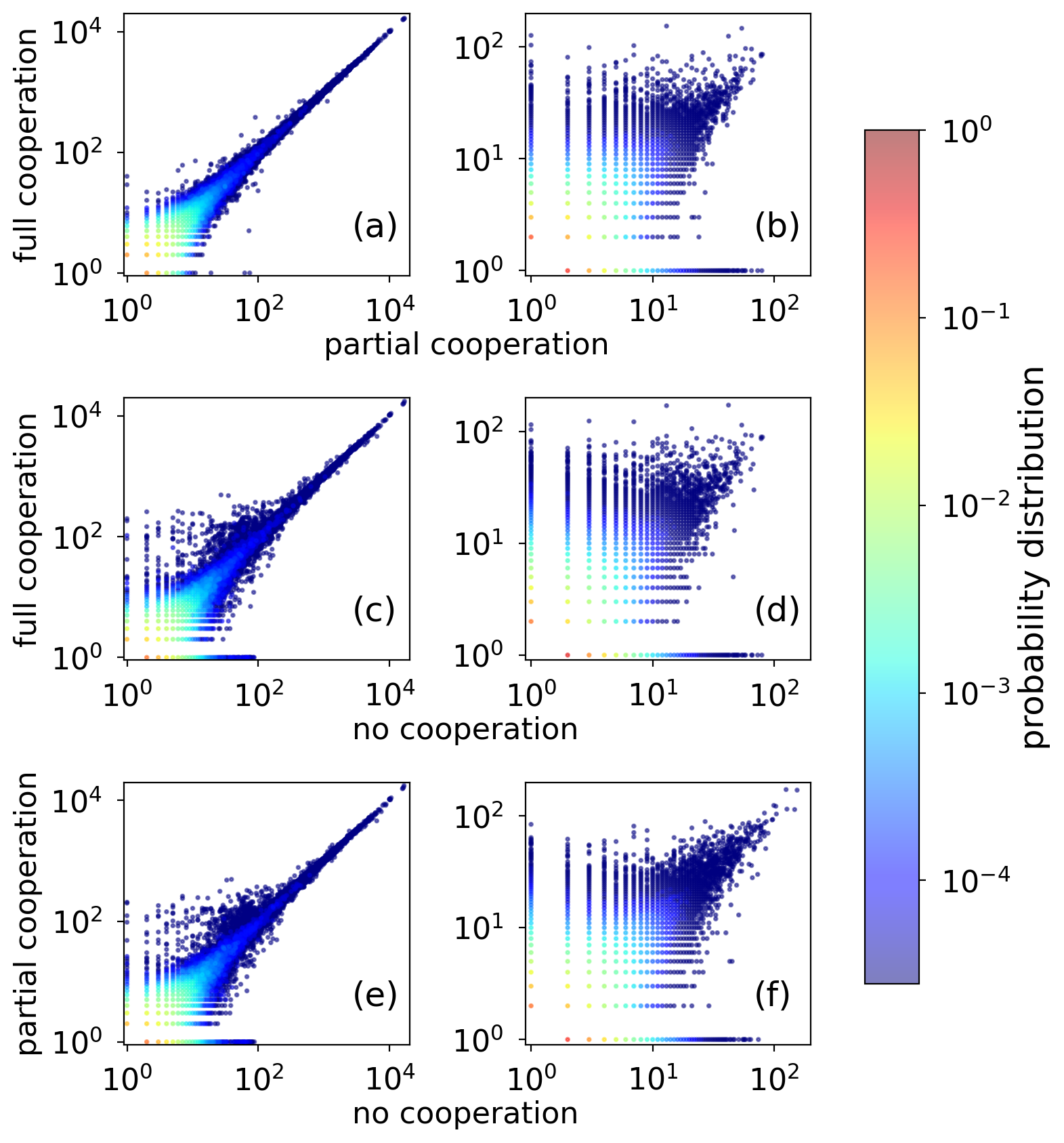}
    \caption{
    {\bf Utilization of the flight-connection network.}
    Same as Figure~\ref{fig_sm:scatter-plots-DB1B-distance-Y2023M4D18}, but the minimum-cost-percolation model utilizes the seat-availability-cost of the itinerary as the cost function to be minimized. 
    }
    \label{fig_sm:scatter-plots-DB1B-seats-Y2023M4D18}
\end{figure*}

\clearpage


\begin{figure*}[!htb]
   \includegraphics[width=0.85\textwidth]{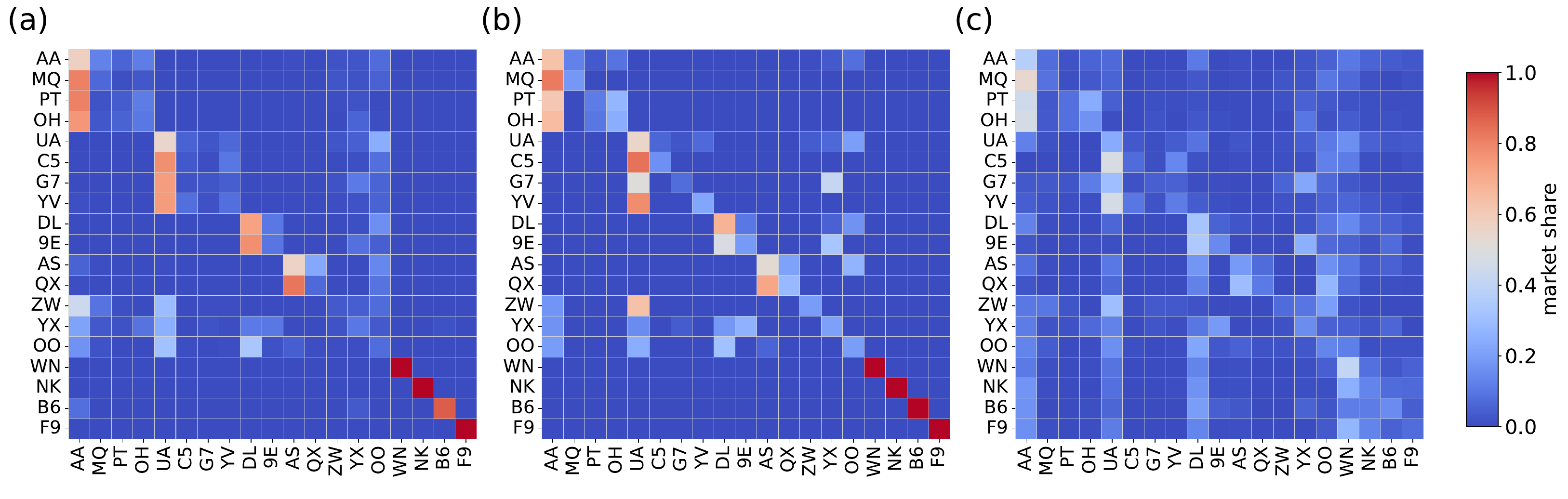}
    \caption{
    {\bf Adjacency matrix of the market-share networks.}
    {\bf (a)} The entry in row $c$ and column $c'$ in the matrix is proportional to the number of two-flight itineraries in which one flight was operated by carrier $c$ and the other by carrier $c'$. Entries are normalized by row, see Eq.~(\ref{eq:market-share}). Data here refer to tickets sold in the second quarter of 2023. Only carriers operating at least a flight on April 18, 2023 are considered in the matrix. {\bf (b)} Same as in (a), but for two-flight itineraries generated by the MCP model. We use the same data as in Figure~\ref{fig:3} where agents in the MCP model optimize the length of the itineraries, considering partial cooperation. \textbf{(c)} Same as in (b), but for FCN considering full cooperation. 
    }
    \label{fig_sm:market-share-adjacency-matrix-gravity-B-Y2023M4D18}
\end{figure*}


\begin{figure*}[!htb]
    \includegraphics[width=0.85\textwidth]{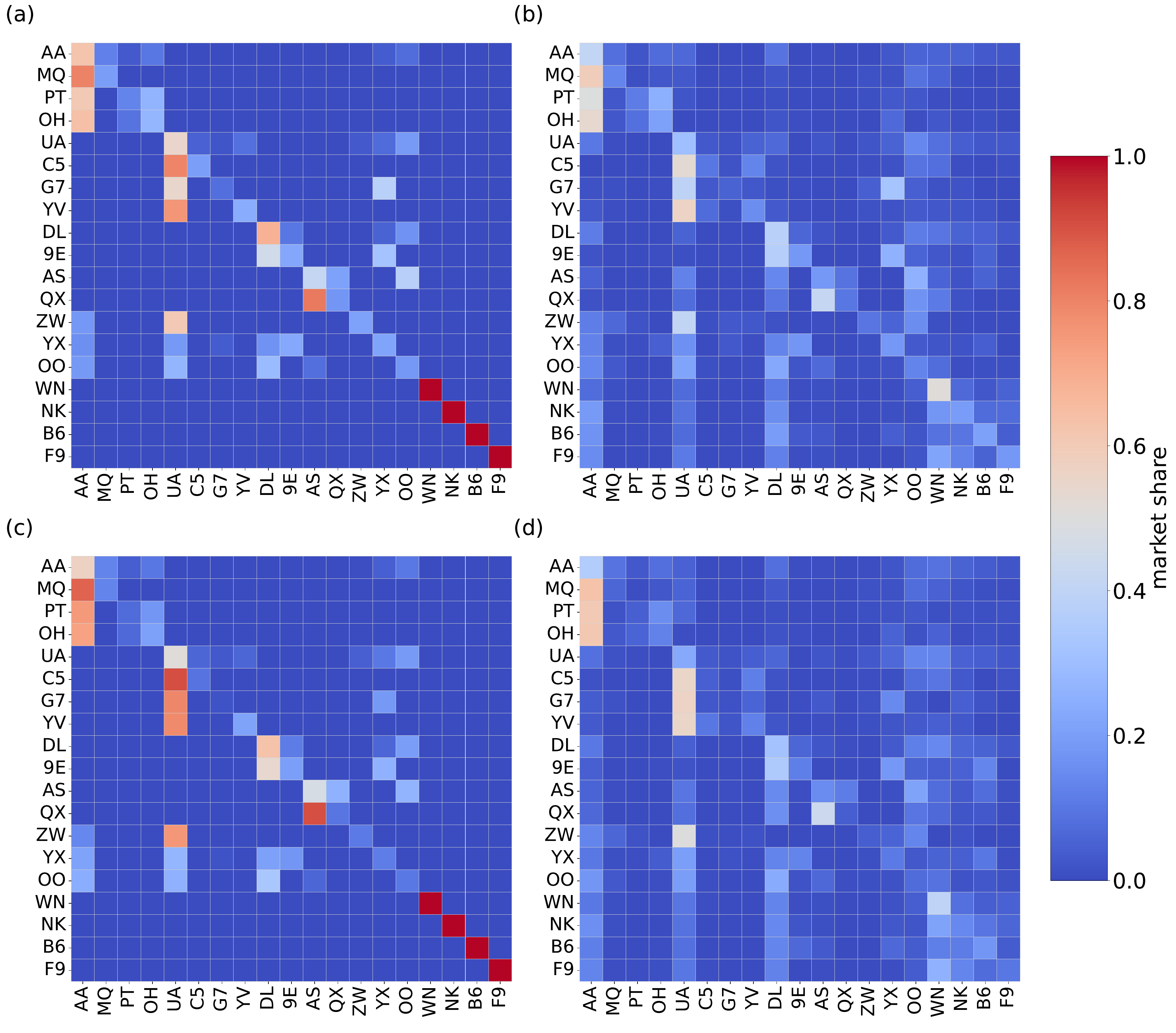}
    \caption{
    {\bf Adjacency matrix of the market-share networks.}
    {\bf (a)}  Same as in Figure~\ref{fig_sm:market-share-adjacency-matrix-gravity-B-Y2023M4D18}(b), but obtained from results of the MCP model where agents optimize the duration of the itineraries. 
    {\bf (b)}  Same as in Figure~\ref{fig_sm:market-share-adjacency-matrix-gravity-B-Y2023M4D18}(c), but obtained from results of the MCP model where agents optimize the duration of the itineraries. 
    {\bf (c)}  Same as in Figure~\ref{fig_sm:market-share-adjacency-matrix-gravity-B-Y2023M4D18}(b), but obtained from results of the MCP model where agents optimize the seat-availability-based cost function. 
    {\bf (d)}  Same as in Figure~\ref{fig_sm:market-share-adjacency-matrix-gravity-B-Y2023M4D18}(c), but obtained from results of the MCP model where agents optimize the seat-availability-based cost function.
    }
    \label{fig_sm:market-share-adjacency-matrix-time-seats-Y2023M4D18}
\end{figure*}


\begin{figure*}[!htb]
    \includegraphics[width=0.85\textwidth]{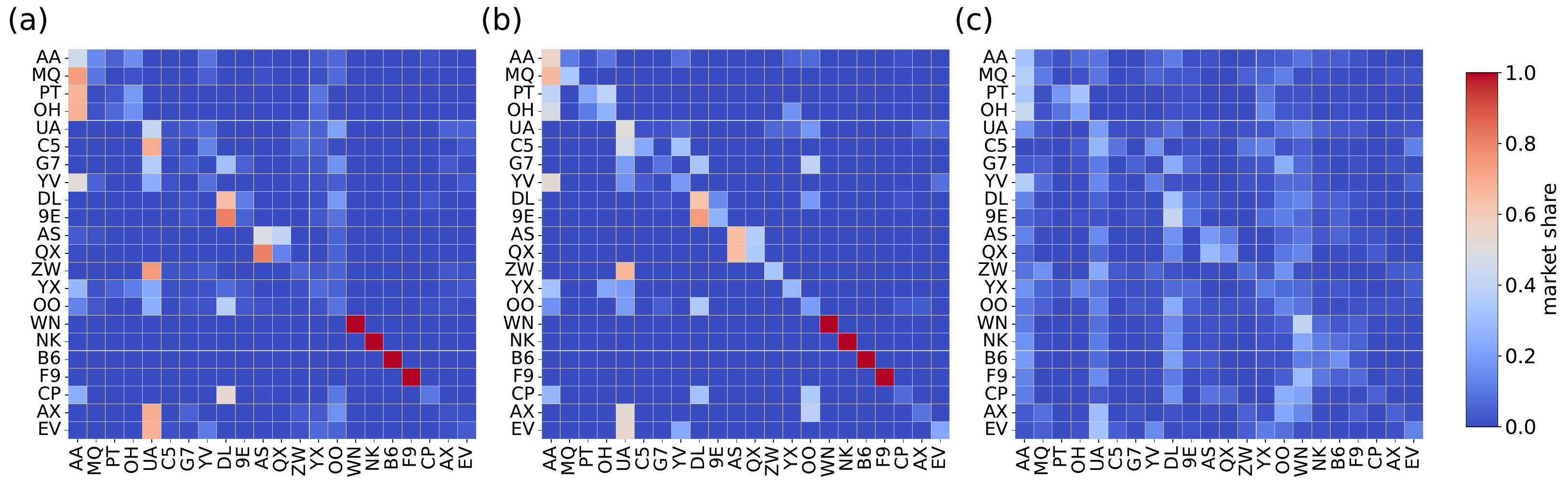}
    \caption{
    {\bf Adjacency matrix of the market-share networks.}
    {\bf (a)} Same as in Figure~\ref{fig_sm:market-share-adjacency-matrix-gravity-B-Y2023M4D18}(a), but obtained using data about sold tickets of the second quarter of 2019. Only carriers operating at least a flight on April 18, 2019 are considered in the matrix. 
    {\bf (b)} Same as in Figure~\ref{fig_sm:market-share-adjacency-matrix-gravity-B-Y2023M4D18}(b), but obtained using data from the MCP model with the flight schedule of April 18, 2019.
    {\bf (c)} Same as in Figure~\ref{fig_sm:market-share-adjacency-matrix-gravity-B-Y2023M4D18} (c), but obtained using data from the MCP model with the flight schedule of April 18, 2019.
    We use the same data as in Figure~\ref{fig_sm:cooperative-fcn-Y2019M4D18} where agents in the MCP model optimize the length of the itineraries. 
    Note that additional carriers were operating flights in 2019, e.g., \textit{Compass Airlines} (CP), \textit{Trans State Airlines} (AX), and \textit{ExpressJet Airlines} (EV). However, these carriers ceased their operations and are not operating any flights in 2023.
    %
    }
    \label{fig_sm:market-share-adjacency-matrix-gravity-B-Y2019M4D18}
\end{figure*}



\begin{figure*}[!htb]
    \includegraphics[width=0.85\textwidth]{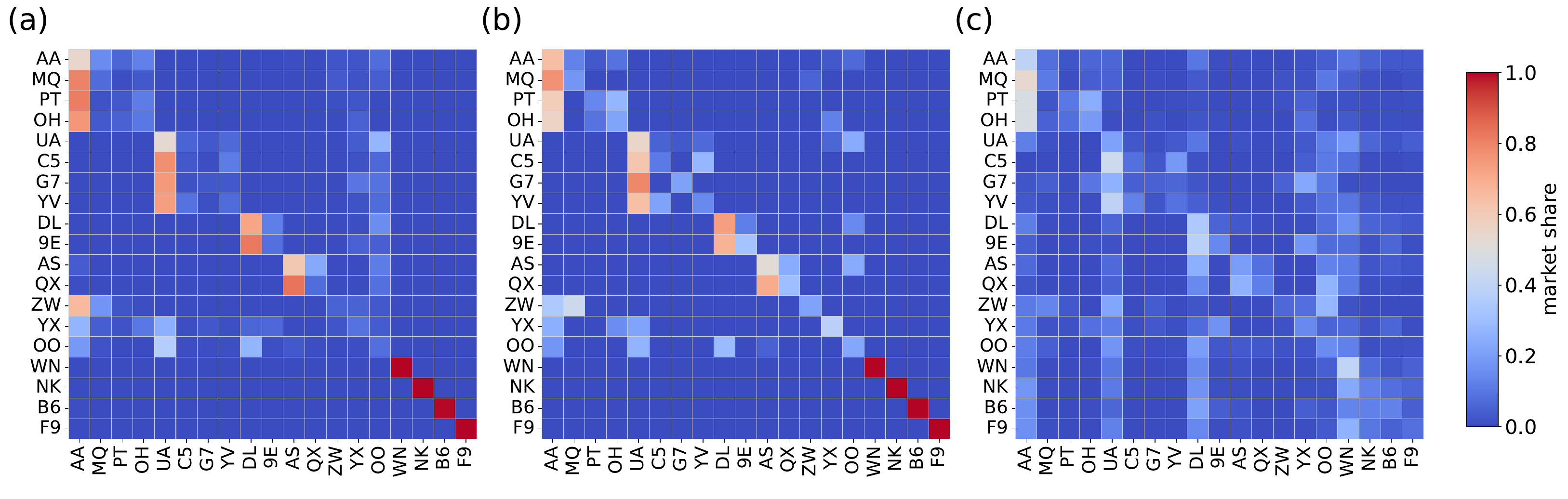}
    \caption{
    {\bf Adjacency matrix of the market-share networks.}
    {\bf (a)} Same as in Figure~\ref{fig_sm:market-share-adjacency-matrix-gravity-B-Y2023M4D18}(a), but obtained using data about sold tickets of the fourth quarter of 2023. Only carriers operating at least a flight on November 22, 2023 are considered in the matrix. {\bf (b)} Same as in Figure~\ref{fig_sm:market-share-adjacency-matrix-gravity-B-Y2023M4D18}(b), but obtained using data from the MCP model with the flight schedule of November 22, 2023. 
    {\bf (c)} Same as in Figure~\ref{fig_sm:market-share-adjacency-matrix-gravity-B-Y2023M4D18} (c), but obtained using data from the MCP model with the flight schedule of November 22, 2023.
    We use the same data as in Figure~\ref{fig_sm:cooperative-fcn-gravity-A-Y2023M11D22} where agents in the MCP model optimize the length of the itineraries. 
    }
    \label{fig_sm:market-share-adjacency-matrix-gravity-B-Y2023M11D22}
\end{figure*}

\end{document}